\documentclass[final]{aa}
\usepackage{epsf}
\def\ave#1{\langle #1 \rangle}
\def\eck#1{\left\lbrack #1 \right\rbrack}
\def\rund#1{\left( #1 \right)}
\def \la {\mathrel{\vcenter
     {\offinterlineskip \hbox{$<$}\hbox{$\sim$}}}}
\def \ga {\mathrel{\vcenter
     {\offinterlineskip \hbox{$>$}\hbox{$\sim$}}}}

\begin{document}

\thesaurus{06(13.07.1, 02.05.1, 08.14.1, 08.02.1, 02.08.1)}

\title{Gamma-ray bursts from accreting black holes in neutron star mergers}

\author{M.~Ruffert\inst{1}\thanks{e-mail: {\tt mruffert@ast.cam.ac.uk}}
\and H.-Th.~Janka\inst{2}\thanks{e-mail: {\tt thj@mpa-garching.mpg.de}}
}
\institute{Institute of Astronomy, Madingley Road, Cambridge~CB3~0HA, U.K.
\and Max-Planck-Institut f\"ur Astrophysik, Postfach 1523, 
D-85740 Garching, Germany} 
\offprints{H.-Th.~Janka}


\titlerunning{Gamma-ray bursts from neutron star mergers}
\maketitle

\begin{abstract}
By means of three-dimensional hydrodynamic simulations with a Eulerian 
PPM code we
investigate the formation and the properties of the accretion torus 
around the stellar mass black hole which originates from the 
merging of two neutron stars. The simulations are performed
with four nested cartesian grids which allow for both a good 
resolution near the central black hole and a large computational volume.
They include the use of a physical equation of state as well as the 
neutrino emission from the hot matter of the torus. The gravity
of the black hole is described with a Newtonian and alternatively with
a Paczy\'nski-Wiita potential. In a post-processing step, we evaluate
our models for the energy deposition by $\nu\bar\nu$ annihilation around
the accretion torus. 

We find that the torus has a mass between several $10^{-2}\,M_{\odot}$
and a few $10^{-1}\,M_{\odot}$ with maximum densities around 
$10^{12}\,{\rm g\,cm}^{-3}$ and maximum temperatures of about 
$10\,$MeV (entropies around 5$\,k_{\rm B}$ per nucleon). Correspondingly,
the neutrino emission is huge with a total luminosity near
$10^{53}\,{\rm erg\,s}^{-1}$. Neutrino-antineutrino annihilation 
deposits energy in the vicinity of the torus at a rate of 
(3--$5)\times 10^{50}\,{\rm erg\,s}^{-1}$. It is most efficient near the
rotation axis where 10 to 30\% of this energy or up to 
a total of $10^{49}\,{\rm erg}$
are dumped within an estimated emission period of 0.02--0.1$\,$s
in a region with a low integral baryonic mass of
about $10^{-5}\,M_{\odot}$. This baryon pollution is still dangerously
high, and the estimated maximum relativistic Lorentz factors
$\Gamma-1$ are around unity. The conversion of neutrino energy into a
pair plasma, however, is sufficiently powerful to blow out the baryons
along the axis so that a clean funnel should be produced within only 
milliseconds. Our models show that $\nu\bar\nu$ annihilation can yield
the energy to account for weak, short gamma-ray bursts, if moderate
beaming is involved. In fact, the barrier of the dense baryonic gas of
the torus suggests that the low-density $e^\pm\gamma$ plasma is beamed
as axial jets 
into a fraction $f_{\Omega} = 2\delta\Omega/(4\pi)$ between $1/100$
and $1/10$ of the sky, corresponding to opening half-angles of roughly
ten to several tens of degrees. Thus $\gamma$-burst energies of 
$E_{\gamma}\approx E_{\nu\bar\nu}/f_{\Omega}\la
10^{50}$--$10^{51}\,$erg seem within the reach of our models
(if the source is interpreted as radiating isotropically),
corresponding to luminosities around $10^{51}\,{\rm erg\,s}^{-1}$ for
typical burst durations of 0.1--1$\,$s. Gravitational capture of 
radiation by the black hole, redshift and ray bending do not reduce 
the jet energy
significantly, because most of the neutrino emission comes from 
parts of the torus at distances of several Schwarzschild radii
from the black hole. Effects associated with the Kerr character of
the rapidly rotating black hole, however, could increase the 
$\gamma$-burst energy considerably, and effects due to magnetic 
fields might even be required to get the energies for long complex 
gamma-ray bursts.

\end{abstract}

\section{Introduction}
In a sequence of preceding papers (Ruffert et al.~\cite{ruf96}, 
\cite{ruf97}; Ruffert \& Janka~\cite{ruf98a,ruf98b})
we have shown that the neutrino emission
associated with the dynamical phase of the merging or collision 
of two neutron stars is powerful, but too short to provide
the energy for gamma-ray bursts by neutrino-antineutrino annihilation.
Significant heating of the coalescing stars occurs only after they
have plunged into each other, and the neutrino luminosities can
rise to several $10^{53}\,{\rm erg\,s}^{-1}$ and even exceed 
$10^{54}\,{\rm erg\,s}^{-1}$ in case of the more violent collisions. 
After a few milliseconds, however, the compact massive remnant of
the merger will most likely collapse to a black hole. If that did
not happen, the remnant's continuing neutrino emission would drive 
a dense baryonic wind off its surface which would lead to a sizable 
mass loss but non-relativistic expansion (Woosley \& Baron \cite{woo92},
Woosley \cite{woo93b}, Hernanz et al.~\cite{her94}, 
Qian \& Woosley \cite{qia96}), a situation
which is not favorable for producing gamma-ray bursts which require
relativistic Lorentz factors $\Gamma \ga 100$ (Paczy\'nski \cite{pac90}).

If the merger remnant
collapses to a black hole, some matter remains in an
accretion disk or torus around the black hole. Our hydrodynamic 
simulations (Ruffert et al. \cite{ruf96}, Ruffert \& Janka~\cite{ruf98b}) 
have given hints that about $0.1\,M_{\odot}$ of matter might obtain
enough angular momentum during the merging of the neutron stars to resist
immediate collapse into the black hole. In this case 
a funnel with low baryon density
can develop along the system axis. On the other hand, the large 
angular momentum ensures that the torus matter is swallowed by the
black hole on a time scale much longer than the dynamical time scale.
Therefore there could be enough time for this material to radiate away
a fair fraction of its gravitational binding energy in neutrinos, 
even if the densities become so high that neutrinos get trapped and can
escape only on a diffusion time scale. 
A similar situation could result from the merging of a neutron star 
with a black hole (Lee \& Klu\'zniak \cite{lee95,lee98}; 
Klu\'zniak \& Lee \cite{klu98};
Eberl \cite{ebe98a}; Eberl et al.~\cite{ebe98b}), 
from the collapse of a very massive, 
rapidly rotating star (Woosley \cite{woo93a}, Popham et al.~\cite{pop98b},
MacFadyen \& Woosley \cite{mac98}), 
or from the coalescence of a neutron star/black
hole with a white dwarf (Fryer et al.~\cite{fry98b}) or with the helium core of 
its red giant companion (Fryer \& Woosley \cite{fry98a}).

All these events are estimated to occur at rates which can account for
the observed frequency of gamma-ray bursts (about one burst per day).
Also, the huge amount of gravitational binding energy released during
the accretion proc\-ess of up to several solar masses of gas
into the black hole is hoped to be able to explain the
energetics of even the most distant cosmological gamma-ray bursts
(e.g., GRB981214, see Kulkarni et al.~\cite{kul98}). 
Moreover, the compactness of the 
stellar-mass black hole could naturally produce the rapid 
variability on time scales of milliseconds observed in many bursts.
For these reasons, massive accretion disks or thick accretion tori
around stellar-mass black holes are considered as possible
cosmological origin of the enigmatic gamma-ray bursts, powered by
neutrino-antineutrino annihilation or by magnetically driven energy
release (e.g., Paczy\'nski \cite{pac86}; Goodman \cite{goo86};
Goodman et al. \cite{goo87}; Eichler et al.~\cite{eic89};
Paczy\'nski \cite{pac91}; Narayan et al. \cite{nar92}; 
M\'esz\'aros \& Rees \cite{mes93};
Woosley \cite{woo93a}; Jaroszy\'nski \cite{jar93,jar96};
Mochkovitch et al.~\cite{moc93,moc95}; 
Thompson \cite{tho94}; Witt et al.~\cite{wit94};
Janka \& Ruffert~\cite{jan96}; M\'esz\'aros \& Rees \cite{mes97};
Popham et al.~\cite{pop98b}; M\'esz\'aros et al.~\cite{mes98}).

In this paper we simulate the formation of the accretion torus 
after two neutron stars have merged, and assume that the compact remnant 
with a baryonic mass of about $3\,M_{\odot}$ has collapsed into a black
hole. The initial model is taken from our merger simulations (Ruffert \&
Janka~\cite{ruf98b}) where the central, massive object is replaced by a vacuum
boundary at a radius equal to twice the Schwarzschild radius of the mass
dumped into the black hole.
We follow the evolution of the left-over material in the surroundings 
of the black hole until it either has settled into the disk or has
been swallowed by the black hole. At the end of our three-dimensional
computations, the torus has reached a quasi-stationary state and its
further evolution is governed by the viscous transport of angular
momentum which depends on the uncertain value of the disk viscosity.
Kerr effects associated with the rotation of the black hole and
magnetic fields are not taken into account in our models.
An extensive, general investigation of relativistic
steady-state accretion from tori around hyper-accreting
stellar black holes for different accretion rates and disk viscosities
in Schwarzschild and Kerr geometry was recently published by 
Popham et al.~(\cite{pop98b}). Our simulations focus on the situation that
emerges from the mergings of compact binary systems of neutron stars and  
black holes, and they are intended to help constrain the large parameter
space and to yield insight into the torus properties and non-stationary
aspects of the evolution.

In particular, our
simulations aim at answering the following questions: How much mass
remains in the accretion torus? What is the relativistic rotation
parameter $a \equiv Jc/(GM^2)$ of the black hole? What are the properties
of the accretion torus, its density, temperature, neutrino luminosity?
How much mass pollutes the surroundings of the accretion torus, in
particular, does an effectively baryon-free funnel form along the system 
axis? How efficient is neutrino-antineutrino annihilation in depositing
energy in the regions with low baryon density? Can we make estimates
of the mass accretion rate into the black hole and the corresponding
lifetime of the torus? What are the implications for producing gamma-ray
bursts by neutrino-antineutrino annihilation? Is pair-plasma ejected
in a jet and how large will its opening angle be? Is there enough
variability of the energy release at the central source to account 
for the observed time structure of the gamma-ray burst light curves?

The paper is organized as follows. In Sect.~\ref{sec:initial} the
computational procedures are summarized which are used in
our simulations of the torus formation after neutron star merging.
The initial model for these simulations is
briefly described and the different investigated cases are introduced
with their distinguishing parameters.
Section~\ref{sec:hydro} contains a description of the
dynamical evolution of the torus from the beginning of the simulations
until a quasi-stationary state was reached. In Sect.~\ref{sec:torus}
the properties of the accretion tori at the end of the computations
are described. The results on the neutrino emission are presented in
Sect.~\ref{sec:neutrino} and those for neutrino-antineutrino annihilation
in Sect.~\ref{sec:annihilation}. In Sect.~\ref{sec:analytical} the
hydrodynamic results are used to estimate the numerical viscosity
which determines our torus models; the general relativistic
effects in the neutrino-antineutrino annihilation
are discussed as well as the importance of neutrino-electron/positron
scattering for the heating of the pair-plasma cloud that is formed by 
$\nu\bar\nu$ annihilation. Section~\ref{sec:summary} concludes the
paper with a summary and a discussion of the implications of our 
results for gamma-ray burst scenarios involving massive accretion tori
around stellar mass black holes.

\begin{table*}
\caption[]{
Some parameters for the torus evolution models with Newtonian 
potential (Models~{\bf B}1, {\bf B}2, {\bf B}4, and {\bf B}10) and 
with Paczy\'nski-Wiita potential (Models~${\cal B}$1, ${\cal B}$2,
and ${\cal B}$10), respectively. The dynamical simulations
were started with initial conditions as given from Model~B64 of
Ruffert \& Janka~(\cite{ruf98b}) at different times $t_{\rm init}$ and
the evolutions were followed over time intervals $\Delta t_{\rm cal}$.
All other quantities refer to the conditions found at the end of the
simulations. $M_{\rho<11}$ is the mass on the grid with density less
than $10^{11}\,{\rm g\,cm}^{-3}$, $M_{\rm v}$ the total gas mass 
on the computational grid, and $M_{\rm d}$ the mass of the gas with 
specific angular momentum larger than the Kepler limit
$j^*_{\rm N}\equiv v_{\rm Kepler}^{\rm N}(3R_{\rm s})3R_{\rm s}
= \sqrt{6}GM/c$ in case of the Newtonian potential or
$j^*_{\rm PW}\equiv v_{\rm Kepler}^{\rm PW}(3R_{\rm s})3R_{\rm s} 
= {3\over 2}j^*_{\rm N}$ for the Paczy\'nski-Wiita potential;
the Kepler velocities were evaluated from Eqs.~(\ref{eq:vkepN})
and (\ref{eq:vkepP}) using for $M$ the sum of black hole and total 
gas mass on the grid at the end of the simulation.
$T_{\rm max}$ is the maximum gas temperature in energy units.
$L_{\nu_e}$ denotes the electron neutrino luminosity near
the end of the simulation, $L_{\bar{\nu}_e}$ the 
corresponding electron antineutrino luminosity, and $L_{\nu_x}$
the luminosity of each individual flavor of heavy-lepton neutrino
($\nu_{\mu}$, $\bar\nu_{\mu}$, $\nu_{\tau}$ or $\bar\nu_{\tau}$).
The sum of all individual neutrino luminosities is given 
by $L_\nu$, and the mean energies of the different
neutrino types by $\langle\epsilon_{\nu_e}\rangle$,
$\langle\epsilon_{\bar{\nu}_e}\rangle$ 
and $\langle\epsilon_{\nu_x}\rangle$.
}
\begin{flushleft}
\tabcolsep=1.5mm
\begin{tabular}{lllcccccccccccc}
\hline\\[-3mm]
model & potential & $t_{\rm init}$ & $\Delta t_{\rm cal}$ & $M_{\rho<11}$ & 
   $M_{\rm v}$ & $M_{\rm d}$ & $T_{\rm max}$ & 
   $L_{\nu_e}$ & $L_{\bar{\nu}_e}$ & $L_{\nu_x}$ & $L_\nu$ & 
   $\langle\epsilon_{\nu_e}\rangle$ & 
   $\langle\epsilon_{\bar{\nu}_e}\rangle$ & 
   $\langle\epsilon_{\nu_x}\rangle$ \\ 
 & & ms & ms &
   {\scriptsize$10^{-2}M_\odot$} & {\scriptsize$10^{-2}M_\odot$} &
   {\scriptsize$10^{-2}M_\odot$} & {\scriptsize MeV} &
   {\scriptsize$10^{52}\frac{\rm erg}{\rm s}$} &
   {\scriptsize$10^{52}\frac{\rm erg}{\rm s}$} &
   {\scriptsize$10^{52}\frac{\rm erg}{\rm s}$} &
   {\scriptsize$10^{52}\frac{\rm erg}{\rm s}$} &
   {\scriptsize MeV} & {\scriptsize MeV} & {\scriptsize MeV} 
\\[0.3ex] \hline\\[-3mm]
{\bf B}1  & Newt & 1.84 & 2.5 & 4.6 & 23.2 &16.3 & 7. 
    &  0.3 & 0.7  &0.002 & 1.0 & 7.  &  13. & 15.  \\  
{\bf B}2  & Newt & 2.60 & 1.7 & 6.8 & 36.2 &28.6 & 8. 
    &  0.4  & 0.8  &0.003 & 1.2 & 8.  &  13. & 14. \\  
{\bf B}4  & Newt & 4.09 & 5.0 & 4.8 & 26.7 &22.0 & 9. 
    &  2.2  & 5.6  &0.033 & 8.0 & 9.  &  14. & 18. \\  
{\bf B}10 & Newt & 10.0 & 4.9 & 6.3 & 26.4 &24.2 & 12.
    &  3.5  & 6.5  & 0.40 & 12. &  9. &  13. & 21. 
     \\[0.7ex]  
${\cal B}$1  & PaWi & 1.84 & 3.2 & 0.5 & 0.5  & 0.16 & 4.
    &  0.02  & 0.12 & 0.0001&0.14& 12. &  12. & 10. \\  
${\cal B}$2  & PaWi & 2.60 & 6.0 & 2.8 & 3.4  & 2.4 & 7.
    &  2.0   & 4.5  & 0.017& 6.5 & 11. &  16. & 17. \\  
${\cal B}$10 & PaWi & 10.0 & 5.2 & 2.0 & 3.5  & 3.1 & 8.
    &  2.5   & 4.0  & 0.04 & 6.7 & 10. &  16. & 15. 
     \\[0.7ex]  
\hline
\end{tabular}
\end{flushleft}
\label{tab:models}
\end{table*}

\section{Computational procedures, initial conditions and different models
\label{sec:initial}}

In Ruffert \& Janka~(\cite{ruf98b}) we have calculated a series of neutron stars
merger models with varied neutron star masses, neutron star mass ratios,
neutron star spins, and initial conditions (temperature, entropy) in
the coalescing stars. In all of these simulations the post-merging 
configuration consisted of a compact central object with a mass of about
$3\,M_{\odot}$ and a typical density of the order of 
$10^{14}\,{\rm g\,cm}^{-3}$ which was surrounded by an extended cloud
of more dilute gas, having a mass of a few $0.1\,M_{\odot}$ and a
characteristic mean density around $10^{11}$--$10^{12}\,{\rm g\,cm}^{-3}$. 
To estimate the gas mass which might be able to stay in an accretion 
torus after the massive object has collapsed into a black hole, we
compared the specific angular momentum $j$ of the matter to the
Keplerian angular momentum $j^*$ for a test particle with non-zero mass 
that orbits around the Schwarzschild black hole on the last stable
circular orbit at 3 Schwarzschild
radii, $3 R_{\rm s} = 6GM/c^2$. Taking for simplicity $M$ as the
total (gas) mass on the grid, we found that between several
$10^{-2}\,M_{\odot}$ and a few $10^{-1}\,M_{\odot}$ fulfill this criterion~:
$j > j^*_{\rm N}\equiv v_{\rm Kepler}^{\rm N}(3R_{\rm s})3R_{\rm s}
= \sqrt{6}GM/c$ where $v_{\rm Kepler}^{\rm N}$ is the Keplerian 
velocity and the superscript (or subscript) N indicates the use
of a Newtonian gravitational potential. The amount of mass which has
sufficiently large angular momentum to resist immediate (i.e., on
a dynamical time scale) accretion into the black hole depends on the
neutron star masses and mass ratio as well as on the neutron star
spins which contribute to the total angular momentum of the binary 
system. For the simulations of the formation of the accretion torus
which we describe in this paper, the neutron star merger Model~B64 of 
Ruffert \& Janka~(\cite{ruf98b}) was used as an initial condition (defined by
the distributions of density, temperature and electron fraction, and 
by the velocity field at a certain chosen time during the evolution of
Model~B64). Due to the assumed corotation of the two $1.6\,M_{\odot}$
(baryonic mass) neutron stars before merging, the angular
momentum was largest in this model and correspondingly, the estimated
possible torus mass was maximal.

The three-dimensional computations of neutron star mergings were performed 
with a Newtonian 
hydrodynamics code based on the Piecewise Parabolic Method (PPM) of 
Colella \& Woodward (\cite{col84}) with at least four levels of nested grids 
(Ruffert \cite{ruf92}) to
ensure both high resolution at the neutron stars and a large computational
volume. The code includes the effects of gravitational-wave emission
and their back-reaction on the hydrodynamic flow according to
Blanchet et al.~(\cite{bla90}) (see Ruffert et al.~\cite{ruf96}). In addition, we 
implemented a calibrated neutrino
leakage scheme (Ruffert et al.~\cite{ruf96}) in order to calculate the energy and
lepton number loss by neutrino emission from the heated neutron star matter. 
The latter is described by the finite-temperature nuclear equation of state
of Lattimer \& Swesty (\cite{lat91}) using the Sk180 nuclear force parameter set
(Swesty et al.~\cite{swe94}).

\begin{figure}
\epsfxsize=8.5cm \epsfclipon \epsffile{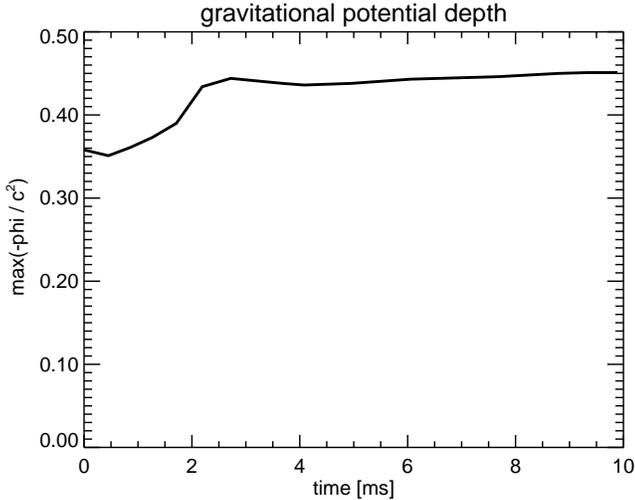}
\caption[]{Maximum absolute value of the
    gravitational potential produced by the mass distribution of
    Model~B64 of Ruffert \& Janka (\cite{ruf98b}) as function of time,
    normalized to $c^2$. This parameter, multiplied by 2,
    is a measure of the importance of relativistic corrections,
    and is very small in the limit of Newtonian gravity. The time
    is measured from the beginning of the simulation of the neutron
    star merger Model~B64.
    }
\label{fig:maxpot}
\end{figure}

The torus simulations presented in this paper are done with the same code
and the same input physics. Because of the spherical symmetry of the mass
in the black hole and the relatively small mass of the accretion torus, 
time derivatives of the quadrupole moment are small and 
grav\-ita\-tion\-al-\-wave
production does not play an important role. Although taken into account in
our simulations, we shall therefore not report data of the
gravitational-wave emission here. 
In one set of our models ({\bf B}1, {\bf B}2, {\bf B}4, and {\bf B}10) 
the black hole potential, which dominates the gravitational field at
the torus (whose self-gravity is only a minor contribution), is 
represented by a Newtonian potential, 
\begin{equation}
\Phi_{\rm N}\,=\,-{GM_{\rm BH}\over r} \ .
\label{eq:phiN}
\end{equation}
In a second sequence of models (${\cal B}$1, ${\cal B}$2, and ${\cal B}$10)
the gravitational potential is described by the Paczy\'nski-Wiita 
expression, 
\begin{equation}
\Phi_{\rm PW}\,\equiv\,-{GM_{\rm BH}\over r-R_{\rm s}}
\label{eq:phiP}
\end{equation}
(Paczy\'nski \& Wiita \cite{pac80}). This allows one to reproduce the 
existence and the effects of a last stable circular orbit at a radius of
$3R_{\rm s} = 6GM_{\rm BH}/c^2$ where the specific angular momentum 
$j_{\rm PW} = r v_{\rm Kepler}^{\rm PW}(r) = 
\sqrt{GM_{\rm BH}r^3/(r-R_{\rm s})^2}$ has an absolute minimum. 
We hope that this approximation, although crude, can give us some
indication of the sensitivity of our results to the inclusion of 
proper general relativity in the modeling.

\begin{figure*}
\tabcolsep=2.0mm
 \begin{tabular}{cc}
   \epsfxsize=8.5cm \epsfclipon \epsffile{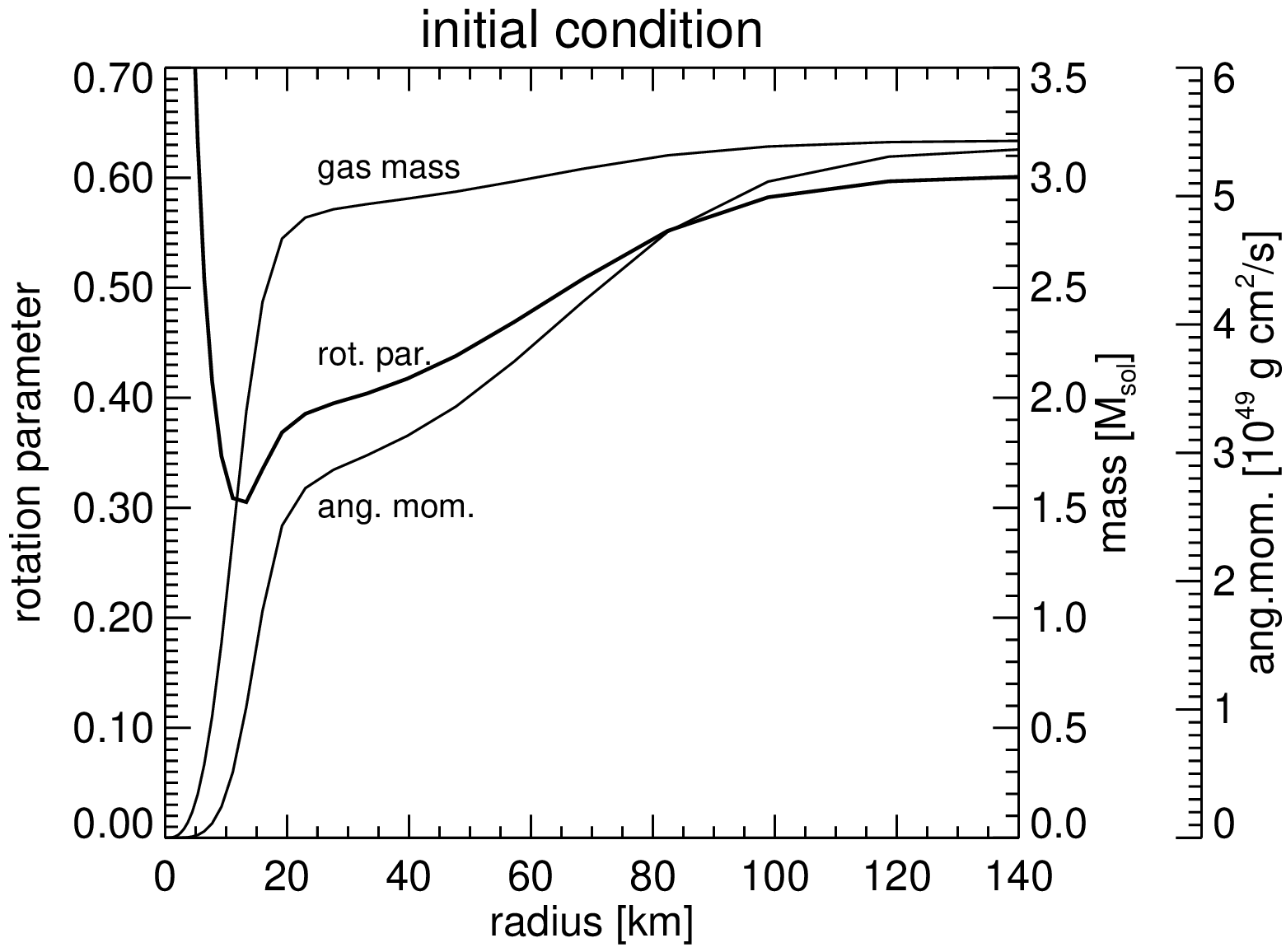} &
   \epsfxsize=8.5cm \epsfclipon \epsffile{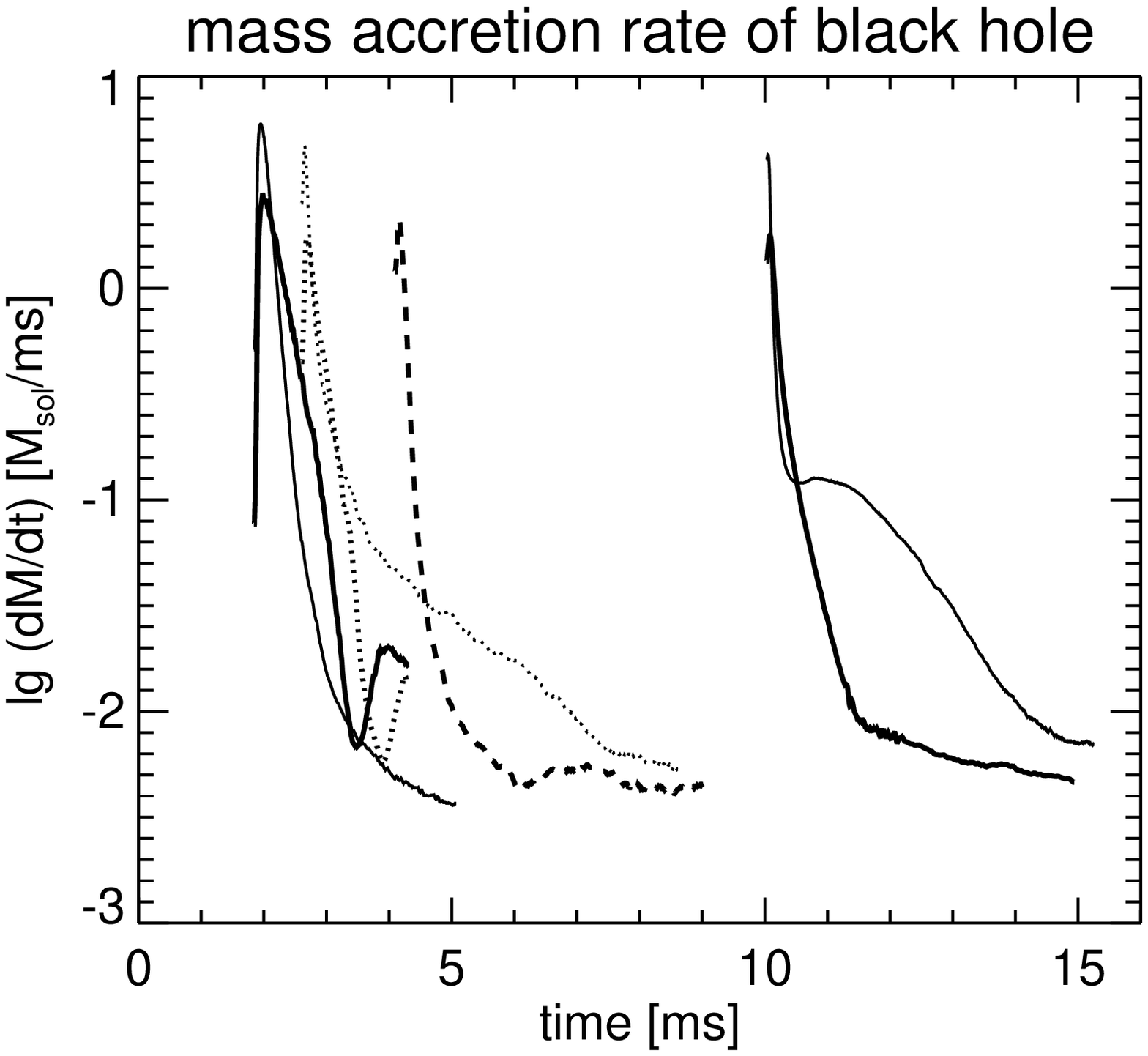} \\
   \parbox[t]{8.7cm}{\caption[]{Distribution of mass and angular
    momentum in the remnant of the neutron star merger of Model~B64 of 
    Ruffert \& Janka (\cite{ruf98b}) at time $t = 10\,$ms. This defines the 
    initial condition used for some of the torus simulations presented
    here. The upper thin solid line gives the cumulative gas mass
    (in solar masses), the lower thin solid line the cumulative angular
    momentum (in $10^{49}\,{\rm g\,cm}^2\,{\rm s}^{-1}$), and the thick
    solid line the relativistic rotation parameter 
    $a(r) \equiv J(r)c/(GM_{\rm gas}^2(r))$ as functions of radius $r$. 
    ($M_{\rm gas}(r)$ is the gas mass inside $r$ and $J(r)$ the 
    corresponding angular momentum perpendicular to the equatorial plane.)}
    \label{fig:rotpar}} &
    \parbox[t]{8.7cm}{\caption[]{Mass accretion rates of the black hole
    (in $M_{\odot}\,{\rm ms}^{-1}$)
    as functions of time for all models listed in Table~\ref{tab:models}.
    The thick lines which start at times $t = 1.84\,{\rm ms}$, $2.60\,{\rm ms}$,
    $4.09\,{\rm ms}$, and $10.0\,{\rm ms}$ correspond to the Newtonian 
    Models~{\bf B}1, {\bf B}2, {\bf B}4, and {\bf B}10,
    respectively, and the thin lines starting at times
    $t = 1.84\,{\rm ms}$, $2.60\,{\rm ms}$, and $10.0\,{\rm ms}$ correspond to 
    Models~${\cal B}1$, ${\cal B}2$, and ${\cal B}10$, respectively.}
    \label{fig:massaccBH}} \\
[35ex]
    \epsfxsize=8.5cm \epsfclipon \epsffile{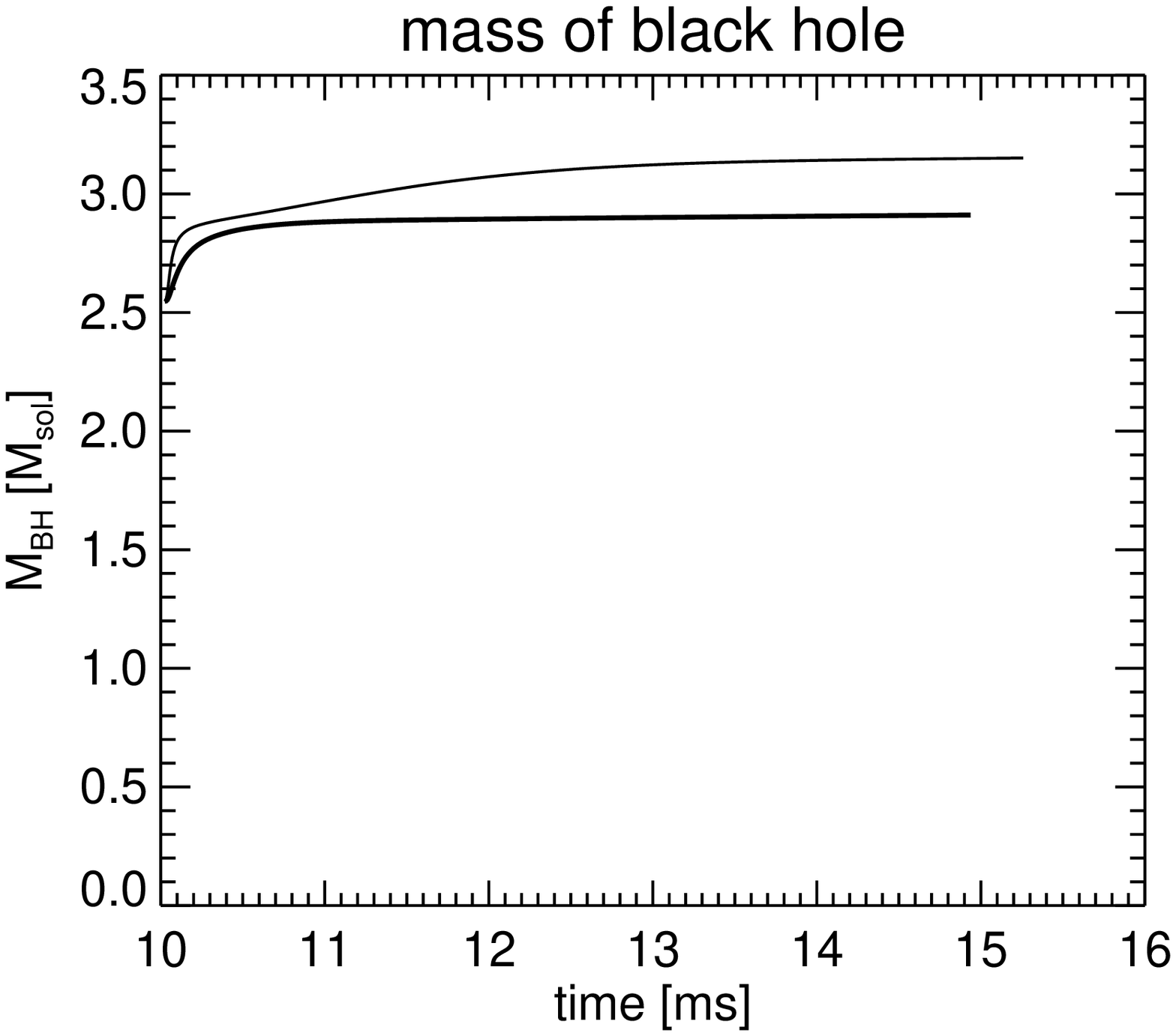} &
    \epsfxsize=8.5cm \epsfclipon \epsffile{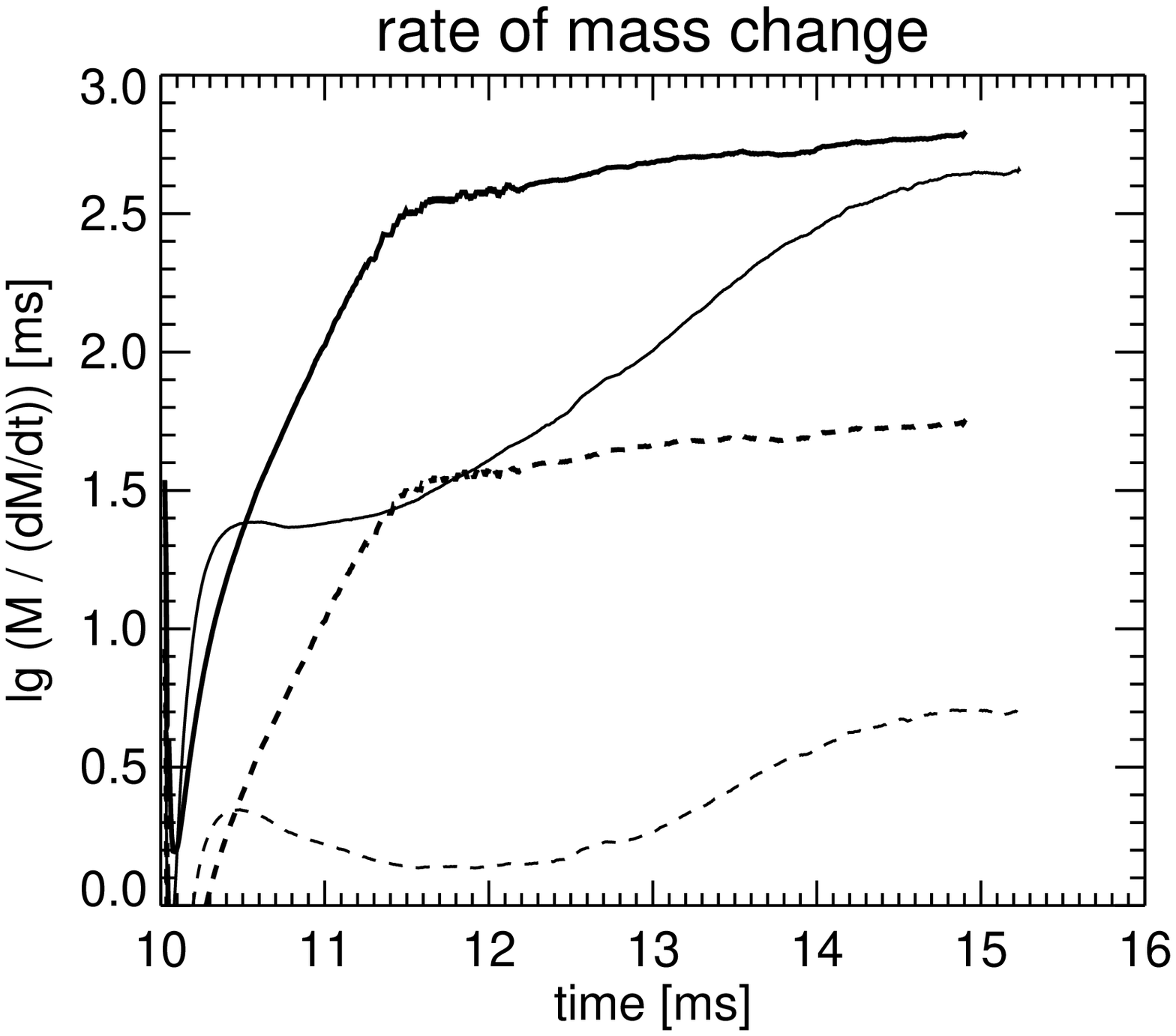} \\
    \parbox[t]{8.7cm}{\caption[]{The mass of the central black hole
    as function of time for Model~{\bf B}10 (bold line) and Model~${\cal B}$10
    (thin line), showing an increase due to accretion of surrounding gas.}
    \label{fig:massBH}} &
    \parbox[t]{8.7cm}{\caption[]{Typical time scales for the changes of black 
    hole mass (solid lines) and torus mass (dashed lines) as functions of 
    time for Model~{\bf B}10 (bold lines) and Model~${\cal B}$10 (thin lines).}
    \label{fig:masstBH}}\\
\end{tabular}
\end{figure*}

\begin{figure*}
\begin{tabular}{cc}
 \epsfxsize=8.5cm \epsfclipon \epsffile{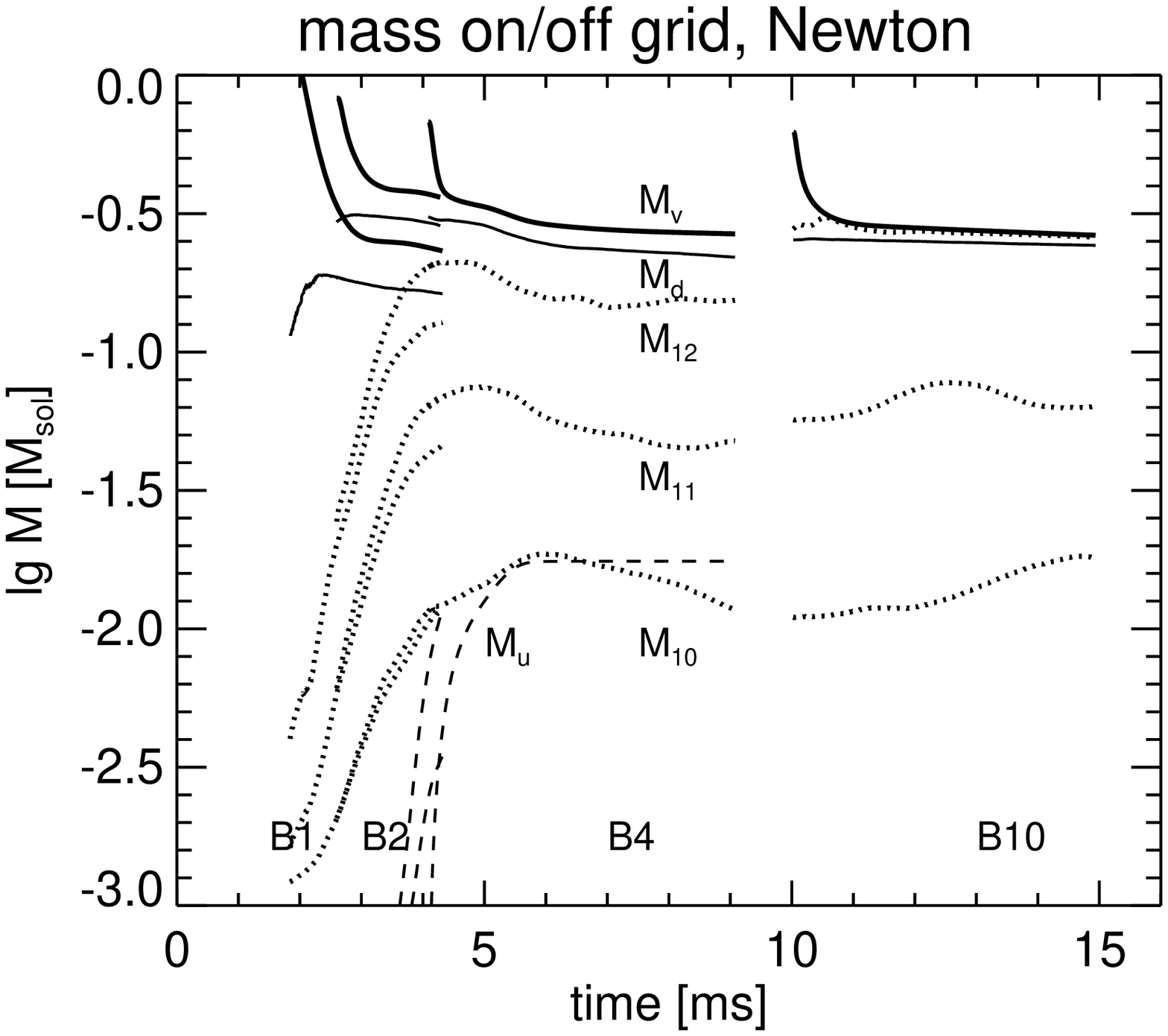} &
 \epsfxsize=8.5cm \epsfclipon \epsffile{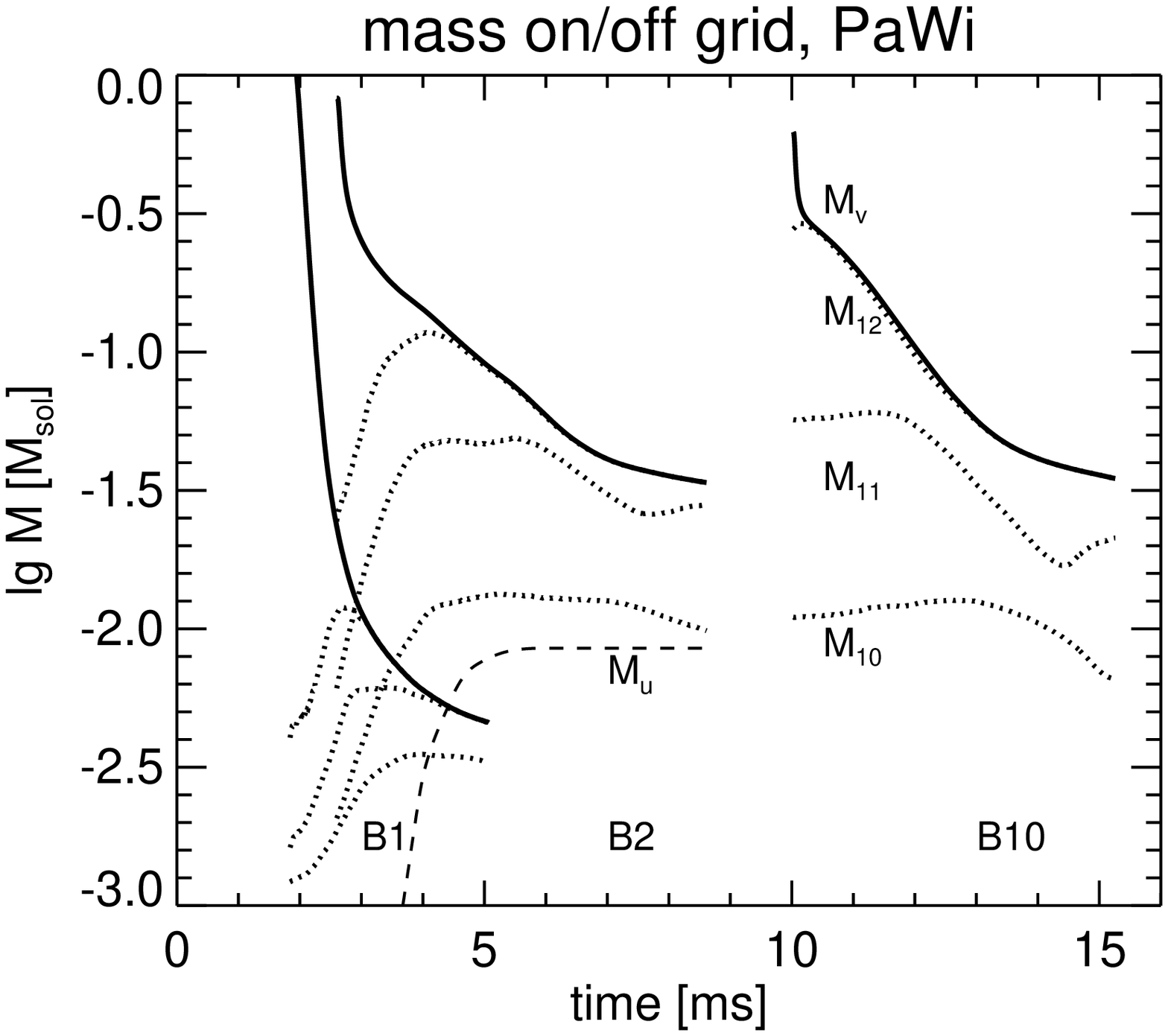} \\
\end{tabular}
\caption[]{Different masses as functions of time
    for the Newtonian Models~{\bf B}1, {\bf B}2, {\bf B}4, and {\bf B}10
    (left plot) and for the different Paczy\'nski-Wiita 
    Models~${\cal B}$1, ${\cal B}$2, and ${\cal B}$10 (right plot)
    (distinguishable by their start and stop times, see Table~\ref{tab:models}).
    $M_{\rm v}$ (bold solid lines) denotes the gas mass on the grid,
    $M_{\rm u}$ (dashed lines) is the cumulative mass which leaves the grid
    during the simulation and becomes unbound, i.e., which fulfils the
    criterion that its total specific energy as the sum of its specific 
    internal, kinetic, and potential energies is positive. The dotted curves
    with labels $M_{10}$, $M_{11}$, and $M_{12}$ show the gas
    masses with densities below $10^{10}\,{\rm g\,cm}^{-3}$,
    $10^{11}\,{\rm g\,cm}^{-3}$, and $10^{12}\,{\rm g\,cm}^{-3}$, respectively.
    In case of the Newtonian potential (left plot), the thin solid lines
    labeled with $M_{\rm d}$ represent the mass on the grid with specific
    angular momentum larger than the Kepler value at 3
    Schwarzschild radii, i.e., $j \ge j_{\rm N}^\ast = \sqrt{6}GM/c$, where
    for $M$ the sum of the black hole mass and gas mass on the grid was used at 
    all times.                                                            
    }
    \label{fig:diskmass}
\end{figure*}

\begin{figure*}
\tabcolsep=2.0mm
 \begin{tabular}{cc}
 \epsfxsize=8.5cm \epsfclipon \epsffile{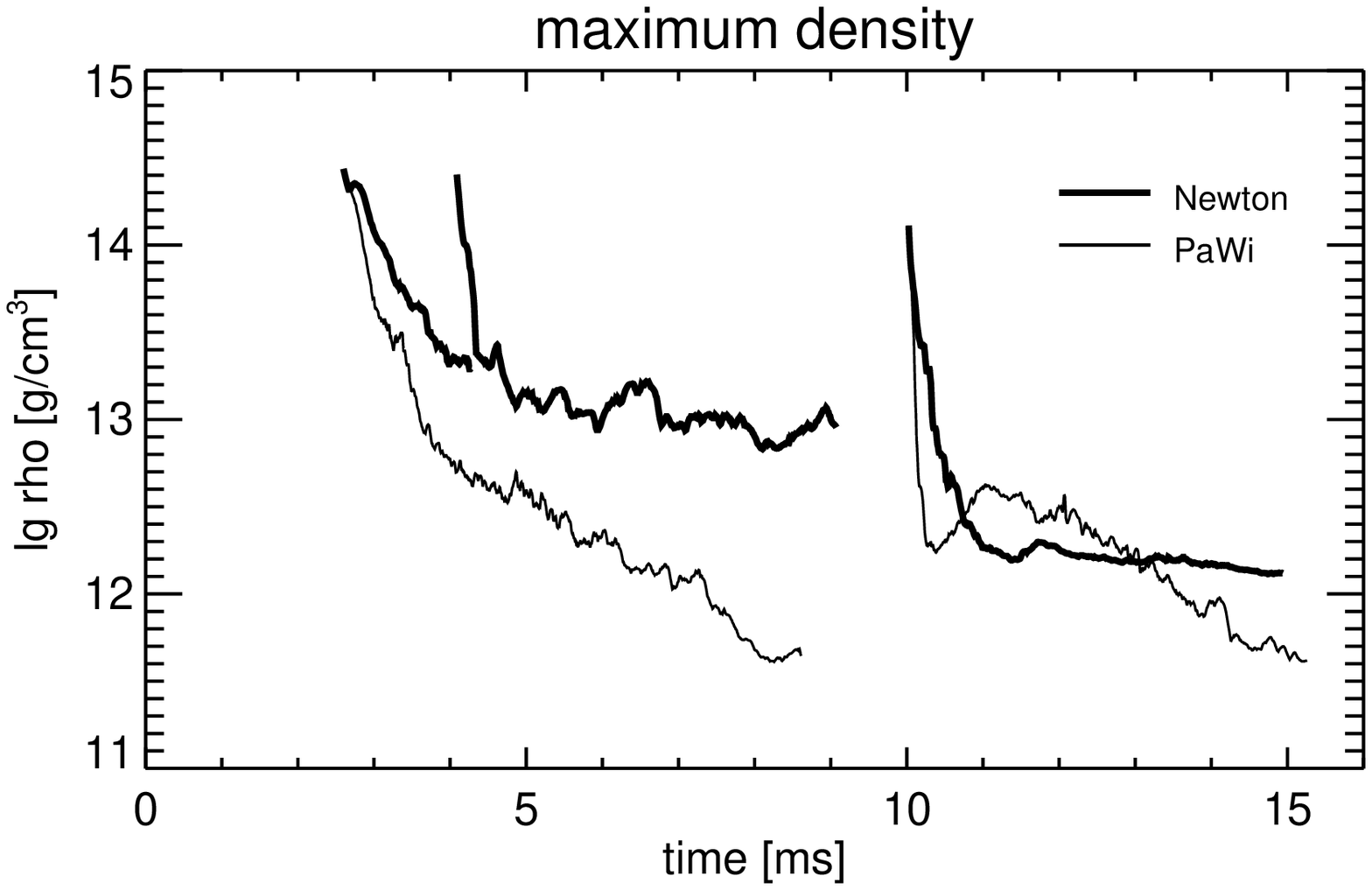} &
 \epsfxsize=8.5cm \epsfclipon \epsffile{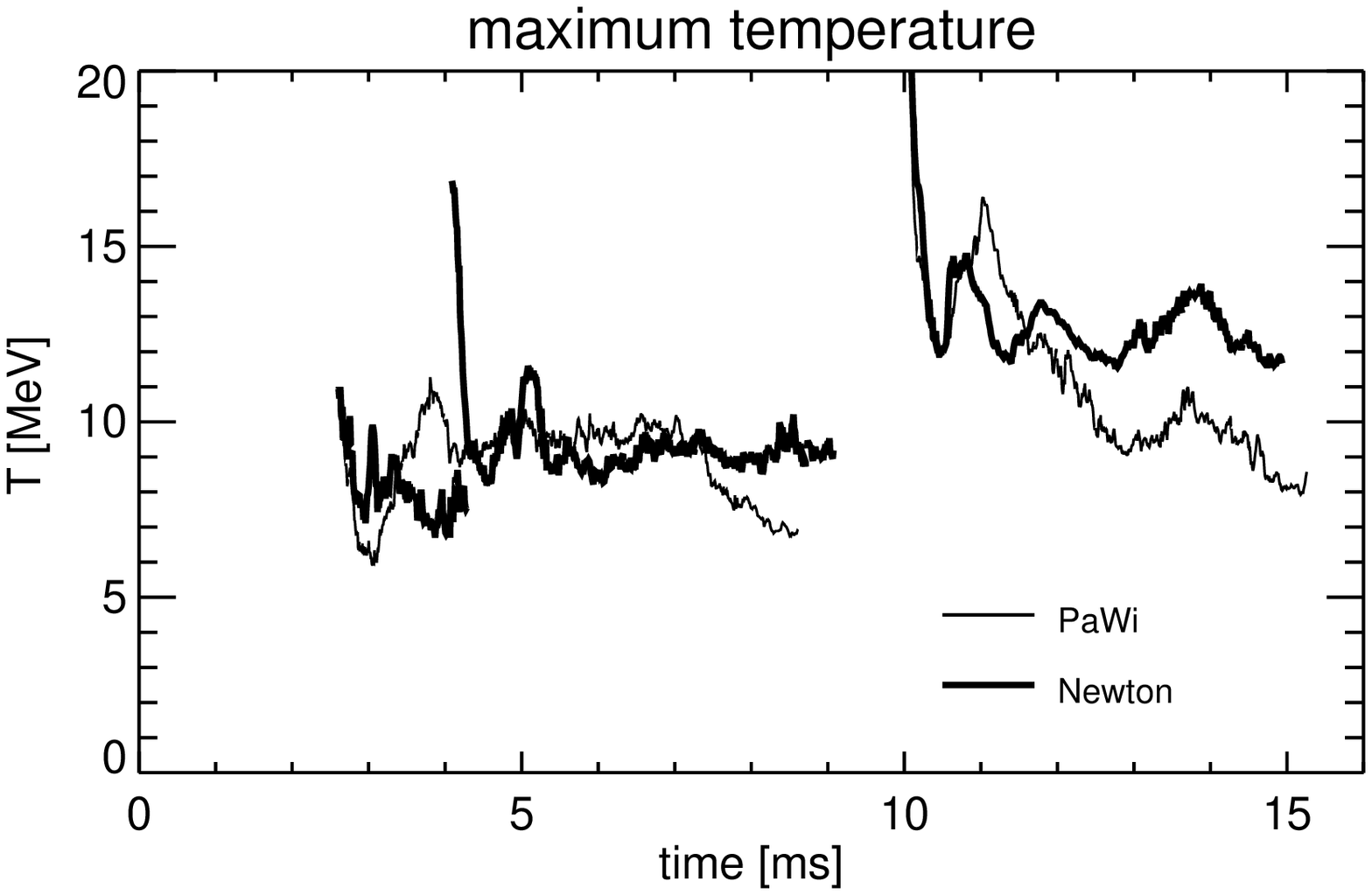} \\
 \parbox[t]{8.7cm}{\caption[]{Maximum density on the grid as
  function of time for the Newtonian Models~{\bf B}2,
  {\bf B}4, and {\bf B}10 (bold lines) and for the
  Paczy\'nski-Wiita Models~${\cal B}$2 and ${\cal B}$10
  (thin lines).}
  \label{fig:maxrhoT}} &
  \parbox[t]{8.7cm}{\caption[]{Maximum temperature on the grid
  as function of time for the Newtonian Models~{\bf B}2,
  {\bf B}4, and {\bf B}10 (bold lines) and for the 
  Paczy\'nski-Wiita Models~${\cal B}$2 and ${\cal B}$10
  (thin lines).}
  \label{fig:maxtempT}} \\
\end{tabular}
\end{figure*}

\begin{figure*}
\tabcolsep=2.0mm
\begin{tabular}{cc}
  \epsfxsize=8.5cm \epsfclipon \epsffile{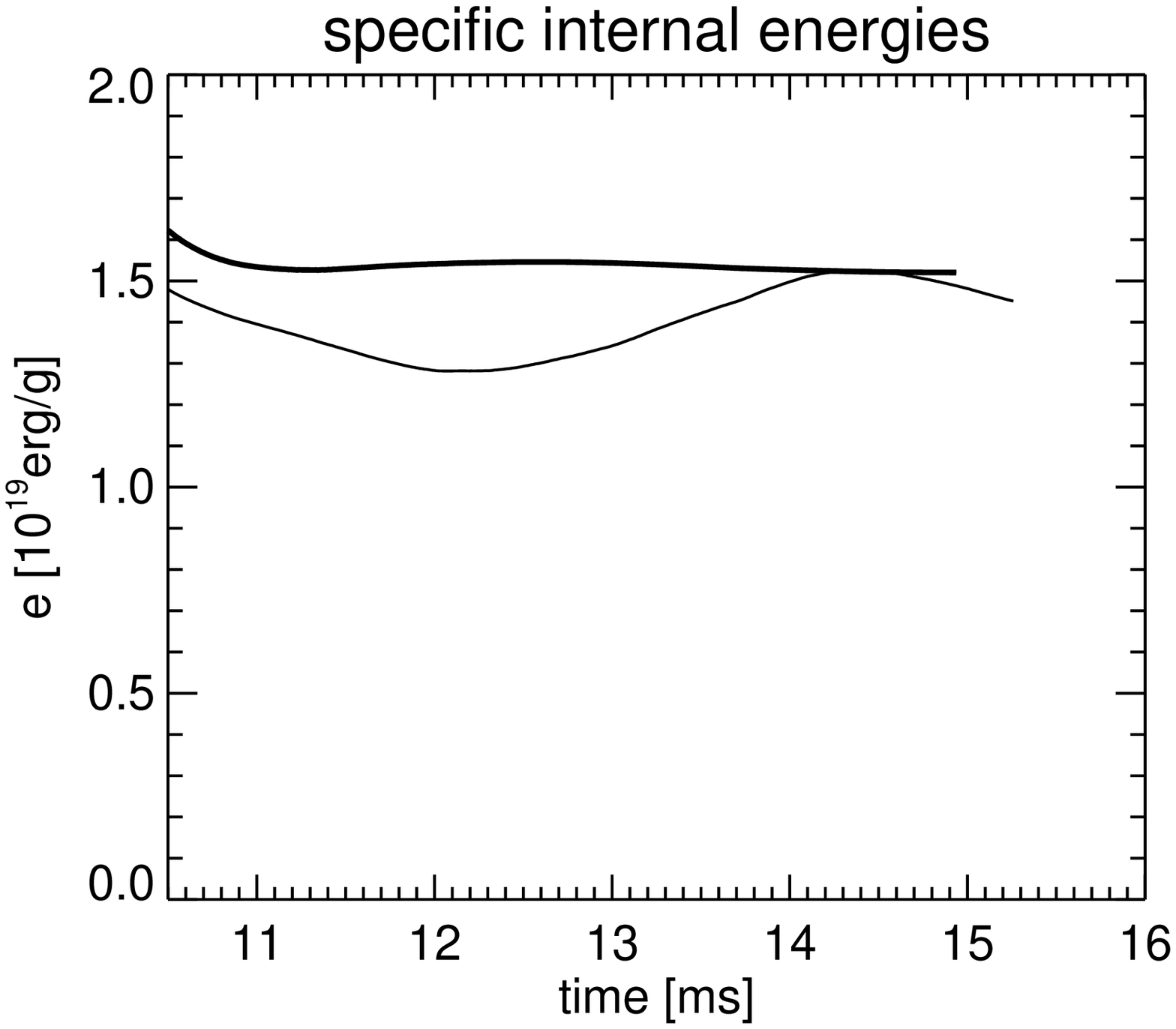} &
  \epsfxsize=8.5cm \epsfclipon \epsffile{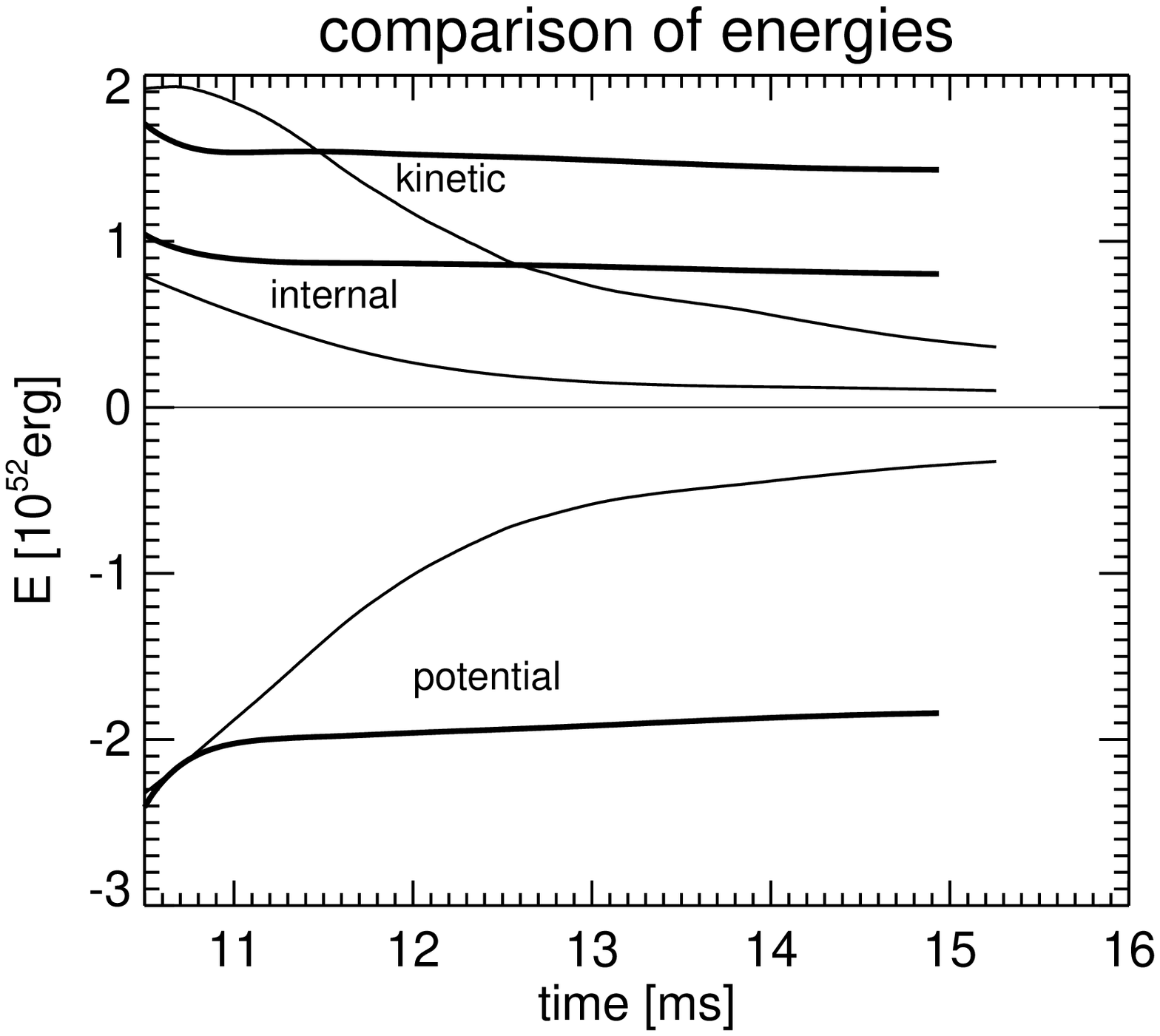} \\
  \parbox[t]{8.7cm}{\caption[]{Average value of the specific internal
  energy of the matter on the grid as function of time
  for Model~{\bf B}10 (bold line) and Model~${\cal B}$10 (thin line).}
  \label{fig:ergspec}} &
  \parbox[t]{8.7cm}{\caption[]{Total kinetic, thermal, and gravitational
   potential energies as functions of time for Model~{\bf B}10
   (bold line) and Model~${\cal B}$10 (thin line).}
  \label{fig:energy}} \\
\end{tabular}
\end{figure*}

The different models of each set, {\bf B}\# and ${\cal B}$\#, 
respectively, are discerned by the different times $t_{\rm init}$
(in milliseconds given by the numbers in the model names) at which
the compact object that formed during the merging of the two neutron stars
in Model~B64 of Ruffert \& Janka (\cite{ruf98b}) was removed and replaced by
a gravitating ``vacuum sphere''. The times $t_{\rm init}$ are measured
from the start of the simulation of Model~B64 and are listed in 
Table~\ref{tab:models}. If a black hole forms from the compact central
body of the merger remnant with a baryonic mass of nearly $3\,M_{\odot}$,
we expect this
to happen at about the time when the most compact state is reached and
thus the gravitational potential becomes strongest. If support by
centrifugal forces or thermal pressure played a role, the collapse 
might be delayed. Figure~\ref{fig:maxpot} shows that the value of the
parameter $\Phi_{\rm N}/c^2$ --- twice of which gives a rough measure for
the importance of general relativistic gravity --- plateaus at about 
$2\,$ms after the simulation of the merging of the neutron stars was started.
Therefore we chose the earliest moment of the possible black hole 
formation and the onset of our torus computations at around this time,
and investigated also models for later collapse times at about $4\,$ms
and $10\,$ms, respectively. The model runs were continued until the accretion
rate into the black hole had reached such a low value that further
changes of the torus properties would happen over a much longer period 
than the dynamical time scale of the system. The computed evolution times
$\Delta t_{\rm cal}$ of all models are also given in 
Table~\ref{tab:models}. The subsequent quasi-stationary
evolution proceeds on the time scale of viscous transport of 
angular momentum which depends on the unknown value of the disk viscosity
and in general is too long to be followed by our three-dimensional, 
hydrodynamic simulations with an explicit code.

The inner vacuum sphere that represents the black hole at the center
of the computational grid is set to a radius of $2R_{\rm s}$.
The Schwarzschild radius $R_{\rm s}$ is initially computed for 
90\% of the total gas mass on the grid and the mass of the gas inside
this radius is collected into the black hole to determine its 
gravitational potential. During the following evolution, the black hole
mass, momentum and angular momentum are updated by adding the 
corresponding values of the matter which is advected through
the sphere at $2R_{\rm s}$. From the current value of the black hole
mass, the new radius of the vacuum sphere and the new gravitational
potential of the black hole are calculated. The loss of mass, momentum,
angular momentum and energy from the gas outside the black hole 
boundary are also monitored during the simulations. In the grid zones
that are located inside the vacuum sphere (these zones are not 
removed from the hydrodynamic grid), the mass density is 
continuously reset to a negligibly small but finite value of 
$10^8\,{\rm g\,cm}^{-3}$ and a correspondingly very small value of 
the pressure is used. The decision to put the vacuum boundary at the
radius $2R_{\rm s}$ was influenced by the facts that on the one hand the 
gravitational potential in the Paczy\'nski-Wiita case diverges when $r$
decreases towards $1R_{\rm s}$, and that on the other hand the
gas velocities come close to the speed of light already near $2R_{\rm s}$
and therefore the nonrelativistic treatment of the
hydrodynamics can definitely not be applied any longer.
Moreover, the Courant-Friedrich-Lewy timestep of the explicit 
computation is limited by the small values in this region on the 
finest grid. From a physics point of view, the choice of the 
black hole boundary at $2R_{\rm s}$ can be justified because 
the generated black hole is not of extreme Kerr type but its
relativistic rotation parameter $a$ is of order 0.4--0.5
(Fig.~\ref{fig:rotpar}). In this case the inner edge of the torus 
should be located at just around two times the horizon radius
(see Usui et al.~\cite{usu98}). Interior to this radius the orbits are
unstable and the infall velocities become very large, i.e., the
matter moves essentially radially inward. Our Newtonian code does not
take into account effects due to the rotation of the black hole on 
the surrounding space and matter, e.g., frame dragging or the 
dependence of the radius of the innermost stable orbit on the 
rotation parameter of the black hole.

The simulations are performed with four levels of nested 
cartesian grids, each having 64 zones per dimension in the orbital 
plane and 16 perpendicular to the orbital plane (here we make use of 
equatorial symmetry and of the smaller extension of the torus in the
vertical direction). A zone on the finest grid has a length of
$\Delta x = \Delta y = \Delta z =0.64\,$km, and the size of the
largest grid is $328\,{\rm km}\times 328\,{\rm km}\times 82\,{\rm km}$.

\begin{figure*}
\begin{tabular}{cc}
  \epsfxsize=8.5cm  \epsfclipon\epsffile{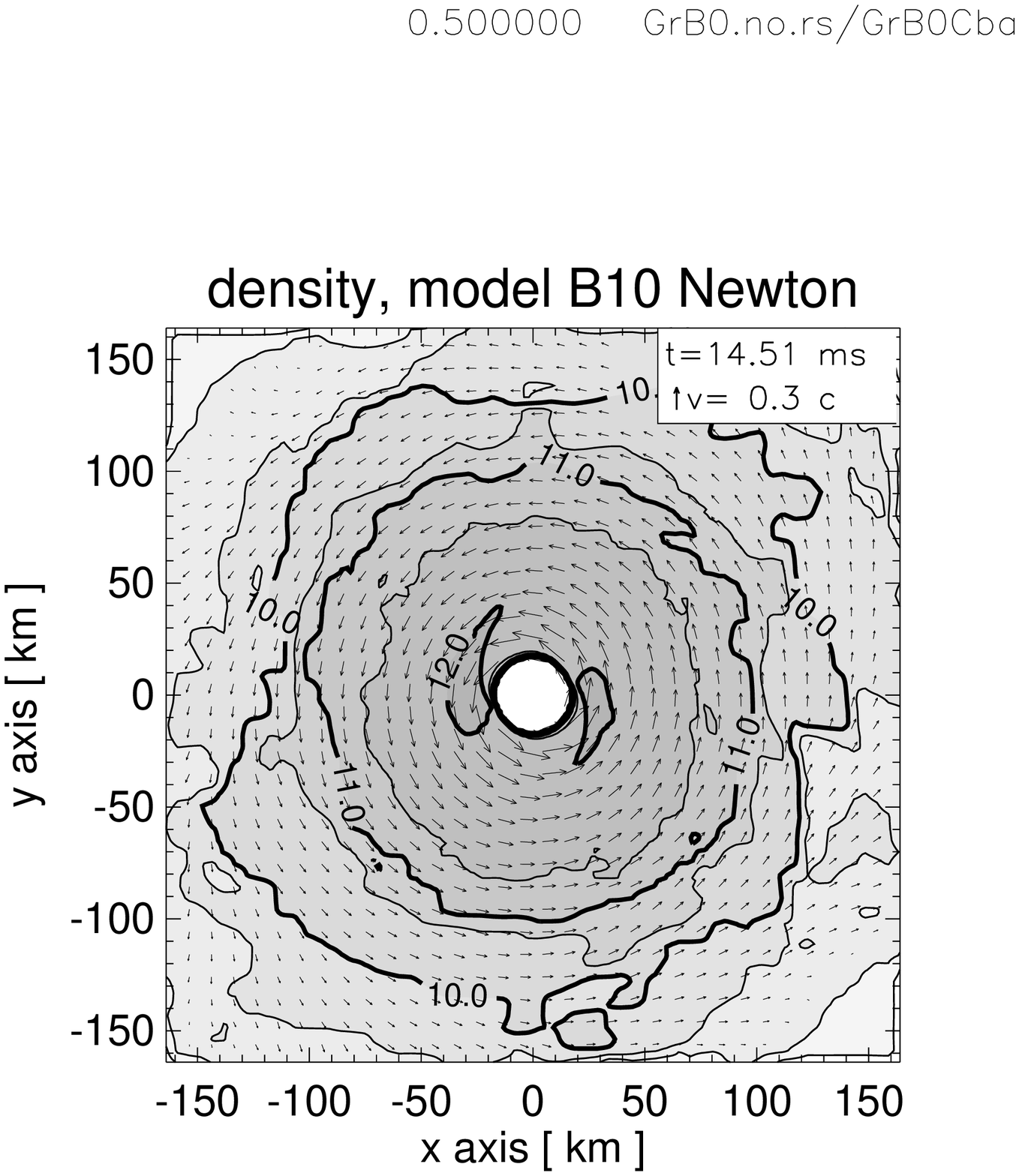}
  \put(-0.55,7.05){{\Large \bf \sf a}}   &
  \epsfxsize=8.5cm  \epsfclipon\epsffile{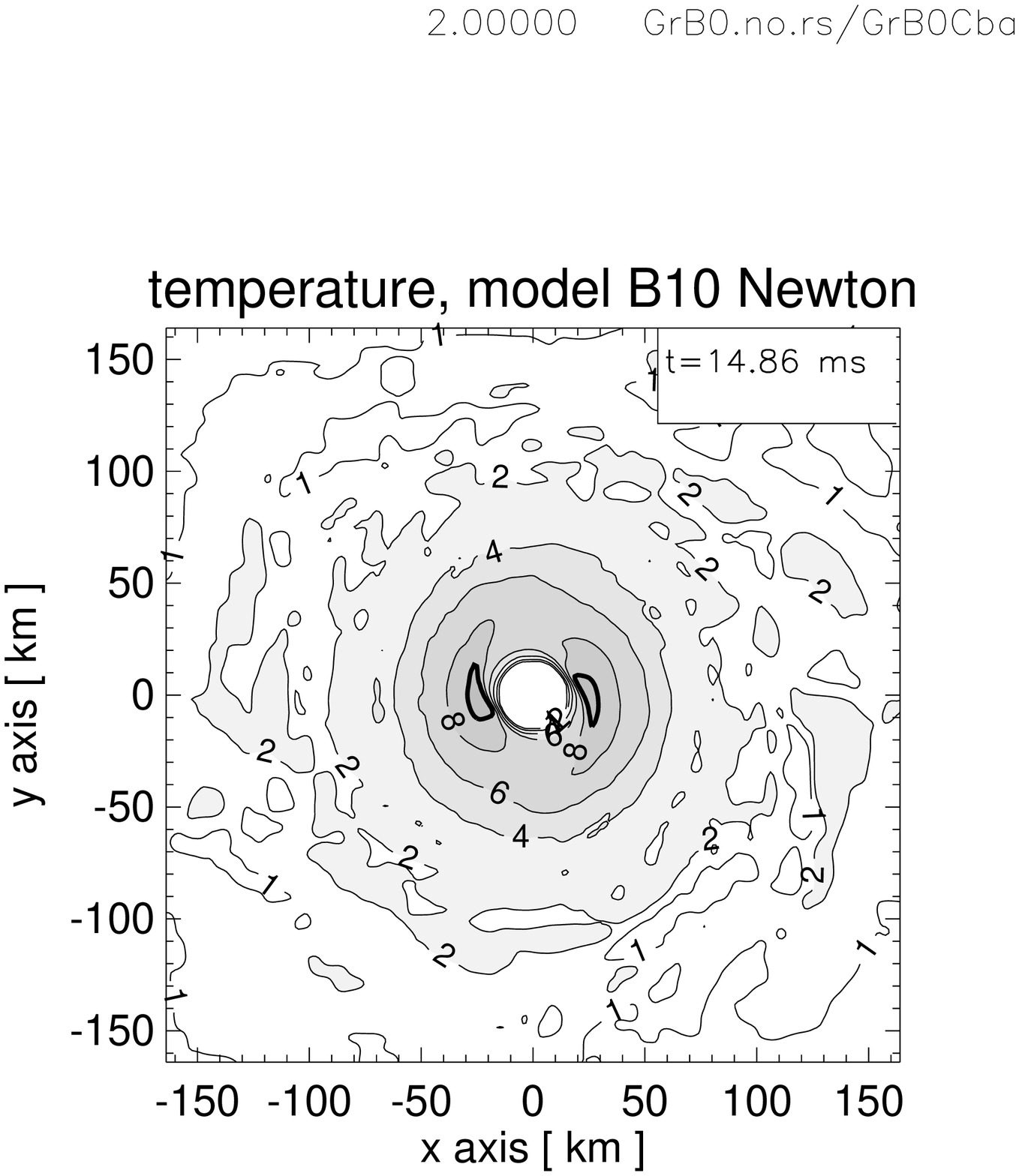}
  \put(-0.55,7.05){{\Large \bf \sf b}}  \\
[2ex]
  \epsfxsize=8.5cm  \epsfclipon\epsffile{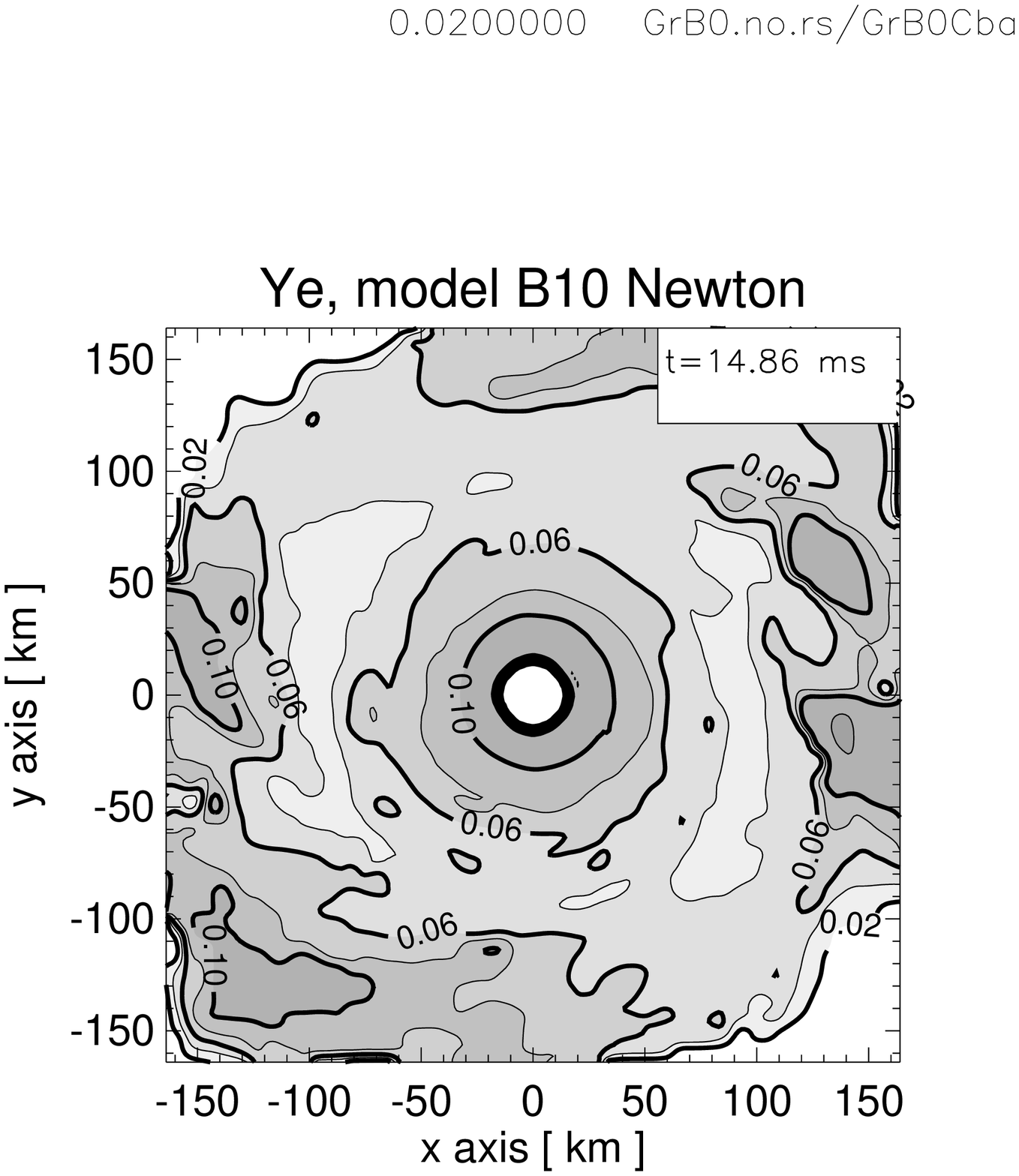}
  \put(-0.55,7.05){{\Large \bf \sf c}}  &
  \epsfxsize=8.5cm  \epsfclipon\epsffile{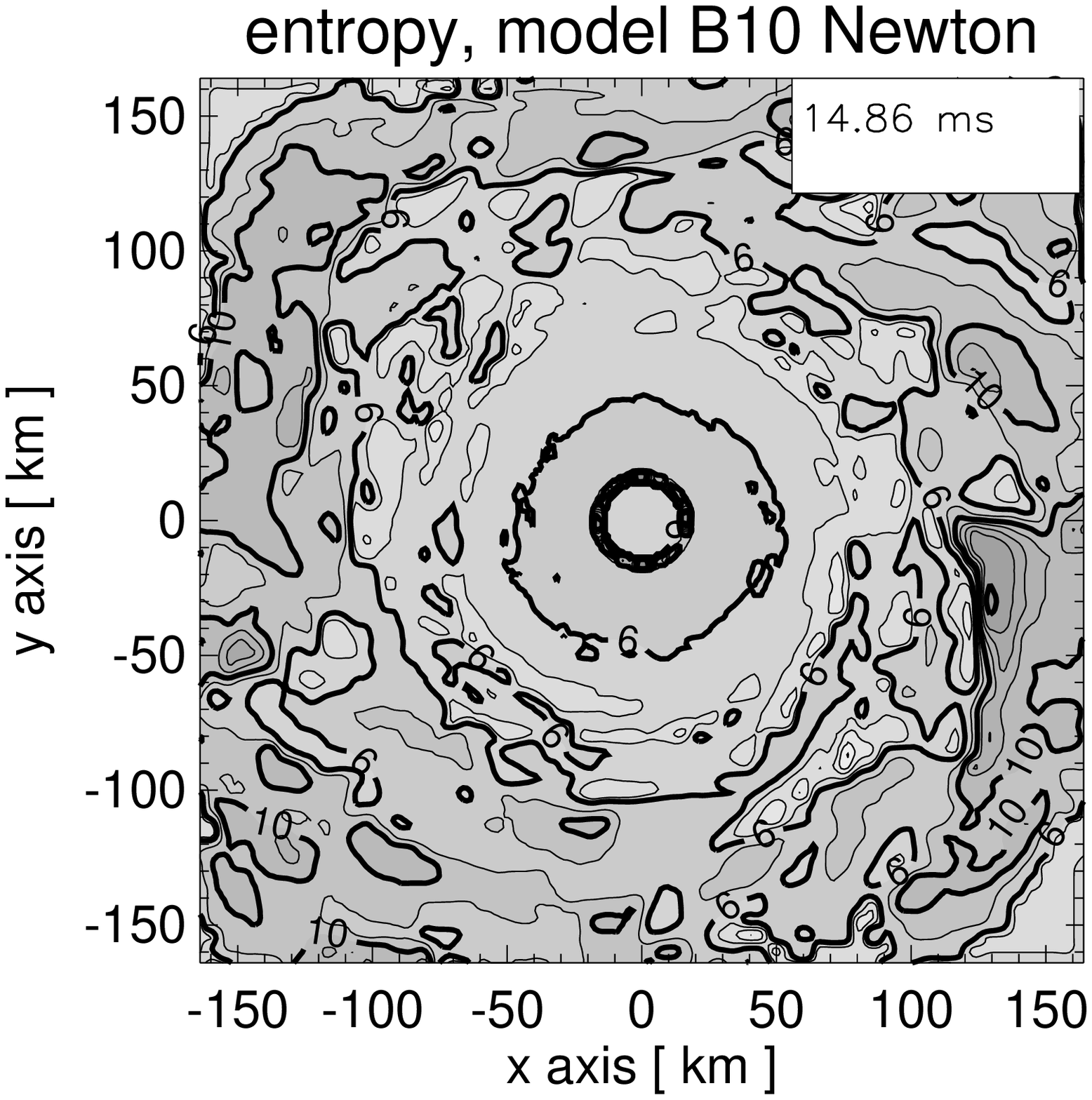}
  \put(-0.55,7.05){{\Large \bf \sf d}}  \\
\end{tabular}
\caption[]{
  Structure of the Newtonian Model~{\bf B}10 in the equatorial plane
  near the end of the simulation. Panel~a shows contours of the baryonic 
  mass density (measured in ${\rm g\,cm}^{-3}$) together with the 
  flow field indicated by velocity vectors, panel~b displays contours
  of the temperature (in MeV), panel~c of the electron fraction $Y_e$,
  and panel~d of the entropy per nucleon.
  The length scale of the velocity vectors and the time elapsed since the
  beginning of the simulation of Model~B64 of Ruffert \& Janka (\cite{ruf98b})
  are given in the insert in the top right corner of the figures.
  The density contours are spaced logarithmically with intervals of
  0.5~dex (bold: $\log(\rho\eck{{\rm g\,cm}^{-3}}) = 10.0,\,11.0,\,12.0$), 
  the temperature contours are spaced linearly, starting with 1$\,$MeV 
  and 2$\,$MeV, and then continuing in steps of 2$\,$MeV 
  (bold: $T = 10\,{\rm MeV}$), the $Y_e$ contours are spaced linearly 
  with intervals of 0.02 (bold: 0.02, 0.06, 0.10, 0.16), and the entropy
  levels were chosen to be 1, 2, 3, 4, 5, 6, 8, 10, 12, 14, 16, and
  20 (bold: 6, 10, 20). Some of the contours are labeled with their 
  respective values, and darker grey shading indicates higher values
  of a quantity. The circle around the center marks the inner ``vacuum'' 
  boundary with semidiameter of twice the Schwarzschild radius for
  the mass which has accumulated in the central black hole.}
  \label{fig:densN}
\end{figure*}

\begin{figure*}
\begin{tabular}{cc}
  \epsfxsize=8.5cm  \epsfclipon\epsffile{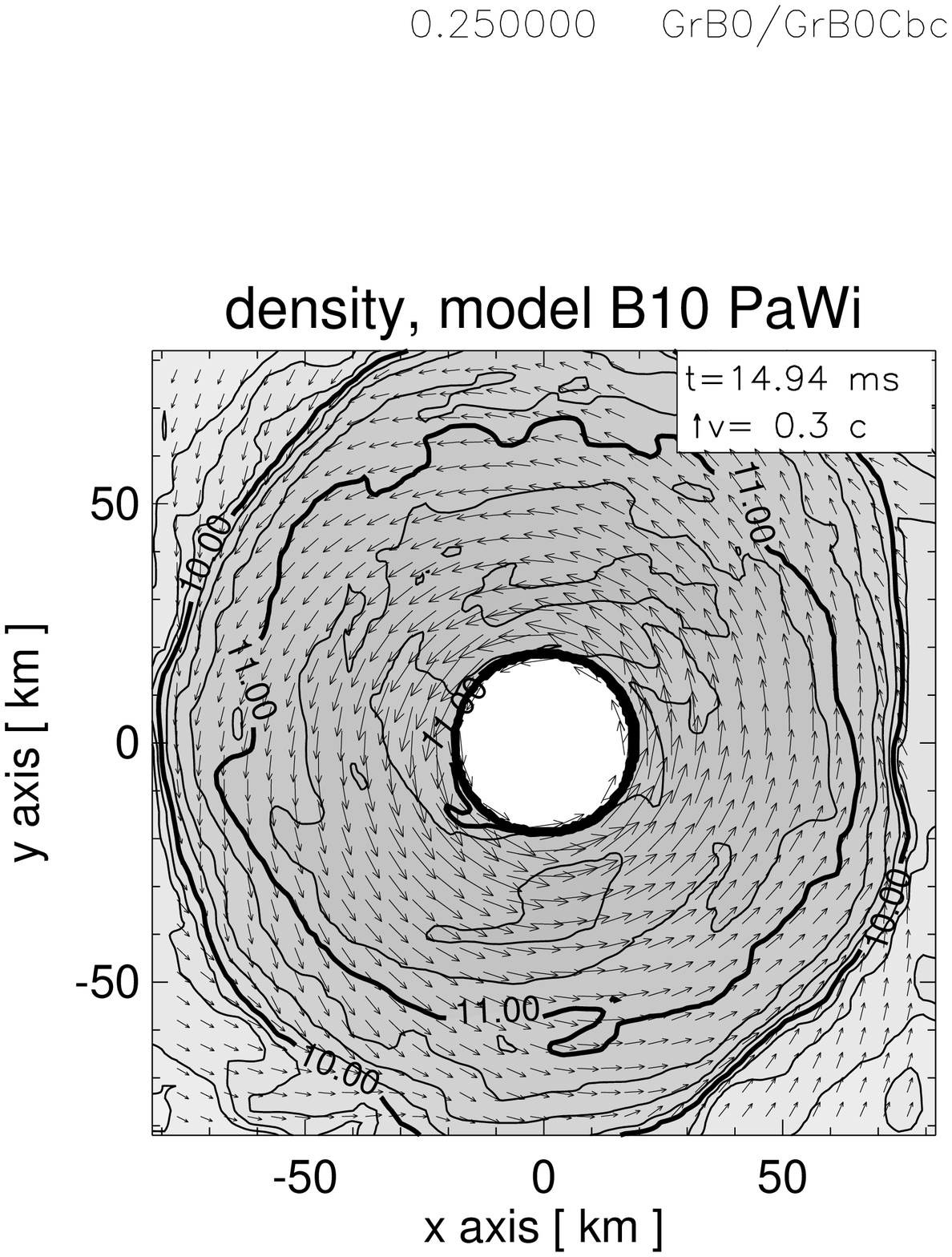}
  \put(-0.55,7.22){{\Large \bf \sf a}}   &
  \epsfxsize=8.5cm  \epsfclipon\epsffile{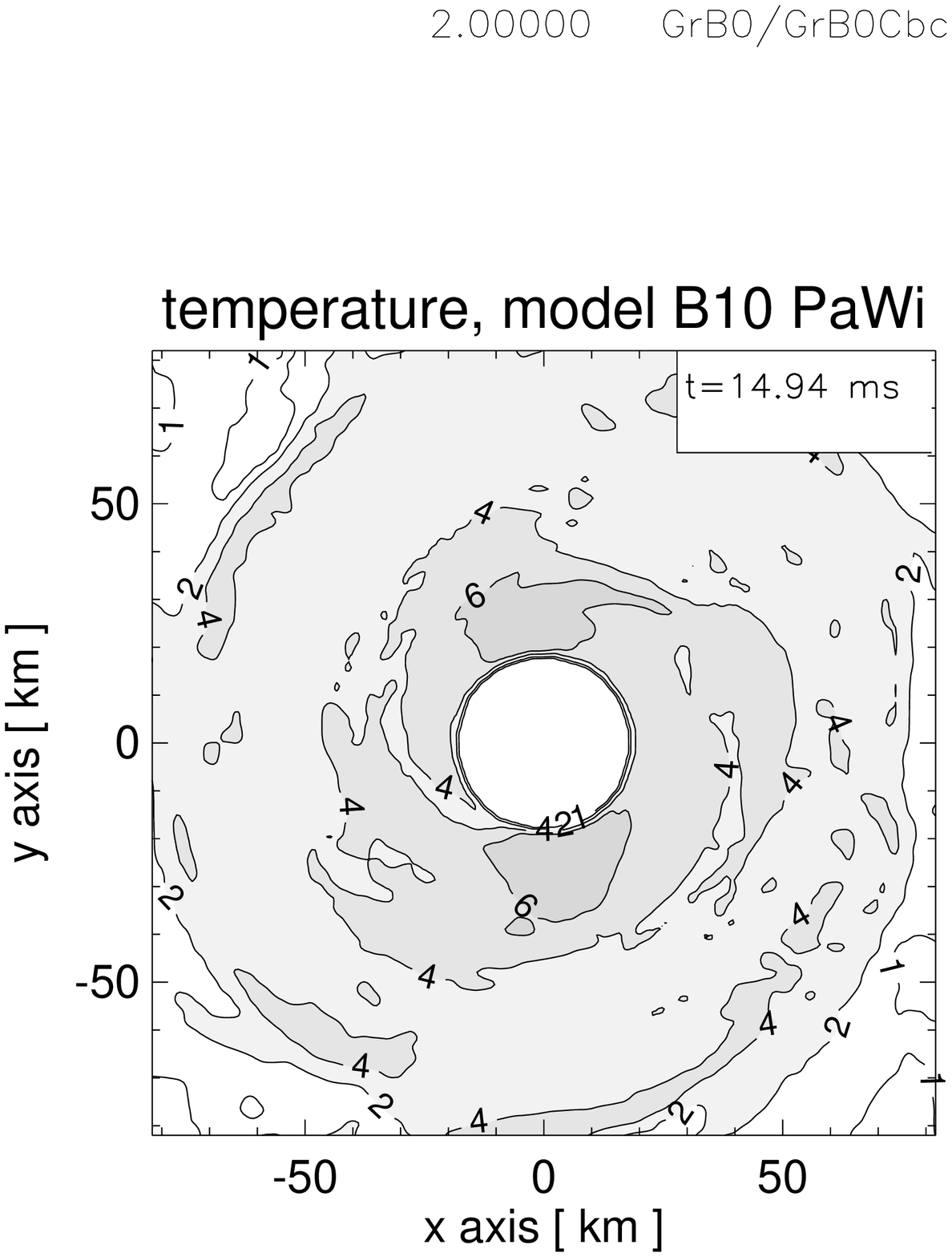}
  \put(-0.55,7.22){{\Large \bf \sf b}}  \\
[2ex]
  \epsfxsize=8.5cm  \epsfclipon\epsffile{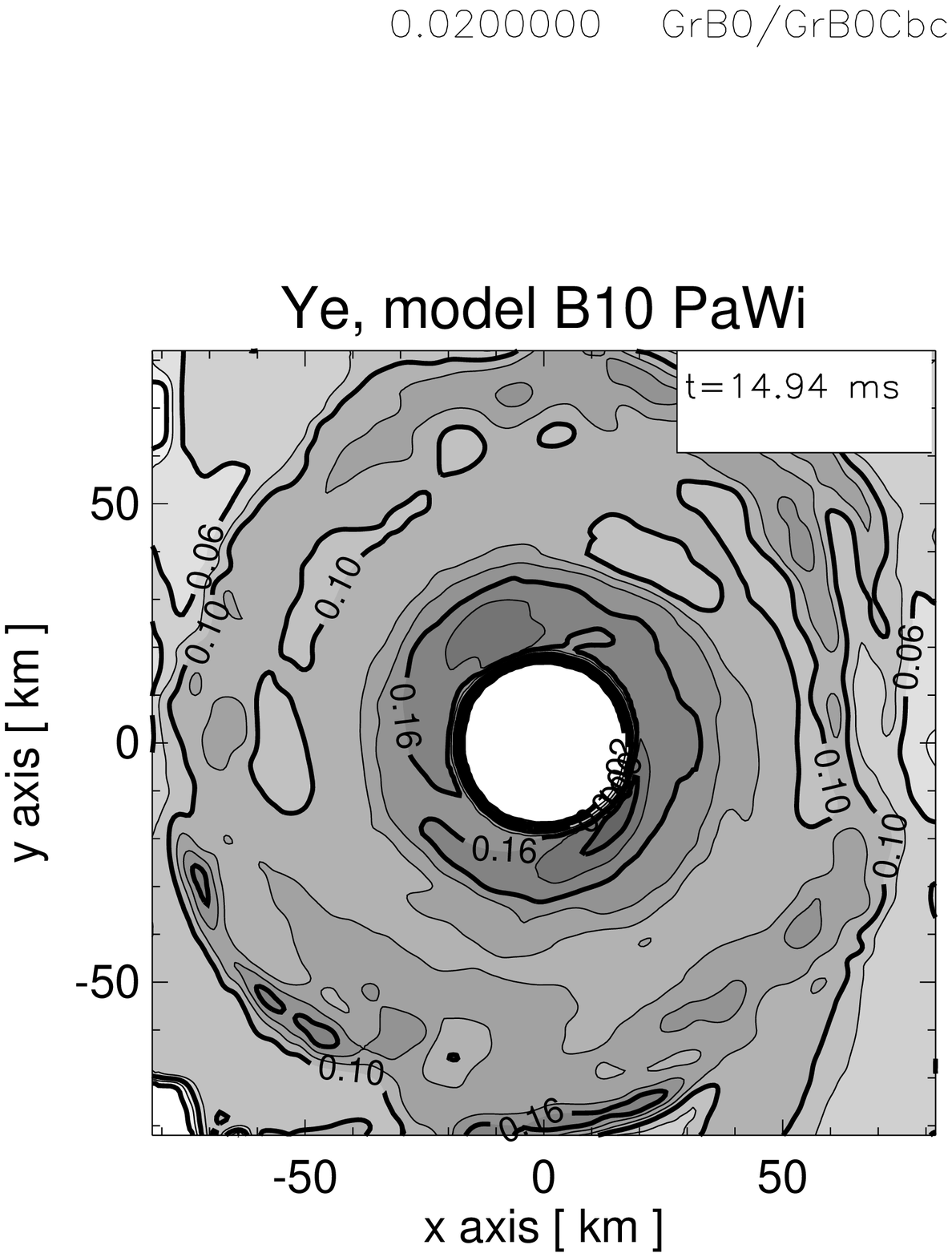}
  \put(-0.55,7.22){{\Large \bf \sf c}}  &
  \epsfxsize=8.5cm  \epsfclipon\epsffile{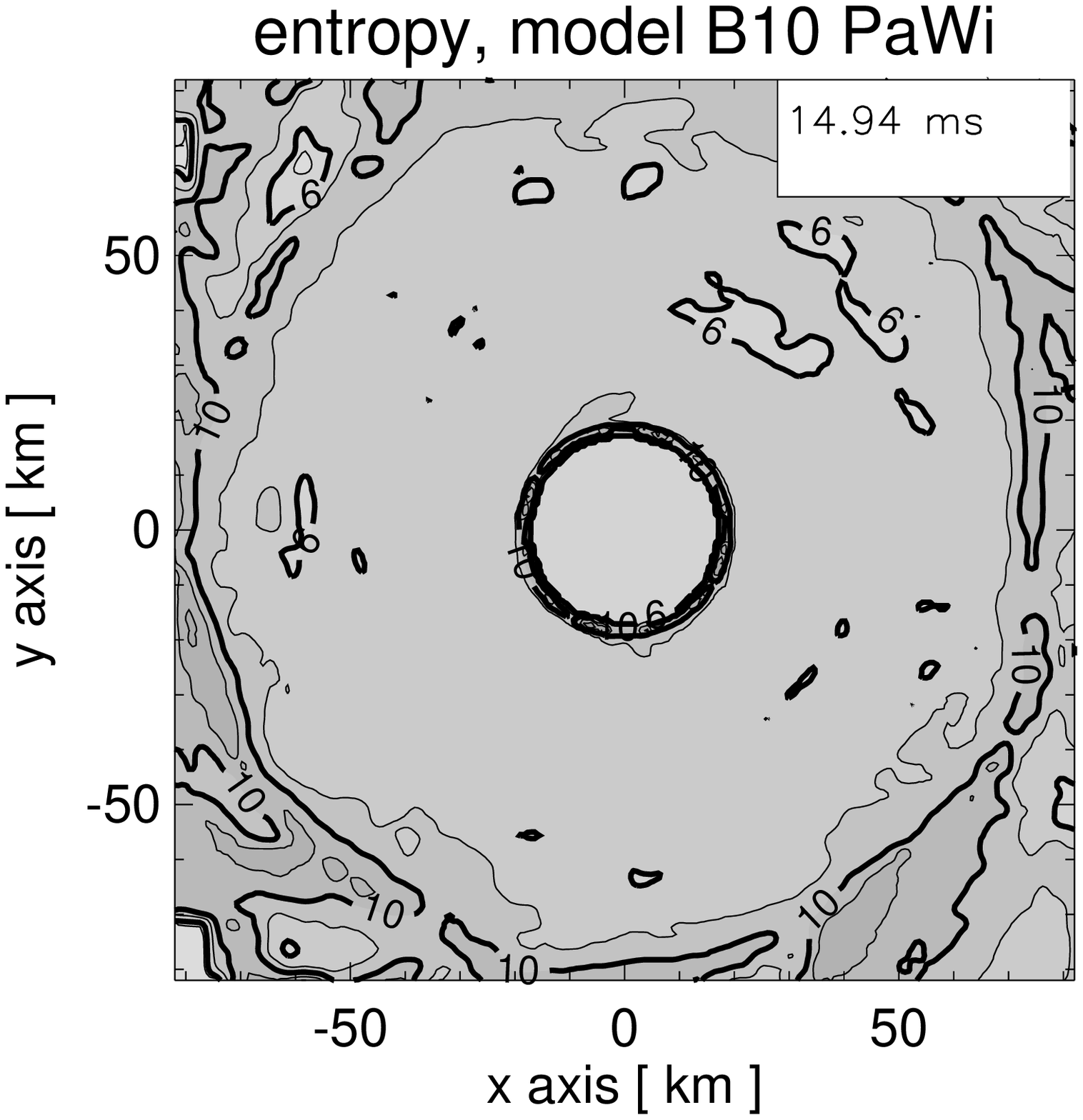}
  \put(-0.55,7.22){{\Large \bf \sf d}}  \\
\end{tabular}
\caption[]{
  Structure of the Paczy\'nski-Wiita Model~${\cal B}$10 in the equatorial plane
  near the end of the simulation. Panel~a shows contours of the baryonic 
  mass density (measured in ${\rm g\,cm}^{-3}$) together with the 
  flow field indicated by velocity vectors, panel~b displays contours
  of the temperature (in MeV), panel~c of the electron fraction $Y_e$,
  and panel~d of the entropy per nucleon.
  The length scale of the velocity vectors and the time elapsed since the
  beginning of the simulation of Model~B64 of Ruffert \& Janka (\cite{ruf98b})
  are given in the insert in the top right corner of the figures.
  The density contours are spaced logarithmically with intervals of
  0.25~dex (bold: $\log(\rho\eck{{\rm g\,cm}^{-3}}) = 10.0,\,11.0$), 
  the temperature contours are spaced linearly, starting with 
  1$\,$MeV and 2$\,$MeV and then continuing in steps of 2$\,$MeV
  (bold: $T = 10\,{\rm MeV}$), the $Y_e$ contours are spaced linearly 
  with intervals of 0.02 (bold: 0.02, 0.06, 0.10, 0.16), and the entropy
  levels were chosen to be 1, 2, 3, 4, 5, 6, 8, 10, 12, 14, 16, and
  20 (bold: 6, 10, 20). Some of the contours are labeled with their 
  respective values, and darker grey shading indicates higher values
  of a quantity. The circle around the center marks the inner ``vacuum'' 
  boundary with semidiameter of twice the Schwarzschild radius for
  the mass which has accumulated in the central black hole.}
  \label{fig:densP}
\end{figure*}

\begin{figure*}
\begin{tabular}{cc}
  \epsfxsize=8.5cm  \epsfclipon\epsffile{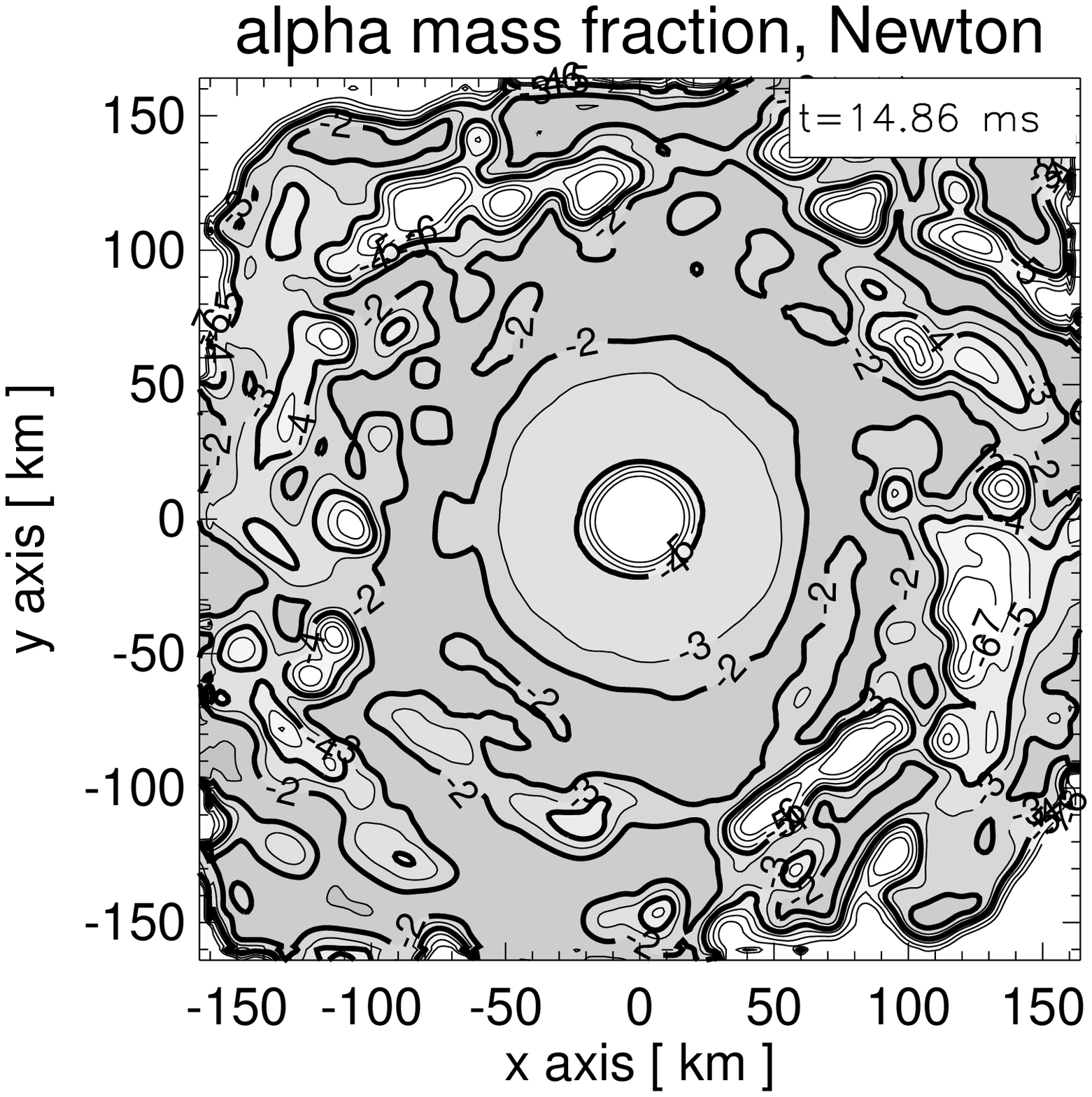}
  \put(-0.85,7.05){{\Large \bf \sf a}}   &
  \epsfxsize=8.5cm  \epsfclipon\epsffile{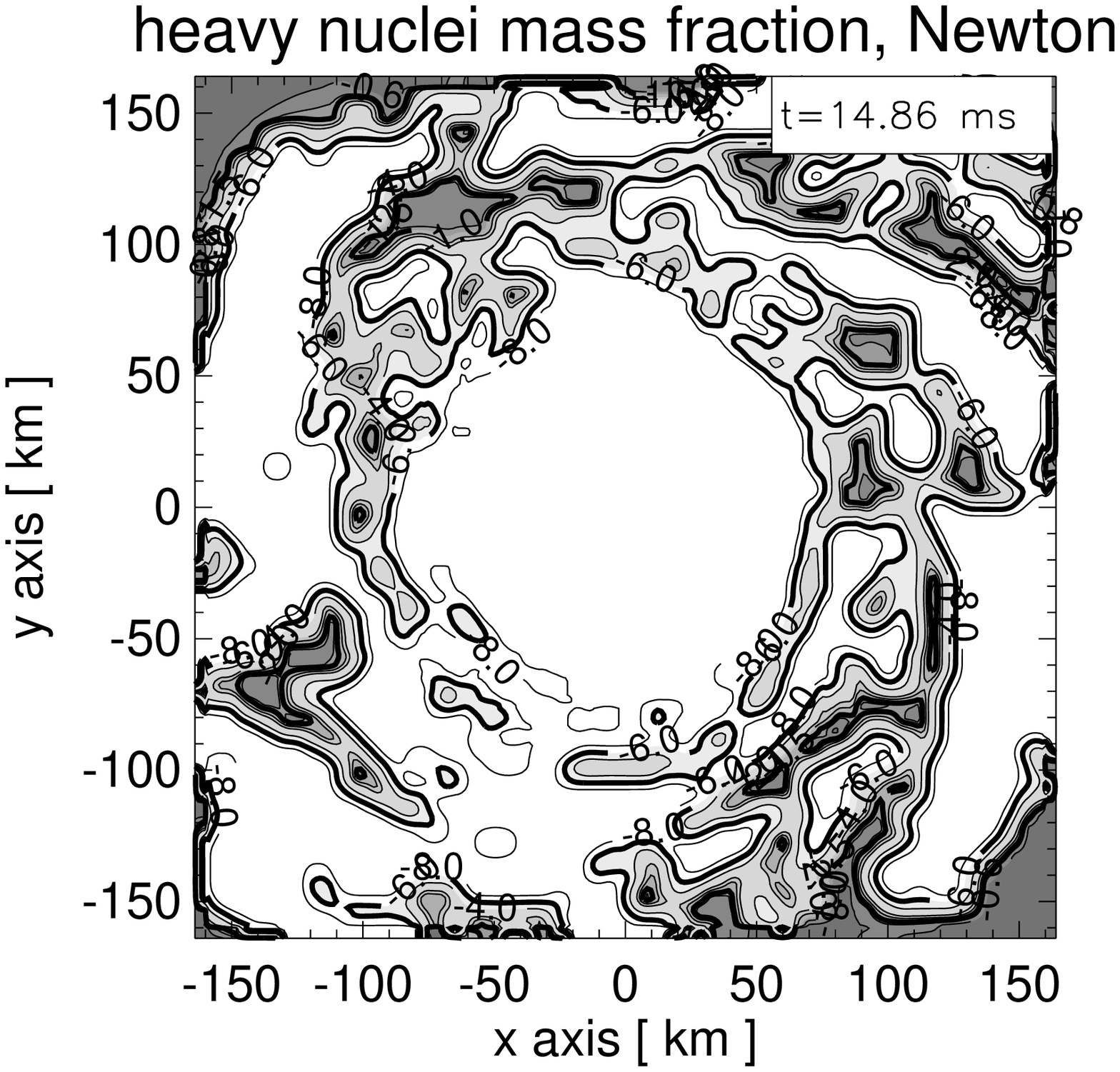}
  \put(-0.85,7.05){{\Large \bf \sf b}}  \\
[2ex]
  \epsfxsize=8.5cm  \epsfclipon\epsffile{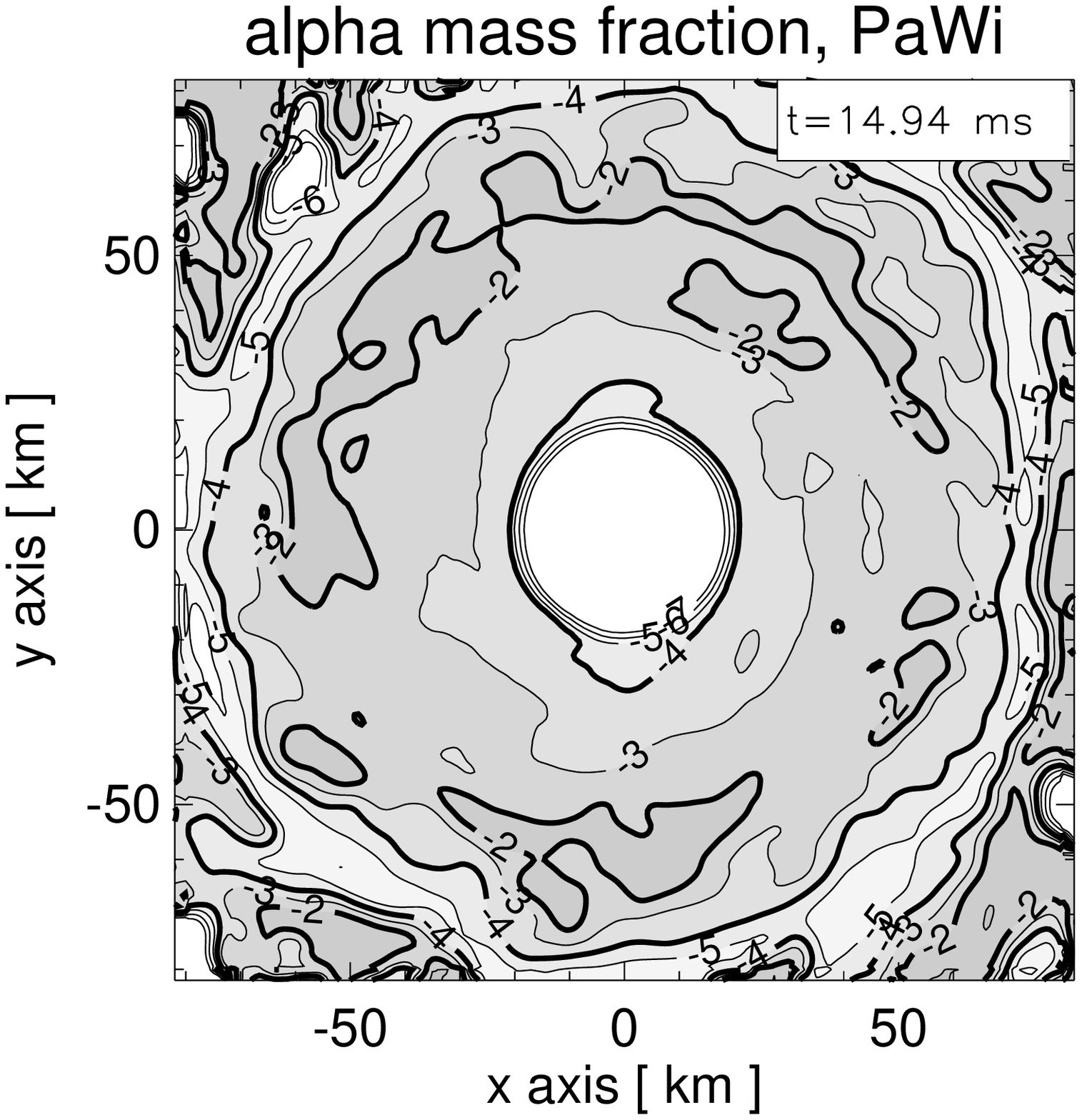}
  \put(-0.85,7.05){{\Large \bf \sf c}}  &
  \epsfxsize=8.5cm  \epsfclipon\epsffile{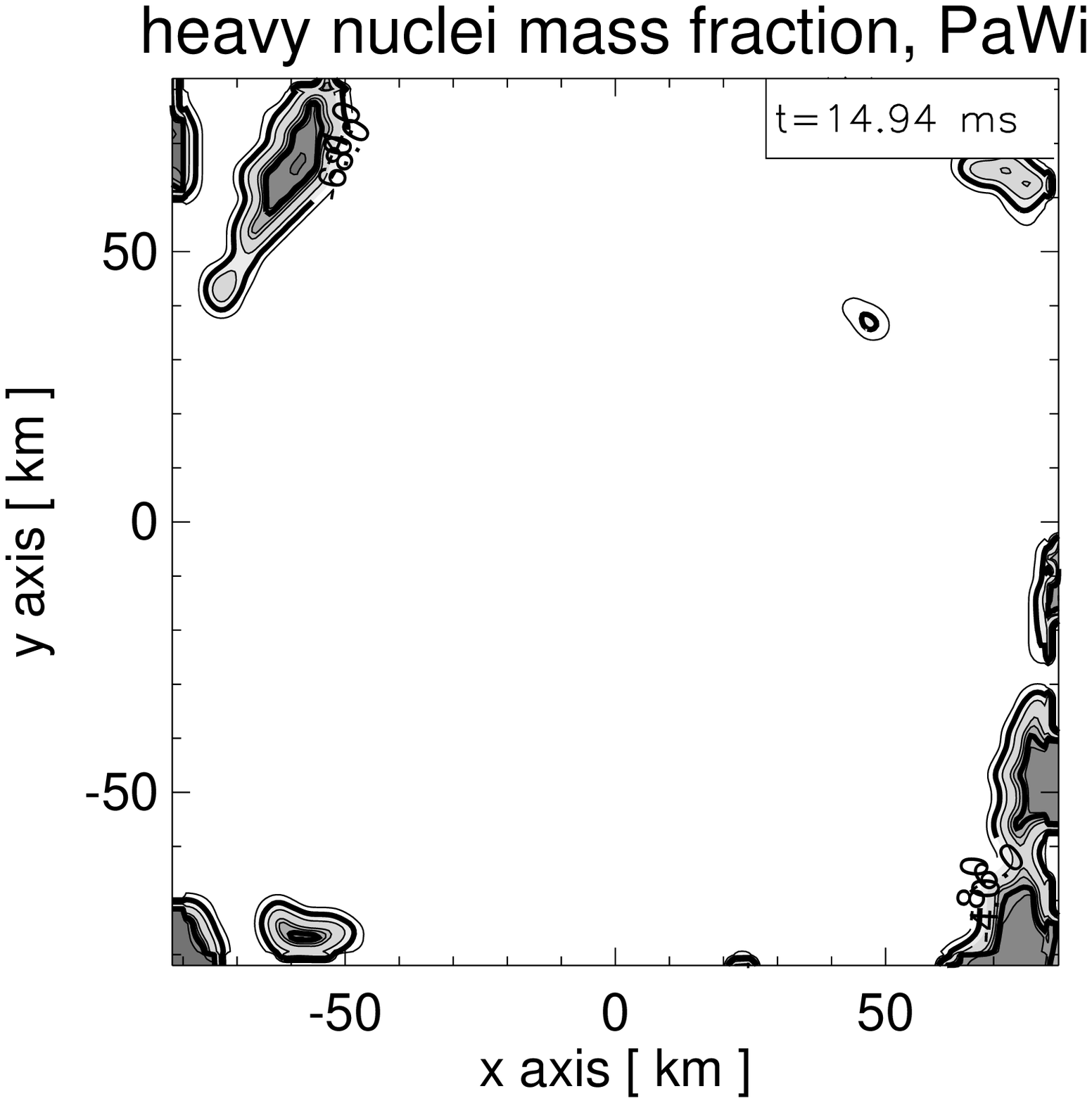}
  \put(-0.85,7.05){{\Large \bf \sf d}}  \\
\end{tabular}
\caption[]{
  Contour plots of the mass fractions of $\alpha$ particles and heavy
  nuclei in the equatorial plane of Model~{\bf B}10 (panels a and b,
  respectively) and of Model~${\cal B}$10 (panels c and d, respectively)
  at the times given in the inserts in the top right corners of the 
  figures. The contours are spaced logarithmically with steps of $-1$
  (starting at $-7$; bold lines for levels $-4$ and $-2$) 
  in case of the $\alpha$ mass fraction and with levels 
  $-8$, $-6$, $-4$, $-2$, $-1.5$, $-1$, $-0.8$, $-0.6$, $-0.4$, $-0.2$ and 0 
  (bold lines for levels $-6$ and $-1$) in case of heavy nuclei.
  Typical mass numbers $A$ of heavy nuclei are around 80--120 with
  charge numbers $Z\approx 25$--50.
  Darker grey indicates higher values of the mass fractions.
  }
  \label{fig:massfrac}
\end{figure*}

\section{Dynamical evolution\label{sec:hydro}}

Figures~\ref{fig:massaccBH}--\ref{fig:energy} show the evolution of 
the remnant of the merged neutron stars after the compact central
object has been replaced by a vacuum sphere to represent the black hole.
The different simulations, discriminated by the use of a Newtonian or
Paczy\'nski-Wiita type of gravitational potential and 
by different times of the assumed black hole formation, are represented
by different line styles. Time is measured from the start of the 
simulation of Model~B64 of Ruffert \& Janka (\cite{ruf98b}).

In Fig.~\ref{fig:massaccBH} the mass accretion rates of the black hole
are given as functions of time for all models, in Fig.~\ref{fig:massBH} 
the evolution of the black hole mass is displayed for 
the Newtonian and Paczy\'nski-Wiita models which were
started at $t = 10\,$ms (Models~{\bf B}10 and ${\cal B}$10, respectively),
and in Fig.~\ref{fig:masstBH} the corresponding time scales for the 
changes of black hole and torus masses are shown. Initially, the black
hole swallows the surrounding mass at rates as high as several solar
masses per millisecond, but within a dynamical time scale of only about
1$\,$ms the accretion rates settle to much lower values between 
$5\,M_{\odot}\,{\rm s}^{-1}$ and $7\,M_{\odot}\,{\rm s}^{-1}$.
The peak rates as well as the rates towards the end of the simulations
are similar in all models. The amount of surrounding gas which is 
dynamically accreted into the black hole is larger for the
Paczy\'nski-Wiita case, and the black hole mass grows correspondingly
faster (Fig.~\ref{fig:massBH}). When the simulations are stopped
after a quasi-stationary situation has been reached, the torus is
therefore about 10 times more massive in the Newtonian models 
(torus mass $M_{\rm v}^{\rm N}\approx 0.25$--$0.35\,M_{\odot}$;
compare Table~\ref{tab:models}) than 
in those with Paczy\'nski-Wiita potential ($M_{\rm v}^{\rm PW} \approx 
0.005$--$0.035\,M_{\odot}$; compare Table~\ref{tab:models}). Note, however,
that the simulations with the Paczy\'nski-Wiita black hole potential might 
well underestimate the torus mass. Since these simulations were started
with an initial model that resulted from a Newtonian computation of neutron
star merging, the initial angular momentum of the gas may be lower than
would have been obtained in a simulation with the stronger relativistic 
potential. Towards the end of the simulations, the
time scales of the changes of black hole mass and accretion torus mass
level off at values much larger than the dynamical time scale.
The accretion time scale of the Newtonian torus grows to about $60\,$ms,
whereas the accretion time scale for the less massive torus in the 
Paczy\'nski-Wiita case is approximately one order of magnitude shorter
(Fig.~\ref{fig:masstBH}).

\begin{figure*}
\tabcolsep=2.0mm
 \begin{tabular}{cc}
 \epsfxsize=8.5cm \epsfclipon \epsffile{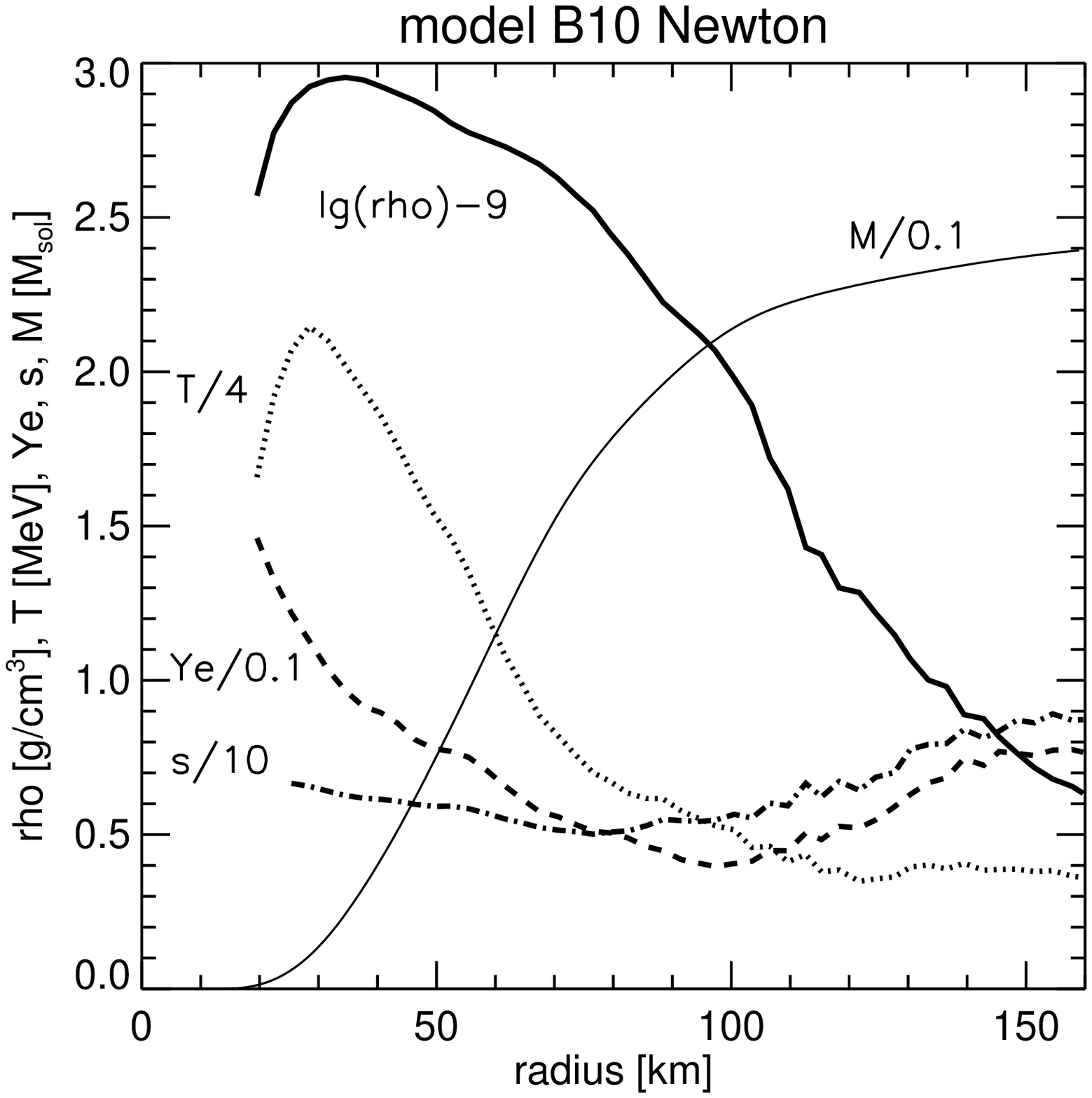} &
 \epsfxsize=8.5cm \epsfclipon \epsffile{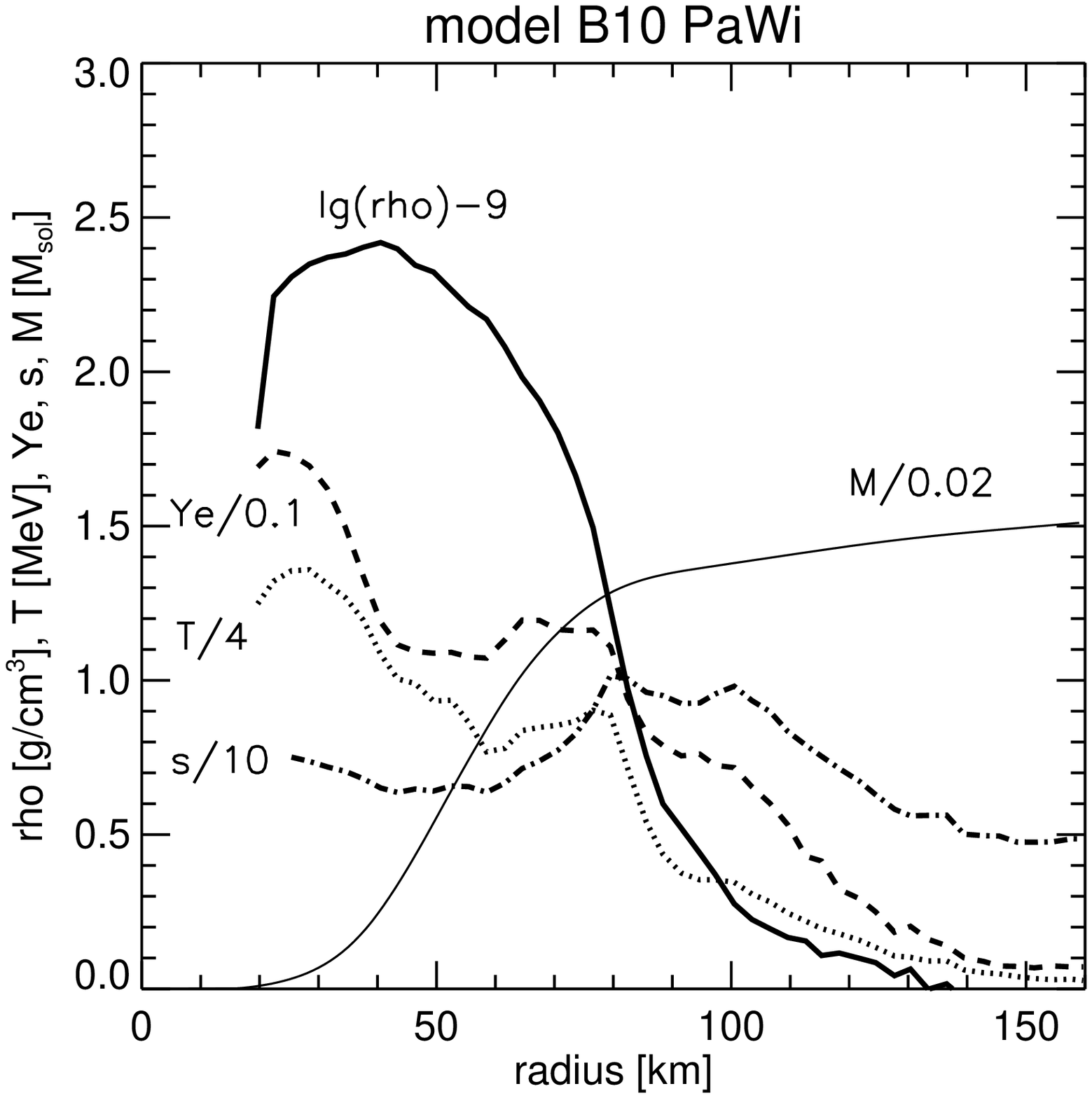} \\
 \end{tabular}
 \caption[]{Azimuthally averaged radial structure of the accretion 
 torus in the equatorial plane of Model~{\bf B}10 (left) and of
 Model~${\cal B}$10 (right) at the end of the simulations. The 
 curves show as functions of the equatorial radius $d$ the logarithm 
 of the density $\rho(d)$ (measured in
 ${\rm g\,cm}^{-3}$), the temperature $T(d)$ (in MeV), the electron
 fraction $Y_e(d)$, the entropy per nucleon $s(d)$, and the total
 cumulative mass in the torus, $M(d)$ (in units of $M_{\odot}$). 
 All curves are normalized to fit into the scale given on the ordinate
 and start at twice the Schwarzschild radius $R_{\rm s}$ of the mass that
 has accumulated in the central black hole. Note that the gas density
 drops significantly inside $d\approx 3R_{\rm s}\approx 27\,$km.} 
 \label{fig:phiav} 
\end{figure*}

Figure~\ref{fig:diskmass} demonstrates that the torus mass --- given
as the gas mass $M_{\rm v}$ on the grid ---
approaches nearly the same value in the quasi-stationary state when the 
moment of black hole formation is $t_{\rm init} > 2\,{\rm ms}$.
In contrast, if the black hole is assumed to form early after
the merging of the neutron stars ($t_{\rm init}< 2\,{\rm ms}$;
Models~{\bf B}1 and ${\cal B}$1), it swallows a lot of the gas which
otherwise ends up in the torus because it has been spun off the surface
of the rapidly rotating and oscillating merger remnant. 
The gas cloud that surrounds the compact central object has an
average density above $10^{12}\,{\rm g\,cm}^{-3}$ initially,
whereas at later times its density is below $10^{12}\,{\rm g\,cm}^{-3}$
because it was heated by shocks and viscous dissipation and becomes
inflated due to the thermal gas pressure. 
If the simulation is started at a moment
when this gas is in a post-merging phase of expansion, some of the matter
(up to $2\times 10^{-2}\,M_{\odot}$) can be ejected to leave the outer 
boundary of the largest grid and to become unbound (determined from the 
criterion that the total energy as the sum of kinetic, internal, and
potential energies is positive). For the Newtonian Model~{\bf B}4 which 
was carried on for a time long enough to follow this gas until it has
reached the outer grid boundary, the value of the lost mass is in 
very good agreement with the mass loss ($0.024\,M_{\odot}$) found in
Model~B64 of Ruffert \& Janka (\cite{ruf98b}). This shows that the evolution of
the gas swept out in large spiral arms becomes independent of the dynamical
evolution of the compact remnant at the center of the merger very early.
In case of the Newtonian potential 
(left plot in Fig.~\ref{fig:diskmass}), the thin
solid lines labeled with $M_{\rm d}$ represent the mass on the grid
with specific angular momentum larger than the Kepler value at 3
Schwarzschild radii, i.e., $j \ge j_{\rm N}^\ast 
\equiv v_{\rm Kepler}^{\rm N}(3R_{\rm s})3R_{\rm s} = \sqrt{6}GM/c$.
For $M$ the sum of the black hole and gas masses on the grid was used at 
all times. Since this overestimates the gravitational potential compared 
to the black hole in the numerical simulation, $j_{\rm N}^\ast$
is too strict a limit on $j$. Therefore $M_{\rm d}$ is systematically
somewhat smaller than $M_{\rm v}$ but yields a reasonably good a priori 
estimate of the mass which can be found in the accretion torus in the
quasi-stationary state.

The maximum density on the grid as a function of time (Fig.~\ref{fig:maxrhoT})
shows a rapid drop from initially 2 to $3\times 10^{14}\,{\rm g\,cm}^{-3}$
to about 30--100 times smaller values within only 1$\,$ms, corresponding to
the catastrophic dynamical accretion after the start of the torus simulations. 
This steep decrease of the density is associated with the sudden replacement
of the compact central object of the merger remnant by the vacuum sphere
to model the black hole. The physical, more gradual generation of the event
horizon will most probably soften this density drop somewhat.
The maximum density then levels off to a value which gradually decreases on the 
much longer evolution time associated with the subsequent accretion of torus 
material onto the black hole (Fig.~\ref{fig:masstBH}). Since the mass remaining
in the accretion torus is larger and its accretion time scale is longer in the
Newtonian simulations, the value of the maximum density at the end of the 
simulations is higher there, and the temporal decrease
of $\rho_{\rm max}(t)$ is slower. The same tendencies can be seen
for the maximum temperatures in the Newtonian and Paczy\'nski-Wiita models 
(Fig.~\ref{fig:maxtempT}). The hottest spots in the accretion tori have 
temperatures around 10$\,$MeV at the end of the simulations with a trend to
somewhat larger values for the models where the black hole formation was
assumed to occur late. In these cases the gas in the surroundings of the compact 
remnant of the merger has experienced additional heating by the shocks and 
compression waves created by the violent oscillations of the central compact
object.

Since the temperatures are similar and the thermal energy of nondegenerate 
baryons contributes a dominant fraction of the internal energy of the gas,
the specific internal energies of the Newtonian and the Paczy\'nski-Wiita tori
(plotted as functions of time in Fig.~\ref{fig:ergspec} for Models~{\bf B}10
and ${\cal B}$10, respectively) are very similar, in both cases around 
$1.5\times 10^{19}\,{\rm erg\,g}^{-1}$ or about 15.5$\,$MeV per nucleon.
In contrast, the kinetic, internal, and potential energies of both models
become very different towards the end of the simulations (Fig.~\ref{fig:energy})
which is explained by the large difference of the torus masses (compare
$M_{\rm v}$ in the two plots of Fig.~\ref{fig:diskmass}).
%
\newpage

\begin{figure*}
\begin{tabular}{cc}
  \epsfxsize=8.5cm  \epsfclipon\epsffile{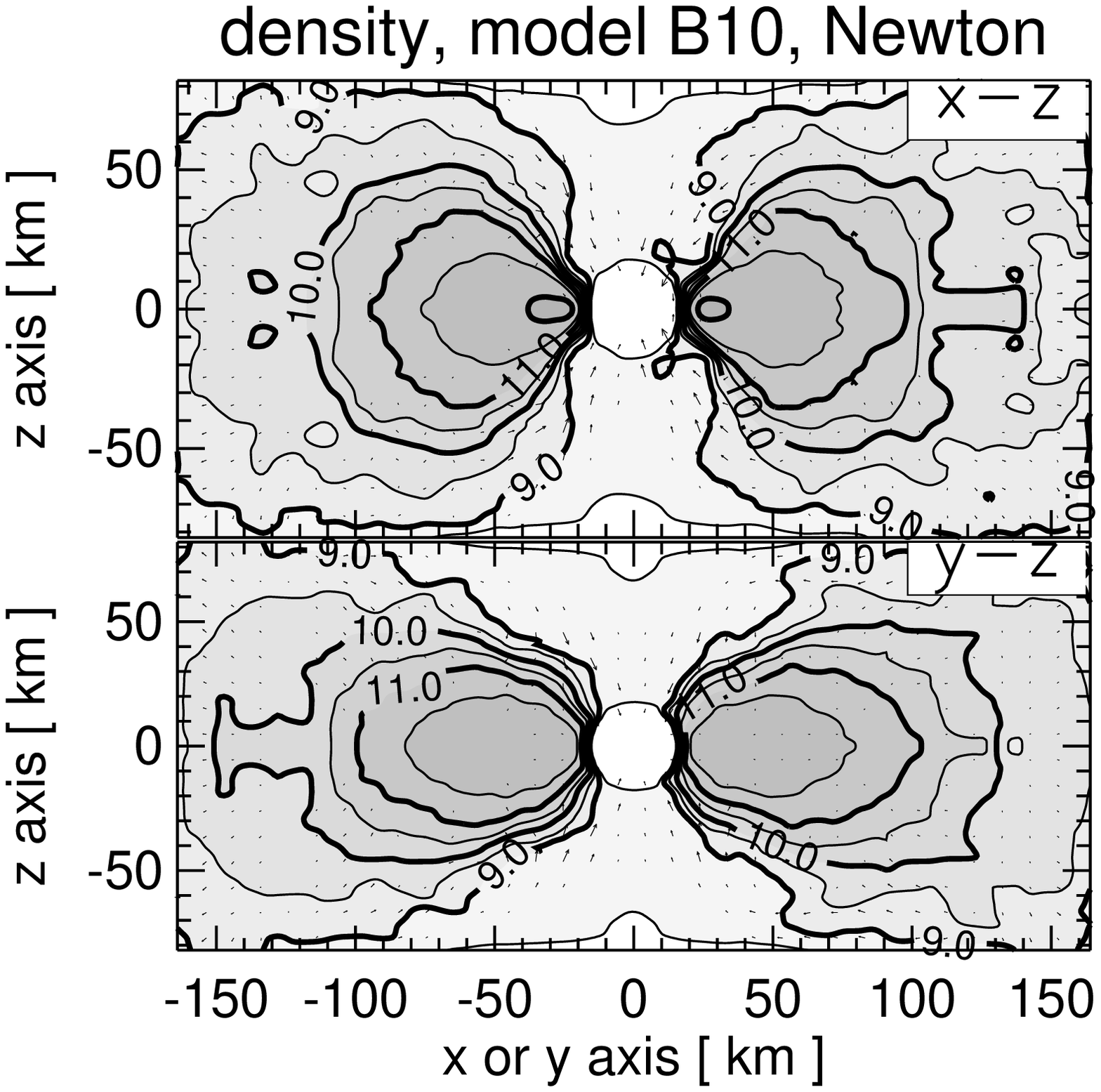}
                                         &
  \epsfxsize=8.5cm  \epsfclipon\epsffile{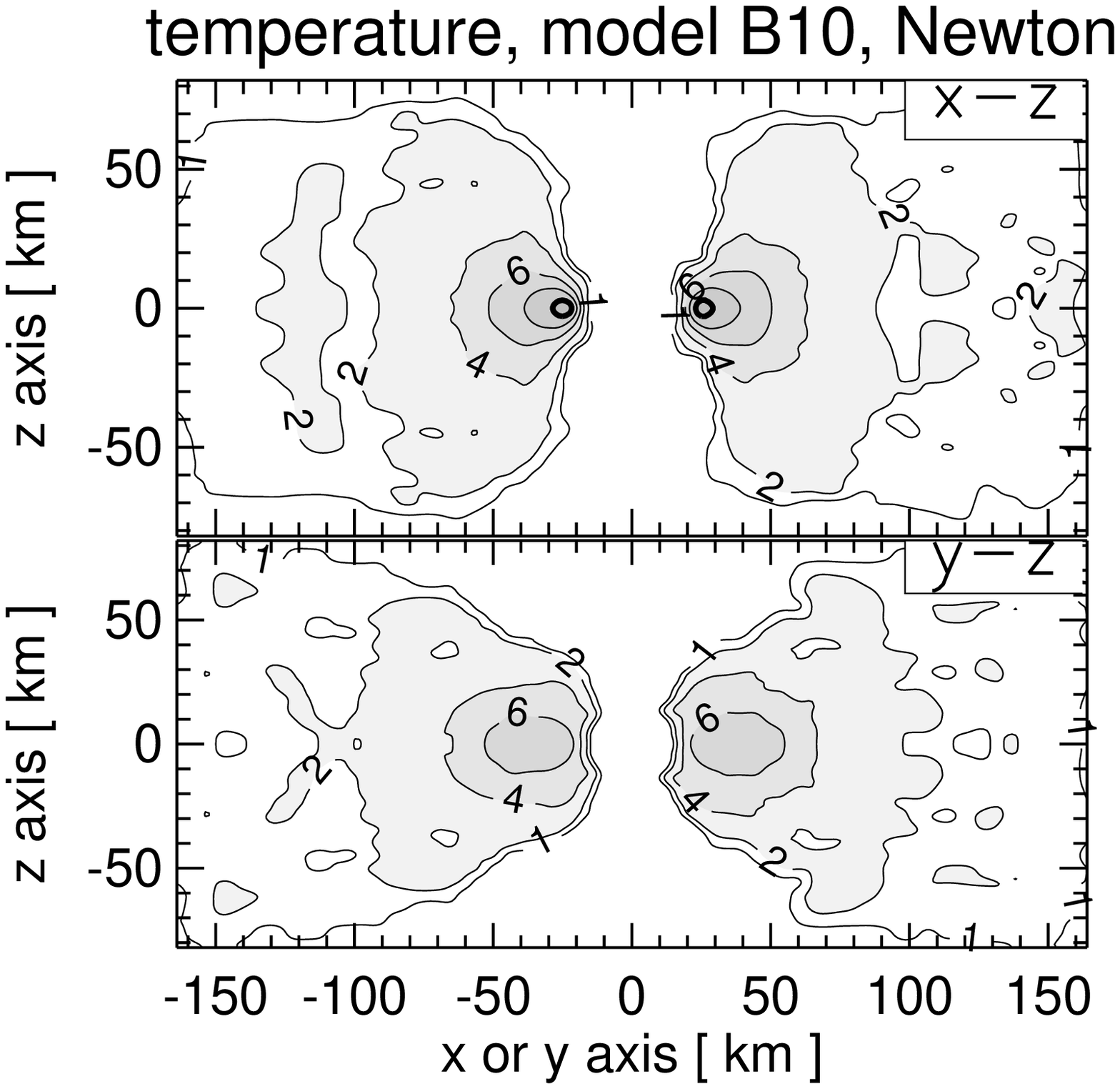}
                                         \\
  \epsfxsize=8.5cm  \epsfclipon\epsffile{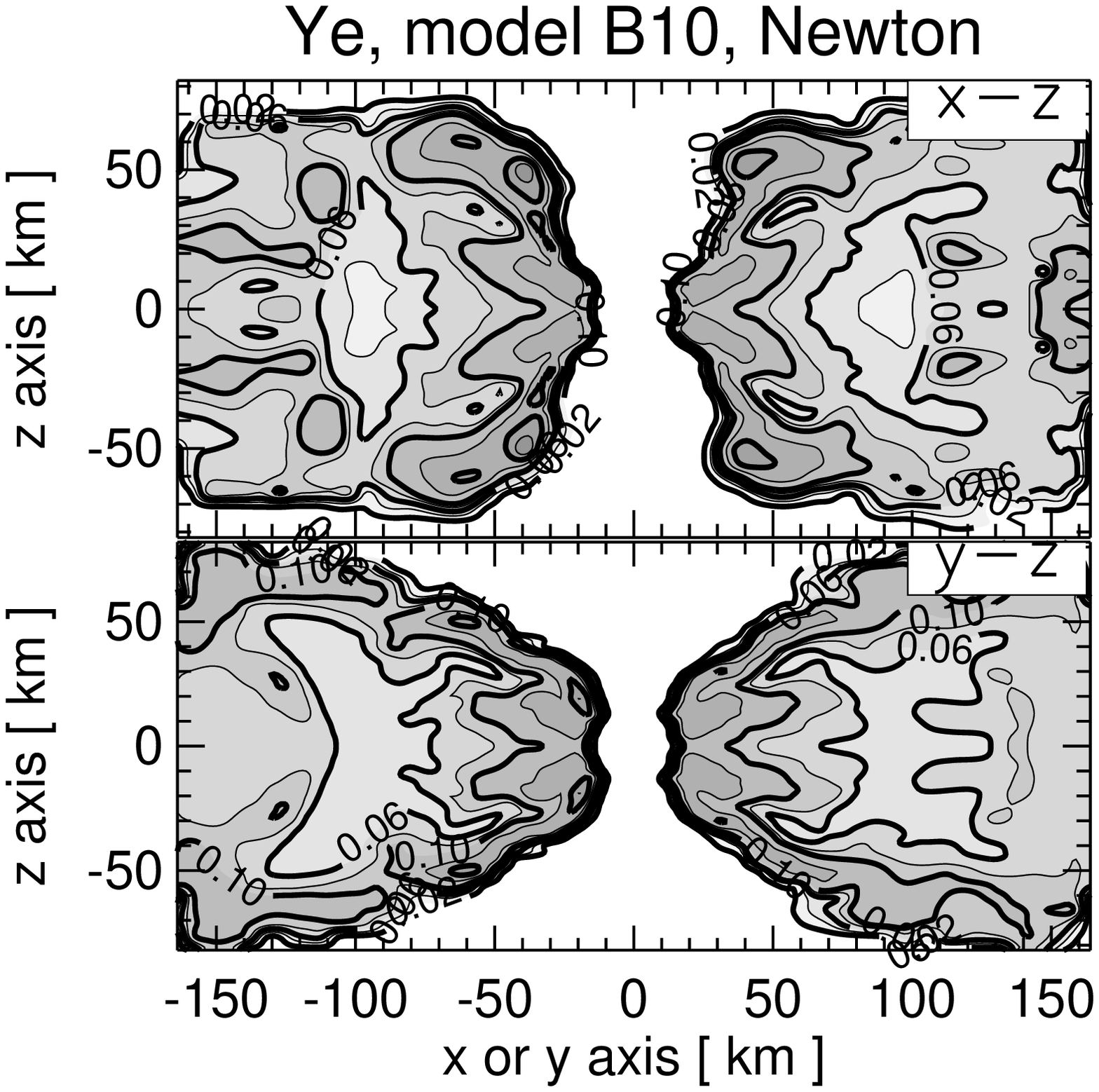}
                                         &
  \epsfxsize=8.5cm  \epsfclipon\epsffile{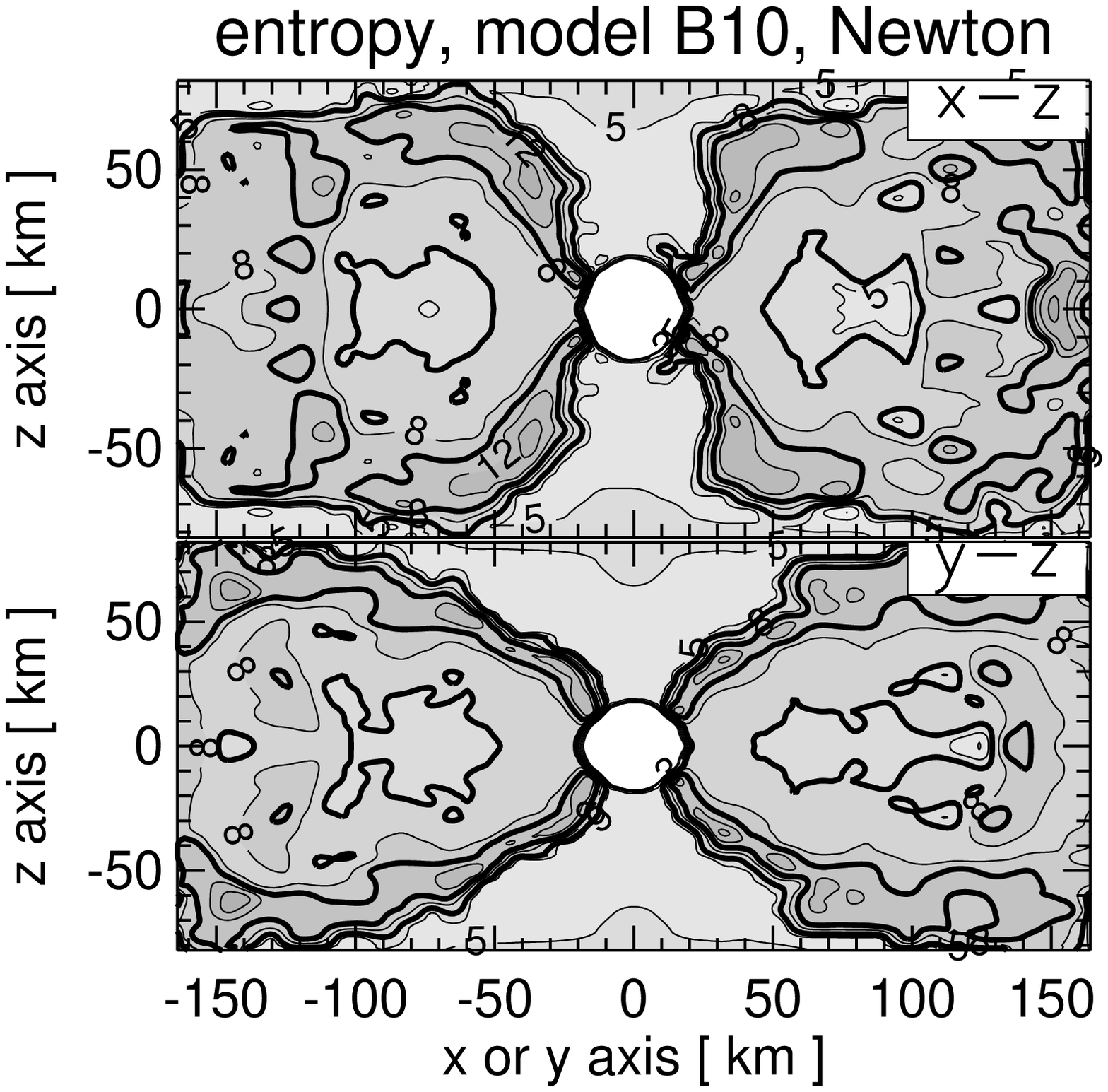}
                                         \\
\end{tabular}
\caption[]{
  Vertical structure of the Newtonian Model~{\bf B}10 in the 
  $x$-$z$-plane and $y$-$z$-plane near the end of the simulation
  ($t \approx 14.5\,$ms).
  The plots correspond to the cuts in the equatorial plane given in
  Fig.~\ref{fig:densN}. The upper left figure shows contours of the baryonic 
  mass density (measured in ${\rm g\,cm}^{-3}$) together with the
  flow field indicated by velocity vectors, the upper right figure
  displays contours of the temperature (in MeV), the lower left figure
  of the electron fraction $Y_e$, and the lower right figure of the
  entropy per nucleon.
  The density contours are spaced logarithmically with intervals of
  0.5~dex (bold: $\log(\rho\eck{{\rm g\,cm}^{-3}}) = 9.0,\,10.0,\,11.0,\,12.0$), 
  the temperature contours are spaced linearly, starting with 1$\,$MeV 
  and 2$\,$MeV, and then continuing in steps of 2$\,$MeV
  (bold: $T = 10\,{\rm MeV}$), the $Y_e$ contours 
  correspond to the levels {\bf 0.02}, 0.04, {\bf 0.06}, 0.08, {\bf 0.10}, 
  0.15, {\bf 0.20}, 0.25, and {\bf 0.30}, and the entropy
  levels were chosen to be 1, 2, 3, 4, 5, 6, 8, 10, 12, 14, 16, and
  20 (bold: 6, 10, 20). Some of the contours are labeled with their
  respective values, and darker grey shading indicates higher values
  of a quantity.}
  \label{fig:vertN}
\end{figure*}

\begin{figure*}
\begin{tabular}{cc}
  \epsfxsize=8.5cm  \epsfclipon\epsffile{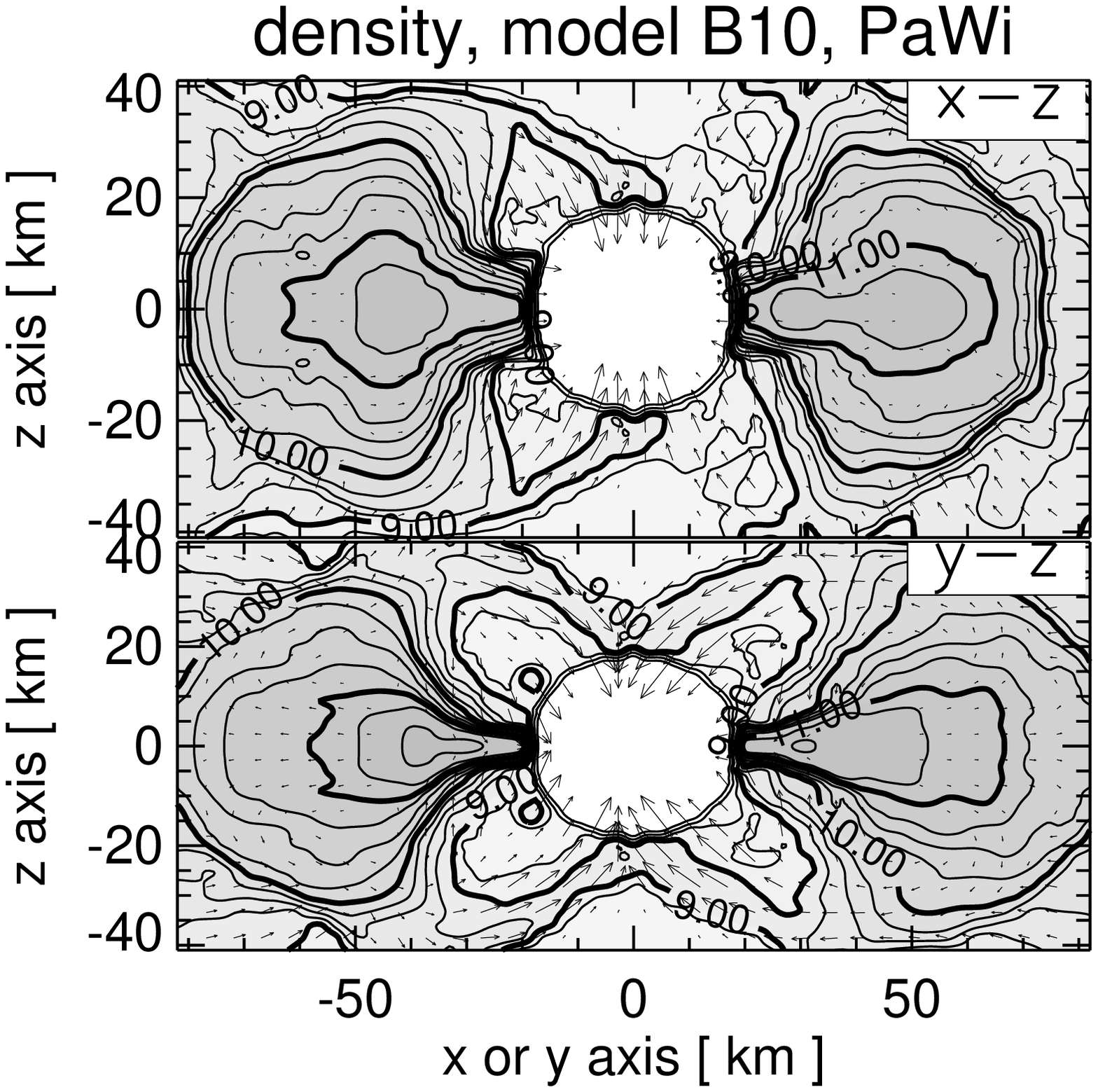}
                                         &
  \epsfxsize=8.5cm  \epsfclipon\epsffile{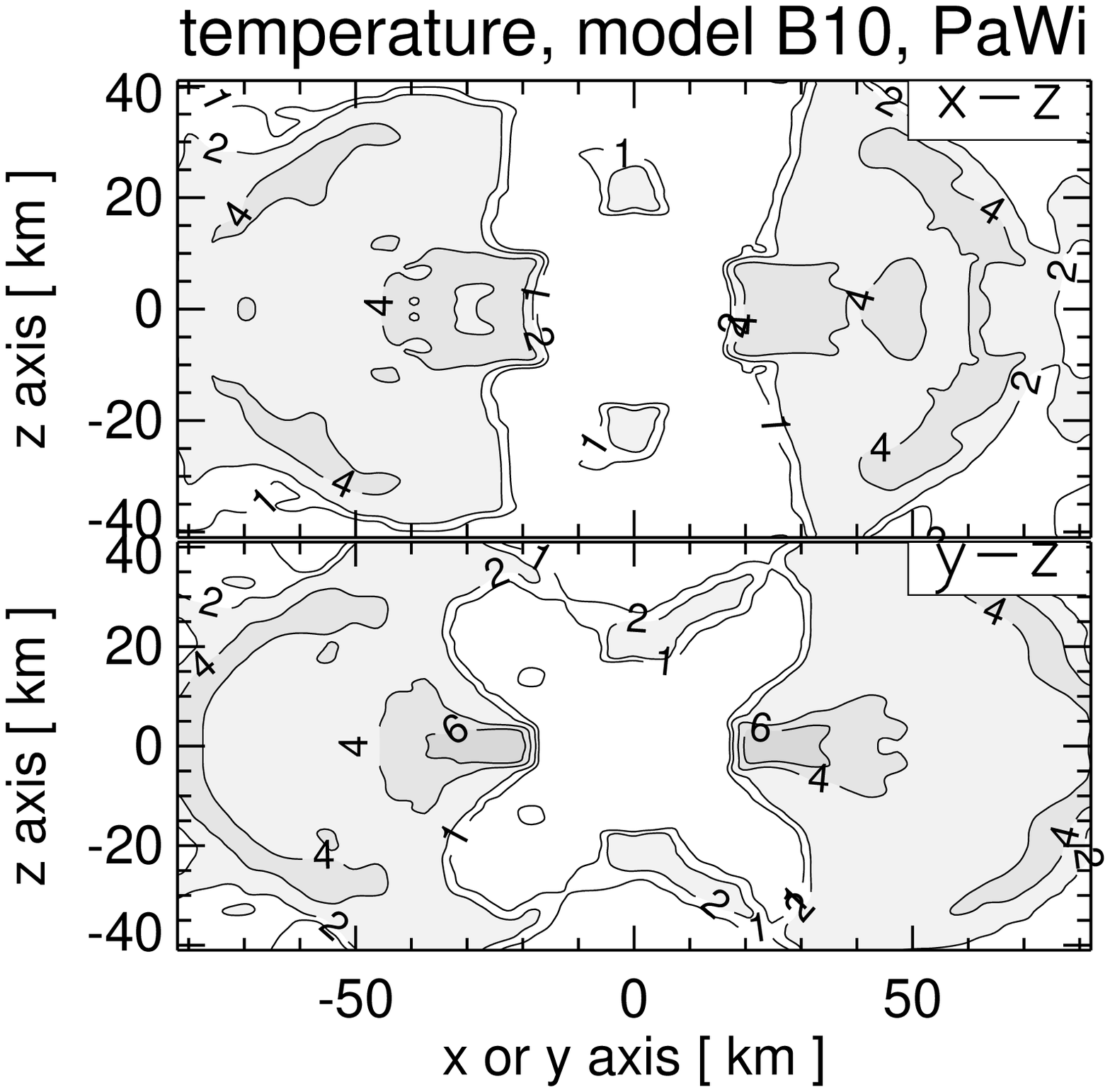}
                                         \\
  \epsfxsize=8.5cm  \epsfclipon\epsffile{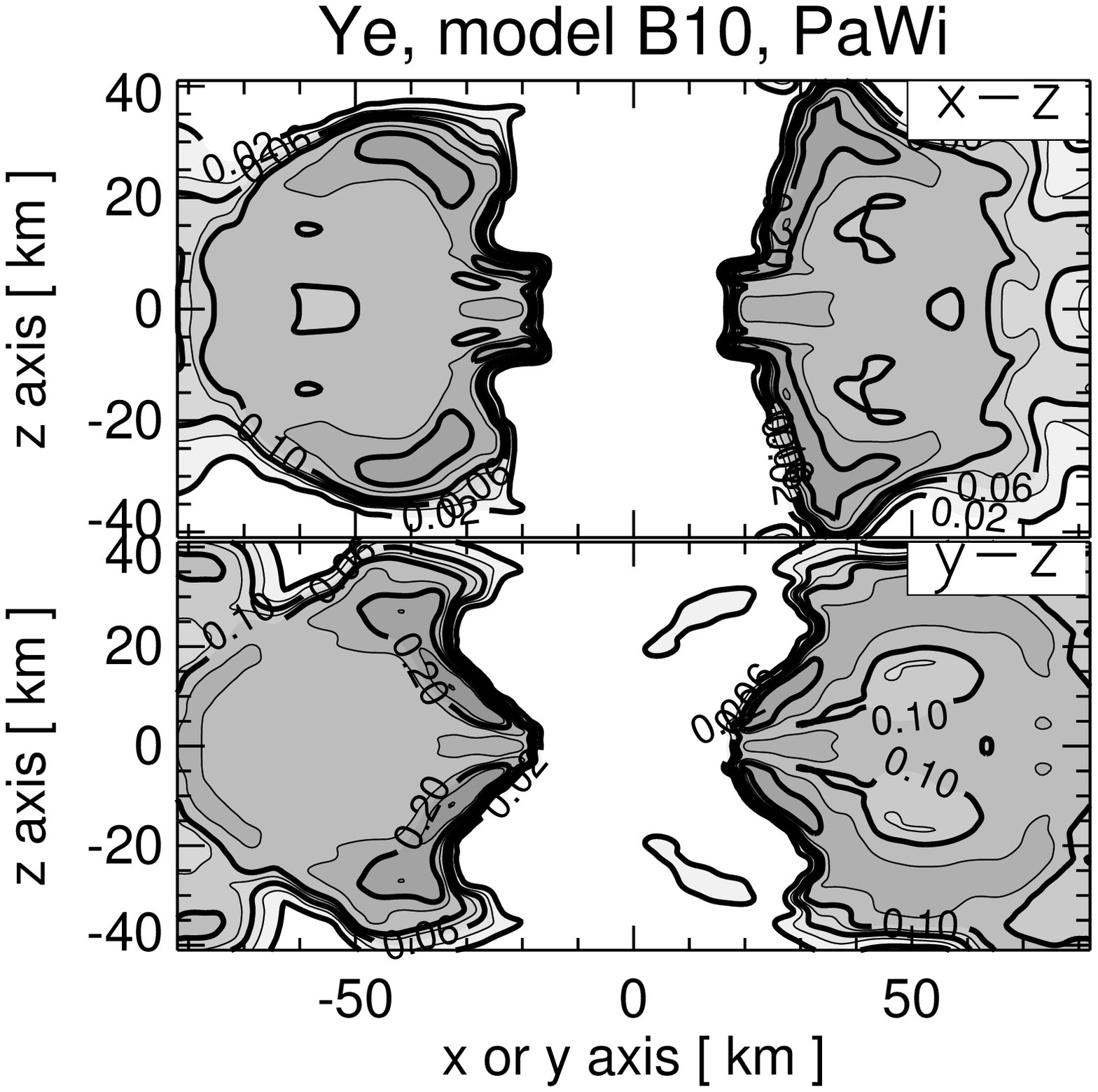}
                                         &
  \epsfxsize=8.5cm  \epsfclipon\epsffile{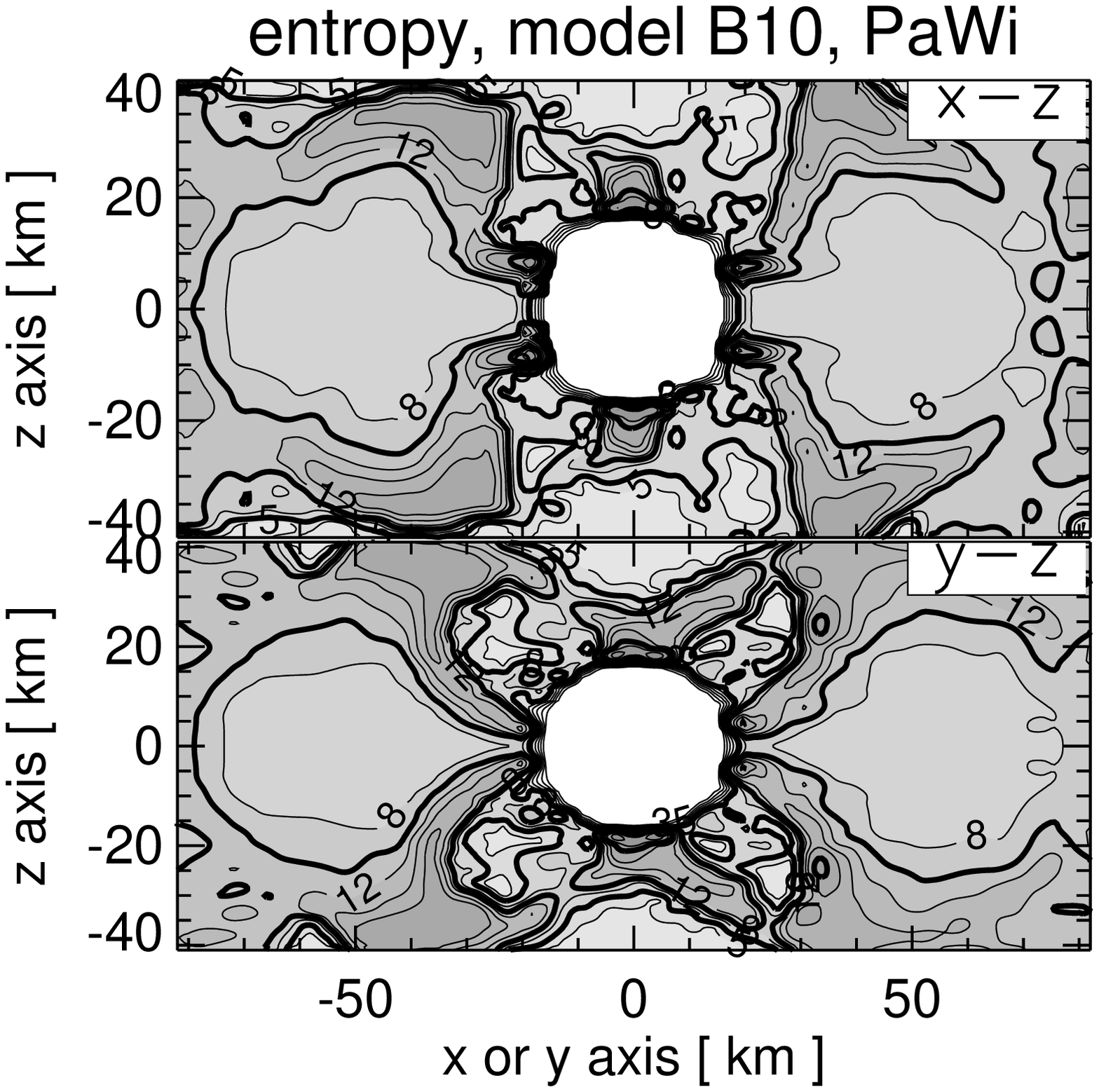}
                                         \\
\end{tabular}
\caption[]{
  Vertical structure of the Paczy\'nski-Wiita Model~${\cal B}$10 in the
  $x$-$z$-plane and $y$-$z$-plane near the end of the simulation
  ($t \approx 14.9\,$ms).
  The plots correspond to the cuts in the equatorial plane given in
  Fig.~\ref{fig:densP}. The upper left figure shows contours of the baryonic
  mass density (measured in ${\rm g\,cm}^{-3}$) together with the
  flow field indicated by velocity vectors, the upper right figure
  displays contours of the temperature (in MeV), the lower left figure
  of the electron fraction $Y_e$, and the lower right figure of the
  entropy per nucleon.
  The density contours are spaced logarithmically with intervals of
  0.25~dex (bold: $\log(\rho\eck{{\rm g\,cm}^{-3}}) = 9.0,\,10.0,\,11.0$),
  the temperature contours are spaced linearly, starting with 1$\,$MeV
  and 2$\,$MeV, and then continuing in steps of 2$\,$MeV
  (bold: $T = 10\,{\rm MeV}$), the $Y_e$ contours
  correspond to the levels {\bf 0.02}, 0.04, {\bf 0.06}, 0.08, {\bf 0.10},
  0.15, {\bf 0.20}, 0.25, and {\bf 0.30}, and the entropy
  levels were chosen to be 1, 2, 3, 4, 5, 6, 8, 10, 12, 14, 16, and
  20 (bold: 6, 10, 20). Some of the contours are labeled with their
  respective values, and darker grey shading indicates higher values
  of a quantity.}
  \label{fig:vertP}
\end{figure*}

\begin{figure*}
\begin{tabular}{cc}
  \epsfxsize=8.5cm  \epsffile{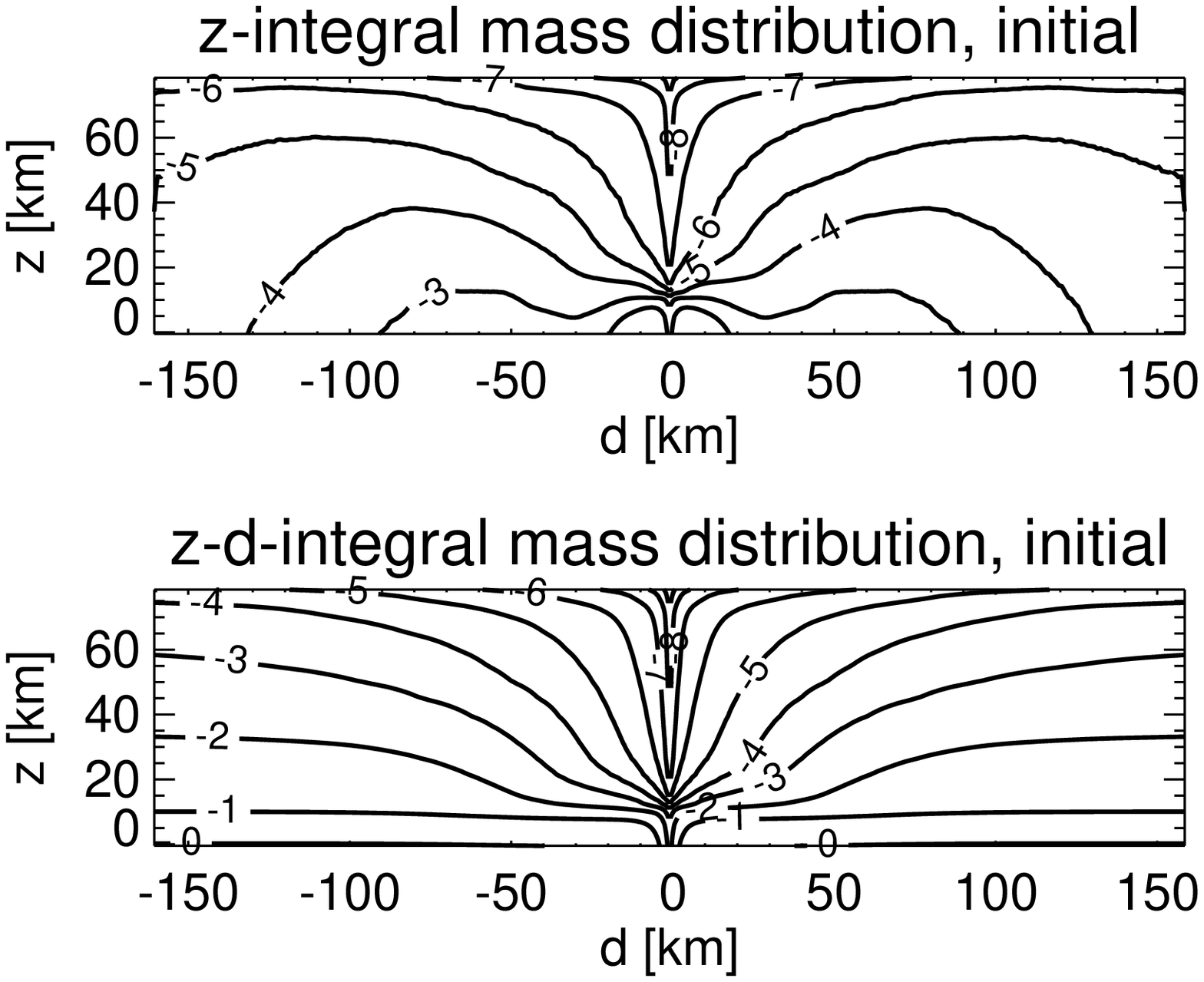} \\
[2ex]
  \epsfxsize=8.5cm  \epsffile{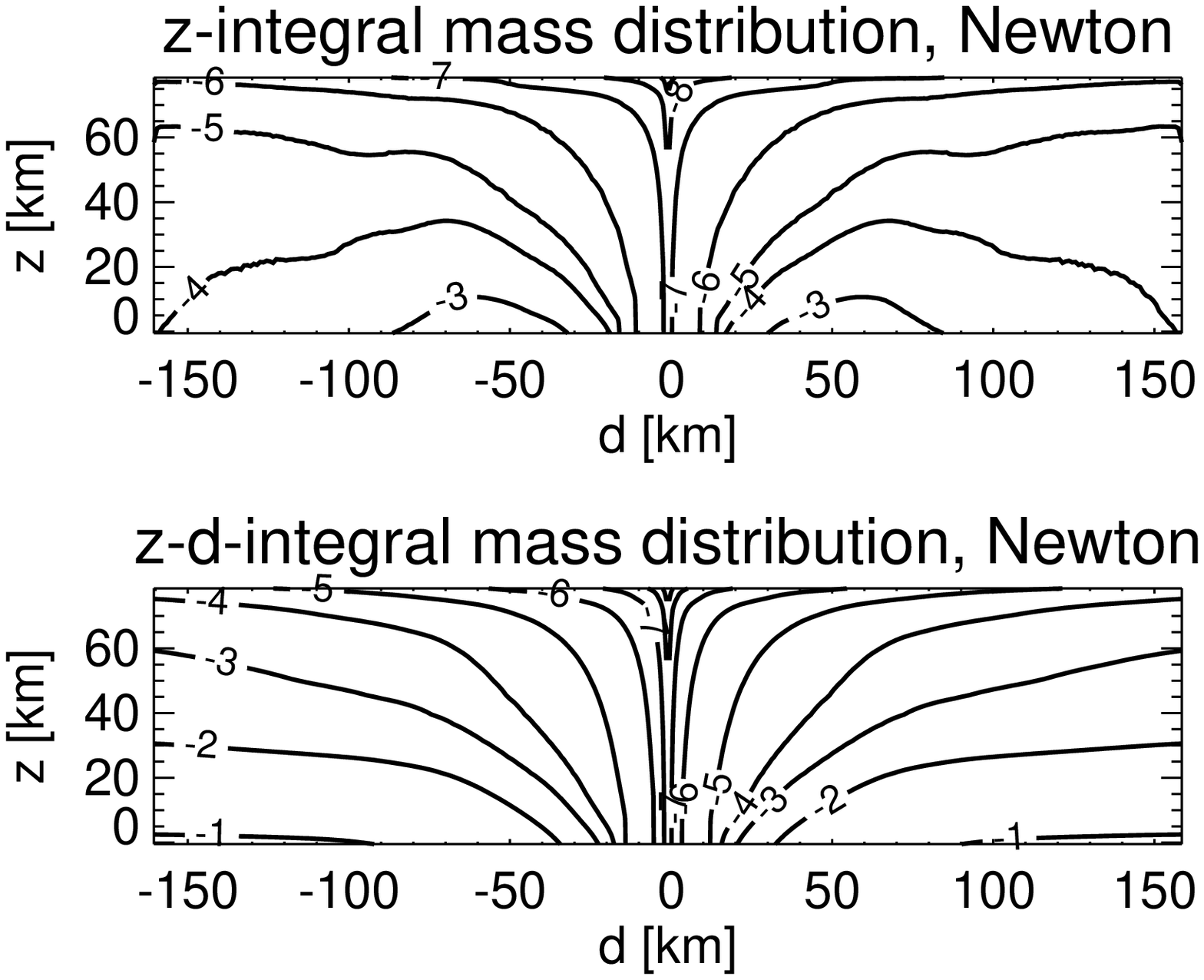} &
  \epsfxsize=8.5cm  \epsffile{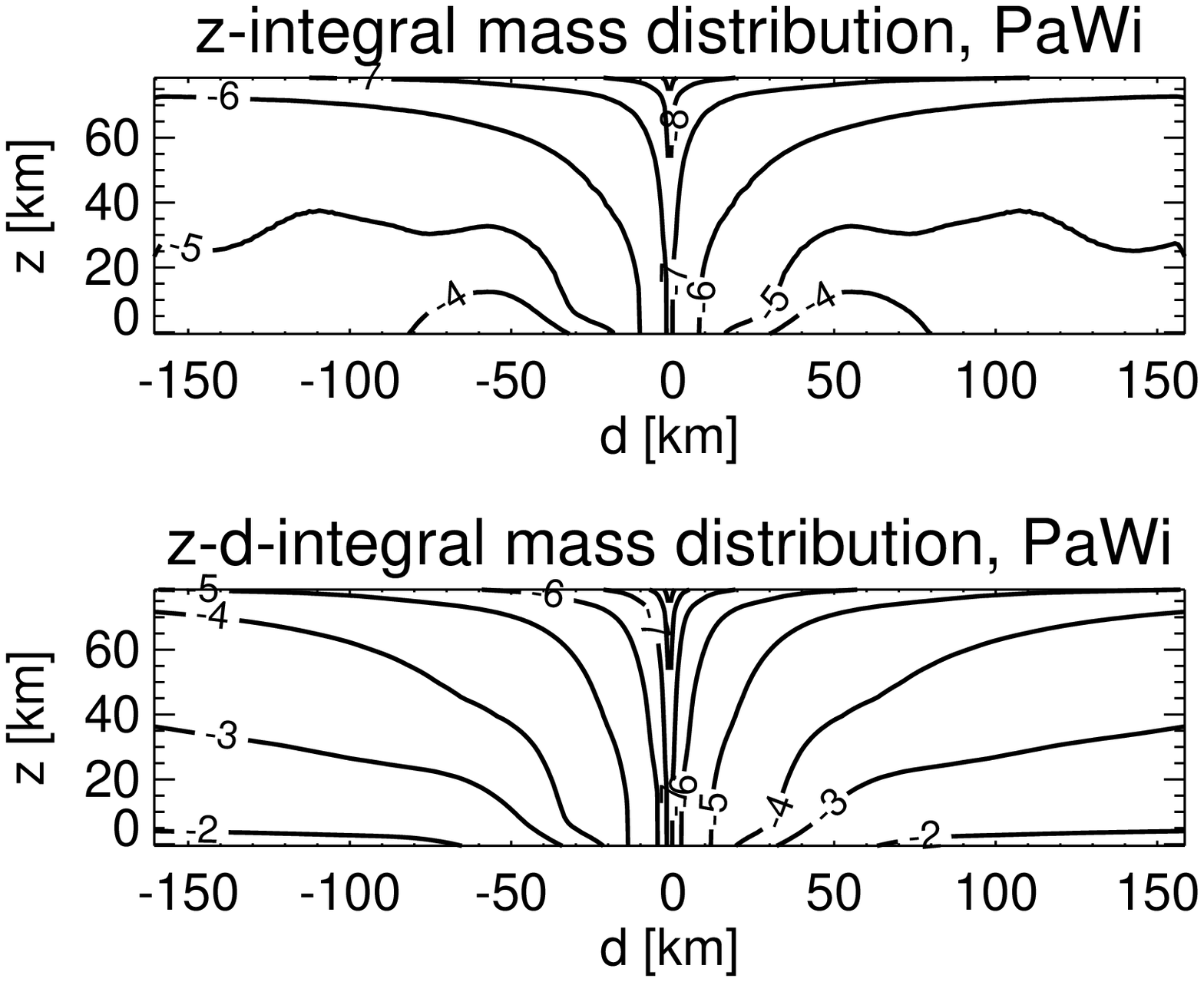} \\
\end{tabular}
  \caption[]{Mass distribution in the Model~B64 of Ruffert \& Janka 
   (\cite{ruf98b}) at $t = 10\,$ms (upper figure),
   in the Newtonian Model~{\bf B}10 at time $t = 14.9\,$ms
   (lower left figure) and in the Paczy\'nski-Wiita
   Model~${\cal B}$10 at time $t = 15.2\,$ms (lower right figure). In the
   upper panels of the figures (``z-integral''), 
   the azimuthally averaged baryon densities were
   integrated along the $z$-direction from a given value of $z$ 
   out to the grid boundary (practically infinity because of the
   rapidly decreasing density) and within 
   radial intervals of width $\Delta d = 1\,$km. 
   In the lower panels (``z-d-integral''), 
   integration in radial direction from $0$ 
   to the value of $d$ on the abscissa was additionally performed.
   The baryonic mass is measured in $M_{\odot}$ and the corresponding
   contours of equal-mass levels are spaced logarithmically in steps 
   of $-1$. Therefore each point $(d,z)$ in the upper panels is associated
   with the mass inside a hollow cylinder between $d-\Delta d$ and $d$
   from $z$ to infinity, and in the lower panels with the mass inside a 
   full cylinder between $d = 0$ and $d$ from $z$ to infinity.
   The contours are mirrored along the $d=0$ axis in all plots
   and show only the mass in the volume on one side of the equatorial 
   plane. The evacuated funnel along the system axis is clearly visible
   in the final stages of Models~{\bf B}10 and ${\cal B}$10.}
  \label{fig:coneNP}
\end{figure*}

\section{Properties of the accretion torus\label{sec:torus}}

\begin{figure*}
\tabcolsep=2.0mm
 \begin{tabular}{cc}
   \epsfxsize=7.5cm \epsfclipon \epsffile{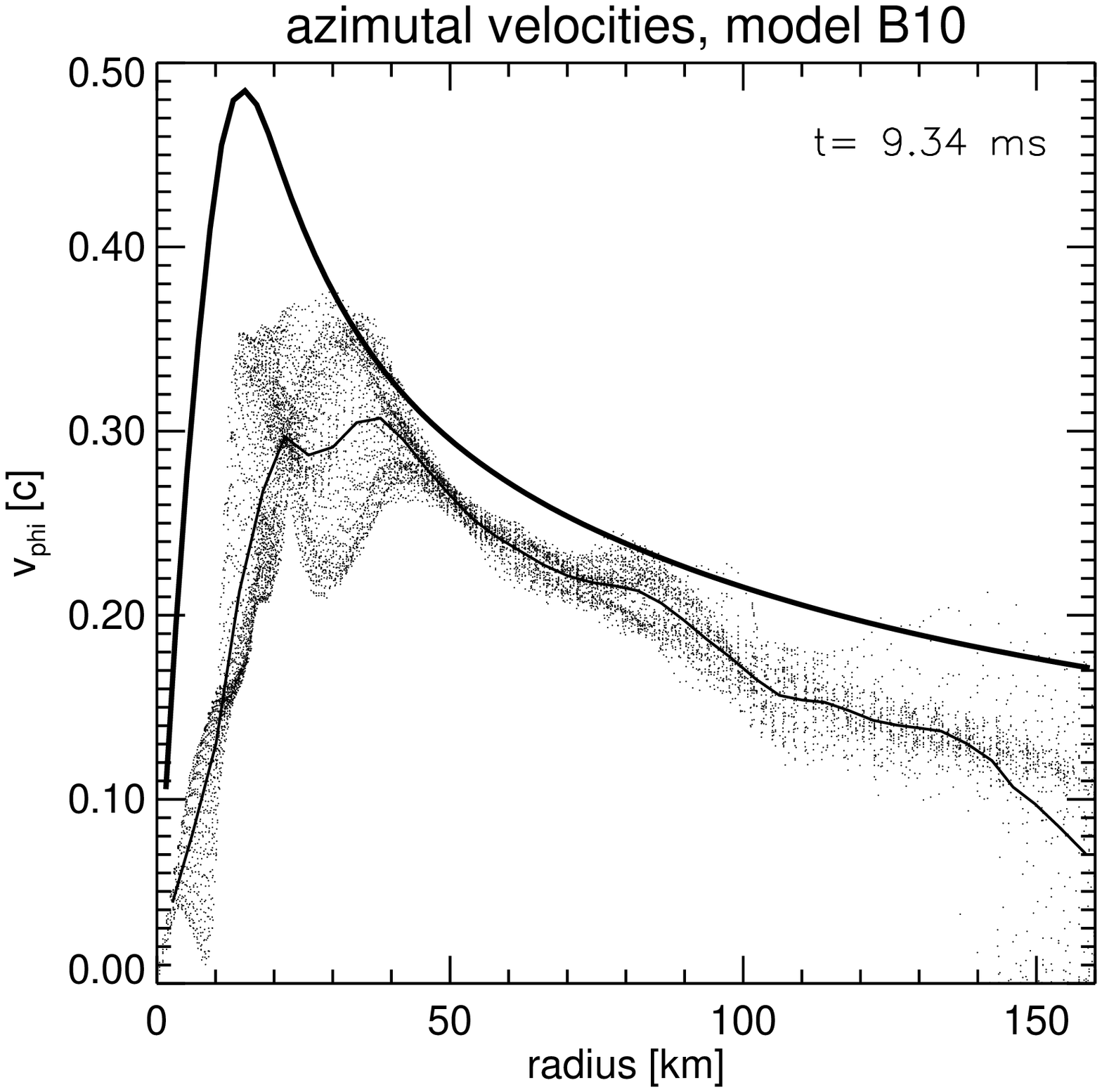} &
   \epsfxsize=7.5cm \epsfclipon \epsffile{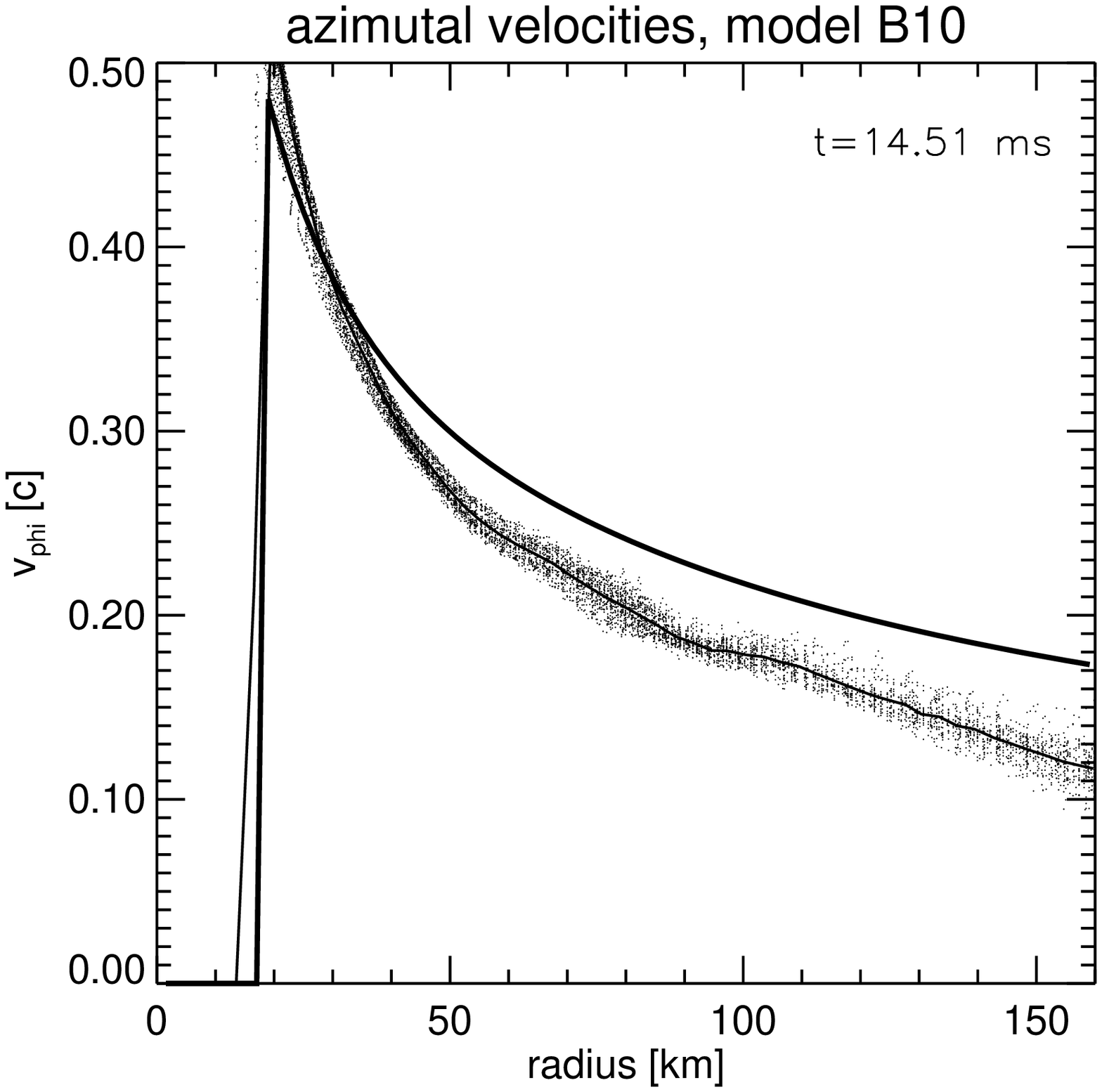} \\
[-2ex]
   \parbox[t]{8.7cm}{\caption[]{The dots represent the azimuthal 
    velocities $v_\varphi(d)$ (normalized to
    the speed of light) of all zones in the equatorial plane
    of the neutron star merger Model~B64 of Janka \& Ruffert (\cite{ruf98b})
    at $t = 9.34\,$ms after the start of that simulation. 
    This model was close to the one (at $t = 10.0\,$ms) used as 
    initial condition for the tori evolution simulations, Models~{\bf B}10
    and ${\cal B}$10. The thin solid line gives the average value of the 
    azimuthal velocities (binned in steps of 3$\,$km), and the bold 
    solid line the local Newtonian Kepler velocity $v_{\rm Kepler}^{\rm N}(d)$
    as a function of the equatorial radius $d$.}
   \label{fig:vphi}} &
   \parbox[t]{8.7cm}{\caption[]{Same as Fig.~\ref{fig:vphi} but for the
    Newtonian torus Model~{\bf B}10 at $t = 14.51\,$ms near the end of the 
    simulation. The kinks of the curves inside the inner vacuum boundary
    at two Schwarzschild radii of the central black hole (about 18$\,$km)
    are unphysical and caused by the interpolation of the plot data.}
   \label{fig:vphiN}} \\
[30ex]
   \epsfxsize=7.5cm \epsfclipon \epsffile{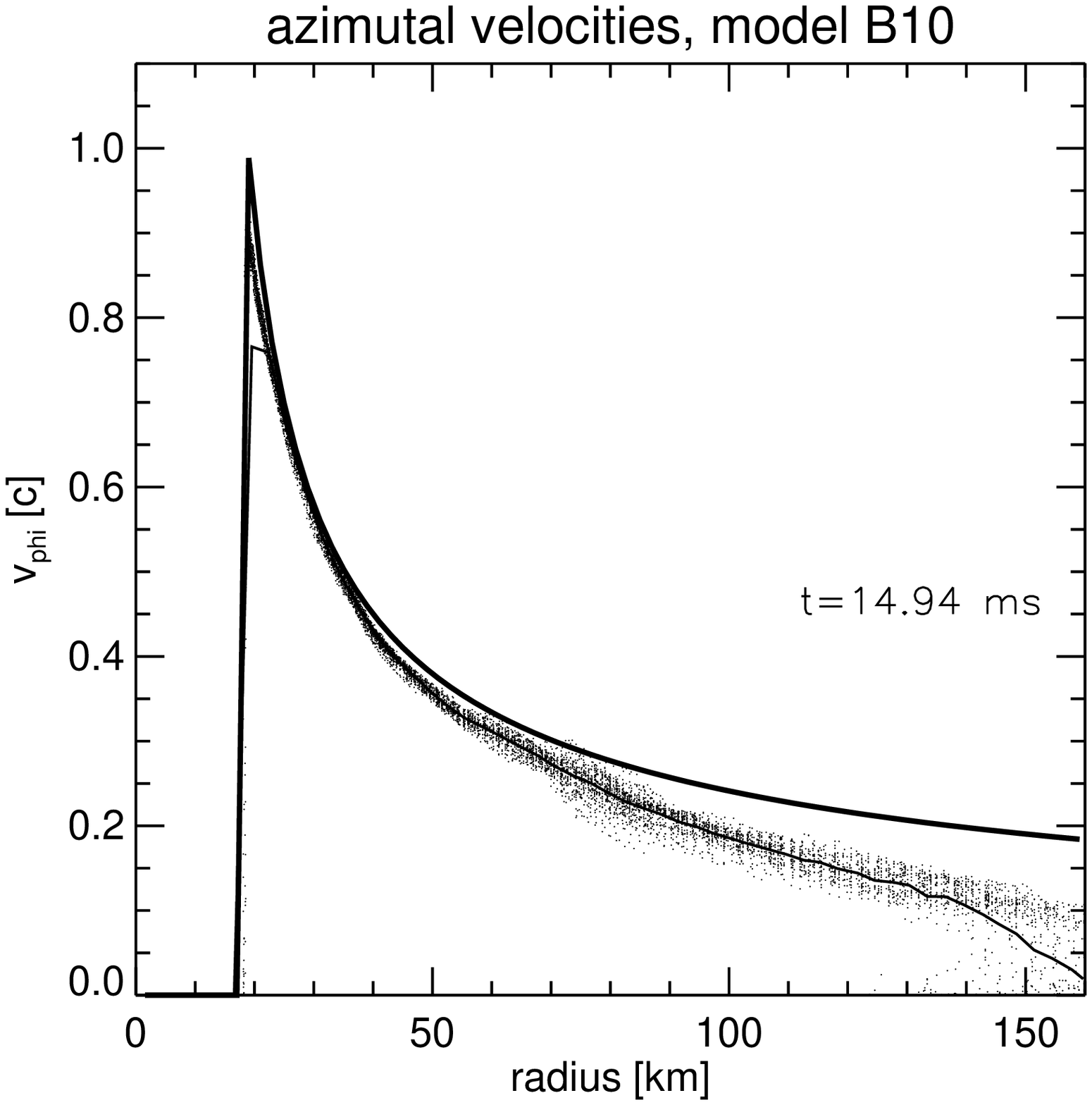} &
   \epsfxsize=7.5cm \epsfclipon \epsffile{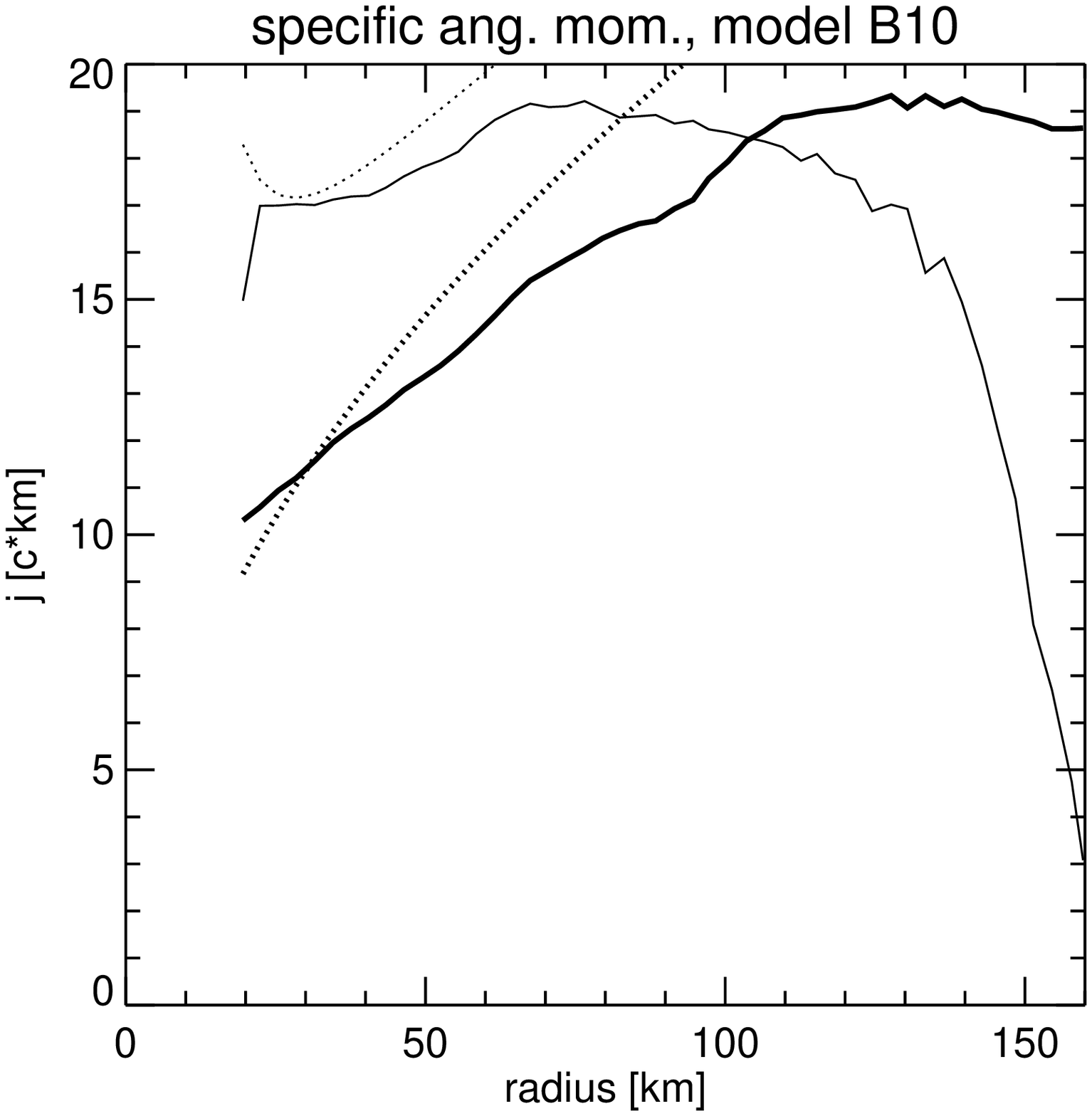} \\
[-2ex]
   \parbox[t]{8.7cm}{\caption[]{The dots represent the azimuthal
    velocities $v_\varphi(d)$ (normalized to the speed of light)
    of all zones in the equatorial plane of the Paczy\'nski-Wiita 
    torus Model~${\cal B}$10 at $t = 14.94\,$ms near the end of the
    simulation. The thin solid line gives the average value of the
    azimuthal velocities and the bold solid line the local Kepler
    velocity $v_{\rm Kepler}^{\rm PW}(d)$ of the Paczy\'nski-Wiita
    potential as a function of the equatorial radius $d$.
    The kinks of the curves inside the inner vacuum boundary
    at two Schwarzschild radii (2$R_{\rm s}$) of the central black hole
    (about 18$\,$km) are unphysical and caused by the interpolation of the
    plot data.}
   \label{fig:vphiP}} &
   \parbox[t]{8.7cm}{\caption[]{Average specific angular momentum 
    $j(d)=v_\varphi(d) d$ (in units of km times the speed of light) of
    the gas in the equatorial plane of the Newtonian torus Model~{\bf B}10
    (bold solid line) at $t = 14.51\,$ms and of the Paczy\'nski-Wiita 
    torus Model~${\cal B}$10 at $t = 14.94\,$ms (thin solid line)
    as function of the equatorial radius $d$. The specific angular
    momentum $j_{\rm Kepler}^{\rm N}(d)$ corresponding to the Kepler velocity
    of the Newtonian gravitational potential is displayed as bold dotted
    line, the equivalent quantity $j_{\rm Kepler}^{\rm PW}(d)$ for the
    Paczy\'nski-Wiita potential as thin dotted line. All curves start at
    the inner vacuum boundary at two Schwarzschild radii of the central
    black hole).} 
   \label{fig:jphi.P}} \\
\end{tabular}
\end{figure*}

\begin{figure*}
\begin{tabular}{cc}
\hskip 0.25truecm
  \epsfxsize=7.5cm \epsfclipon \epsffile{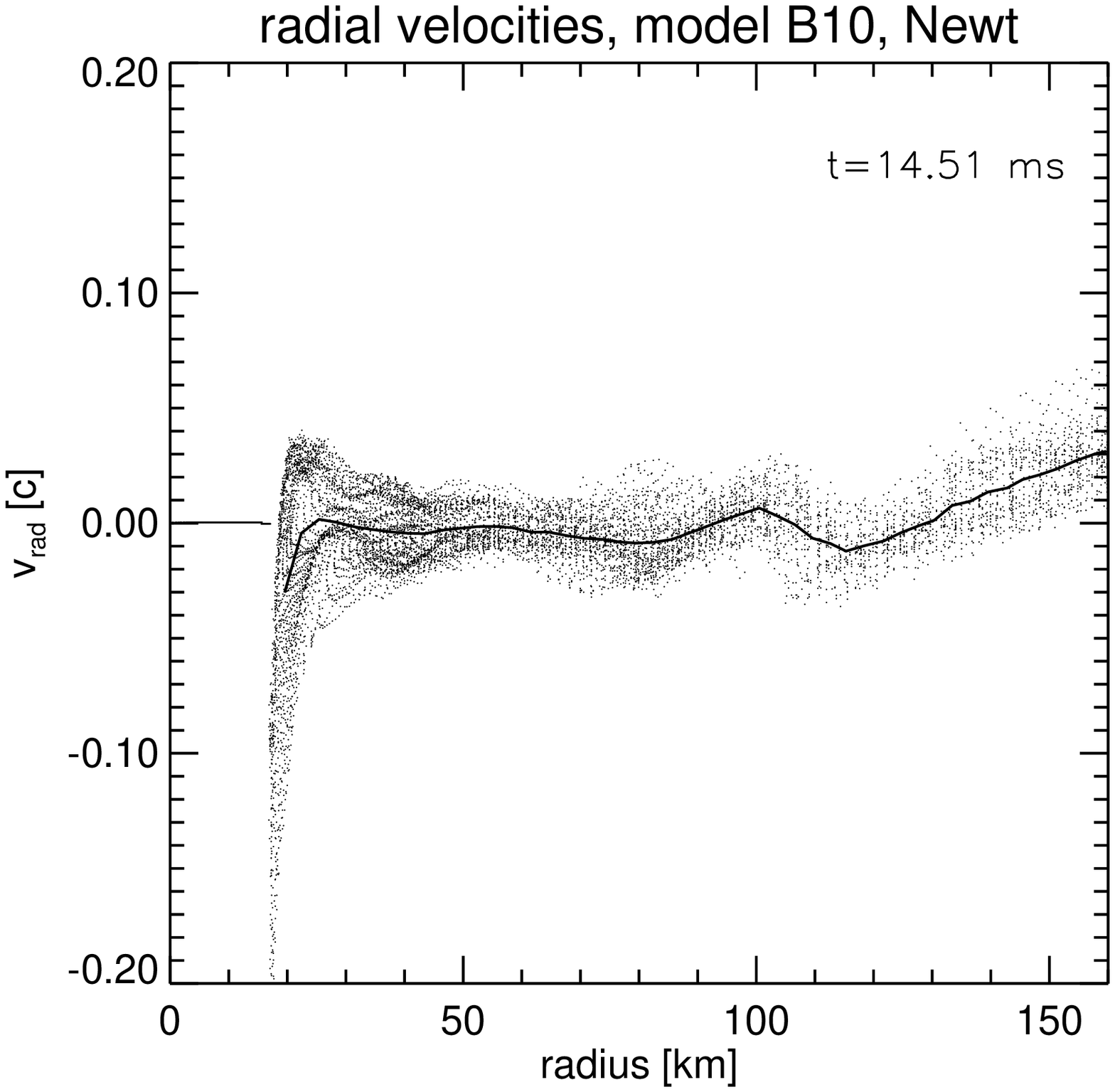} &
\hskip 1.25truecm
  \epsfxsize=7.5cm \epsfclipon \epsffile{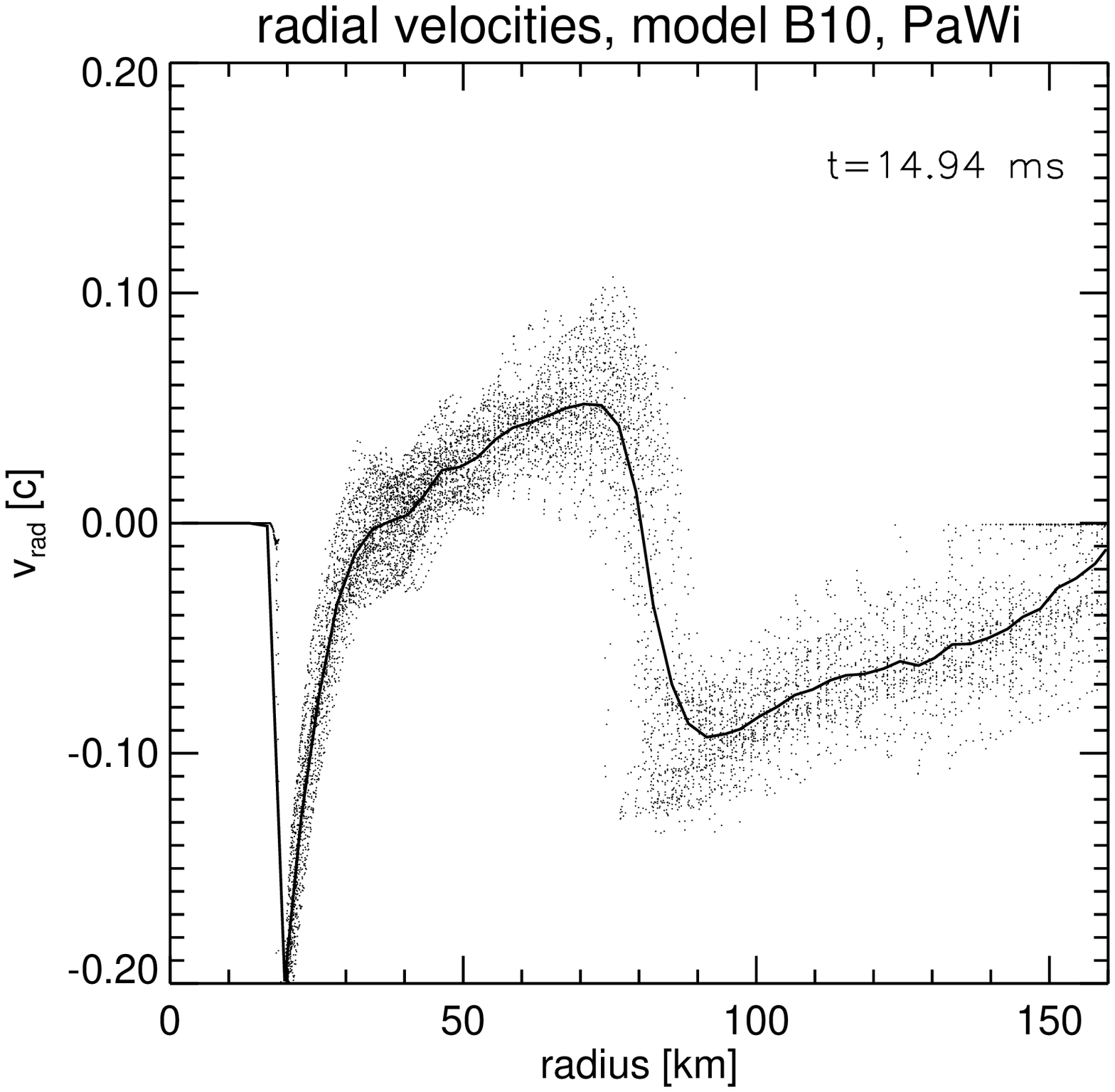} \\
\end{tabular}
  \caption[]{The dots give the radial velocities $v_r(d)$ 
(normalized to the speed of light) as a function of the equatorial
radius $d$ for all grid zones in the equatorial plane
of the Newtonian Model~{\bf B}10 at $t = 14.51\,$ms (left) and of 
the Paczy\'nski-Wiita torus Model~${\cal B}$10 at time $t = 14.94\,$ms 
(right). The solid lines represent the mean values of all zones within
binning intervals of 3$\,$km. In the Newtonian model the velocities
are very small for $d \ga 25\,{\rm km}$ with a radial 
average of about $-10^8\,{\rm cm\,s}^{-1}$, indicating that the model
has relaxed to a quasi-stationary state near the end of the simulation.
In the Paczy\'nski-Wiita torus the gas collapses rapidly inside the
last stable circular orbit at $d = 3R_{\rm s}\approx 27\,{\rm km}$,
but expands in the region $40\,{\rm km}\la d \la 80\,{\rm km}$ due
to ongoing viscous heating and outward transport of angular momentum.
Note that for $d \ga 100\,{\rm km}$ the gas is very dilute whereas
the Newtonian torus extends beyond $d = 160\,{\rm km}$ 
(see Figs.~\ref{fig:vertN} and \ref{fig:vertP}).
In both models the plotted values at $d \le 20\,{\rm km}$ are not 
physical due to the interpolation of the data.}
\label{fig:vrad}
\end{figure*}

The structure of the accretion torus in the quasi-stationary
state is shown in Figs.~\ref{fig:densN}--\ref{fig:vrad}.
In Figs.~\ref{fig:densN} and \ref{fig:densP} contour plots of the 
density $\rho$, temperature $T$, electron fraction $Y_e$, 
and entropy per nucleon $s$ in the
equatorial plane of the Newtonian Model~{\bf B}10 and of the 
Paczy\'nski-Wiita Model~${\cal B}$10, respectively, are given at
the end of the simulations. Figure~\ref{fig:massfrac} presents the
corresponding information for the mass fractions of $\alpha$ particles
and heavy nuclei in both accretion tori. In Fig.~\ref{fig:phiav} 
the azimuthally averaged radial structure of the tori in the 
equatorial plane (quantities $\rho$, $T$, $Y_e$, and $s$) is 
displayed, together with the cumulative torus mass $M(d)$ as
function of the equatorial radius $d$. Contour plots for    
$\rho$, $T$, $Y_e$, and $s$ in the $x$-$z$- and $y$-$z$-planes
perpendicular to the equatorial plane are shown in 
Figs.~\ref{fig:vertN} and \ref{fig:vertP}. Integral mass
distributions perpendicular to the equatorial plane for the 
initial model of our simulations as well as for the final
states of Models~{\bf B}10 and ${\cal B}$10 are presented in
Fig.~\ref{fig:coneNP}. Finally, Figs.~\ref{fig:vphi}--\ref{fig:vrad}
give information about the azimuthal and radial velocities and 
the specific angular momentum of the gas in the equatorial plane
of the initial model and of the evolved tori.

At the end of the simulations the tori have become nearly axially
symmetric with only minor deviations (Figs. \ref{fig:densN} and 
\ref{fig:densP}). Two hot spots can be seen close to
the inner grid boundary at 2$R_{\rm s}$ which coincide with density maxima.
They continue to carry the memory of the two very prominent spiral-like 
arms which are formed during the neutron star merging, grow right afterwards, 
and are
wound up into the toroid around the black hole during the subsequent evolution.
The torus of Model~${\cal B}$10 in Fig.~\ref{fig:densP} is smaller
--- the $\rho = 10^{10}\,{\rm g\,cm}^{-3}$ contour is at 70--80$\,$km ---
than the torus of Model~{\bf B}10 in Fig.~\ref{fig:densN} where the
$\rho = 10^{10}\,{\rm g\,cm}^{-3}$ contour extends out to 120--130$\,$km.
There are two reasons for that. On the one hand, the gas mass which
remains around the black hole is smaller in the former model
(see Table~\ref{tab:models}), on the
other hand the gravitational potential is stronger in the
Paczy\'nski-Wiita case. This is the reason why despite of a difference
of a factor 7 in the torus mass, both models have average densities
which differ only by a factor 3 (about $3\times 10^{11}\,{\rm g\,cm}^{-3}$
for Model~{\bf B}10 compared to approximately $10^{11}\,{\rm g\,cm}^{-3}$
in case of Model~${\cal B}$10, see Fig.~\ref{fig:phiav}).
The entropies in both models are nearly the same (between 5 and 
10$\,k_{\rm B}$ per nucleon) and the entropy profiles are very similar 
in the region where most of the mass is sitting (Fig.~\ref{fig:phiav}).
The Newtonian torus is somewhat hotter, its average temperature is around 
6$\,$MeV compared to 4$\,$MeV for Model~${\cal B}$10. It is also 
more neutron rich with a mean value of $Y_e\approx 0.05$--0.15.
Model~${\cal B}$10 has typical $Y_e$ values between 0.1 and 0.17 
because of its lower density. This has the consequence that the
neutrino opacities are smaller and thus the increase of the proton
abundance by the emission of electron antineutrinos 
(and the cooling by the loss of neutrinos and antineutrinos of all flavors) 
proceeds faster. Since the hot, expanded neutron star matter radiates
predominantly electron antineutrinos (see Sect.~\ref{sec:neutrino}),
the initially very neutron rich state evolves towards higher electron
and proton number fractions. Due to the high temperatures, electrons
are not degenerate in the tori and therefore positrons are abundant.
in considerable numbers. Typical electron degeneracy parameters 
$\eta_e = \mu_e/T$ ($\mu_e$ is the electron chemical potential) are
around 2, some regions have values of about 4, and in large regions
one finds $\eta_e < 2$. For the conditions in the tori, in particular due to
the rather high entropies, the nucleons are mostly unbound and $\alpha$ 
particles and heavy nuclei are present only in small numbers. The
maximum mass fractions $X_{\alpha}$ of $\alpha$ particles are around a few
per cent, and heavy nuclei appear in significant abundances only where 
the temperature drops below 1$\,$MeV (Fig.~\ref{fig:massfrac}).

The perpendicular cuts (Figs.~\ref{fig:vertN} and \ref{fig:vertP}) confirm 
the nearly axially symmetric structure of the tori in Models~{\bf B}10
and ${\cal B}$10 at the end of the simulations. Again, some differences
between the $x$-$z$- and $y$-$z$-cuts reflect the last remainders of the
spiral arms which have been inflated and dissolved into the tori. 
While the temperature and density contours show a rather regular shape,
primarily determined by the balance of pressure gradients and 
gravitational and centrifugal forces,
the electron fraction and entropy are more irregular and patchy because
both quantities carry information about the whole preceding evolution,
in particular about the integral effects of neutrino emission and 
non-adiabatic hydrodynamic processes.
Near the poles of the black hole and along the system axis, the density
has decreased to values below $5\times 10^8\,{\rm g\,cm}^{-3}$. 
This is only one order of magnitude above the lower density limit which
is set to $5\times 10^7\,{\rm g\,cm}^{-3}$ in the surroundings of the 
torus for numerical reasons, but nevertheless it is more than 3--4 orders 
of magnitude below the average densities inside the tori.
In this sense we see the formation of an ``evacuated'', cylindrical funnel 
along the rotational axis of the black hole-torus system. 
Material which was swept into the polar regions during and immediately 
after the merging of the neutron stars falls into the newly formed black hole
very quickly within a free-fall time scale because it is not supported by
centrifugal forces. A comparison of the 
initial condition for the torus simulations of Models~{\bf B}10 and 
${\cal B}$10 (corresponding to the situation at $t\approx 10\,$ms in Model~B64
of Ruffert \& Janka \cite{ruf98b}) with the quasi-stationary states about
5$\,$ms later reveals this rapid cleaning of the axial funnel
(see Fig.~\ref{fig:coneNP}). In the upper panels of the three figures in
Fig.~\ref{fig:coneNP} the contours contain all points $(d,z)$
where the cumulative gas mass $M_{\rm gas}^\prime(d,z)$ given by
\begin{equation}
M_{\rm gas}^\prime(d,z)\,\equiv\,\int_z^\infty{\rm d}\chi\int_d^{d+\Delta d}
{\rm d}\xi\,\xi\int_0^{2\pi}{\rm d}\varphi\,\rho({\bf r})
\label{eq:zmass}
\end{equation}
is constant (the labels at the contours represent logarithmic values of the 
mass measured in $M_{\odot}$). The integration is done on a cylindrical
grid with coordinates $(d,z,\varphi)$. The integral of Eq.~(\ref{eq:zmass}) 
therefore sums up all the mass within a hollow cylinder of thickness
$\Delta d = 1\,$km extending from $z$ to infinity (in praxi: the upper
grid boundary). The contours in Fig.~\ref{fig:coneNP} are mirrored 
along the system axis at $d = 0$. Note that only the mass on one side of
the equatorial plane is added up. The lower panels in Fig.~\ref{fig:coneNP}
show the contours that correspond to constant values $M_{\rm gas}(d,z)$ 
according to the integral
\begin{equation}
M_{\rm gas}(d,z)\,\equiv\,\int_z^\infty{\rm d}\chi\int_0^d
{\rm d}\xi\,\xi\int_0^{2\pi}{\rm d}\varphi\,\rho({\bf r})
\label{eq:zdmass}
\end{equation}
which sums up the gas mass in a cylinder with radius $d$ that
extends from $z$ to infinity around the system axis. The panels in 
Fig.~\ref{fig:coneNP} give detailed information about the mass 
distribution in the surroundings of the accretion torus. Initially,
about $10^{-3}\,M_{\odot}$ of gas were distributed above the compact massive
object at the center of the merger (see second panel from the top in the 
left column of Fig.~\ref{fig:coneNP}). However, with increasing vertical 
distance $z$ from the equatorial plane the mass drops extremely rapidly
already in the post-merging configuration. In the final states, the 
total mass inside a cylinder with radius $d\approx 15\,$km is only a
few $10^{-4}\,M_{\odot}$, most of this gas is very close to the 
equatorial plane. Around the rotation axis the quasi-stationary
states of Models~{\bf B}10 and ${\cal B}$10 look very similar. 
The larger torus mass of the Newtonian computation, however, leads
to a higher mass concentration near the equatorial plane at radii 
$d\ga 20\,$km. Differences are therefore visible in this region for 
the contours corresponding to cumulative gas masses above
$10^{-5}\,M_{\odot}$. We note here that the simulations described 
in this paper do not include the effects of neutrino energy deposition
in the vicinity of the black hole. For this reason and because we do
not allow the gas densities to drop below 
$5\times 10^7\,{\rm g\,cm}^{-3}$ on the grid, our models most likely
overestimate the mass of the gas surrounding the torus.

Information about the motion of the gas in the initial and final models
of the torus simulations is provided by 
Figs.~\ref{fig:vphi}--\ref{fig:vrad}. In Fig.~\ref{fig:vphi} the 
azimuthal velocities $v_{\varphi}(d)$ are plotted for all grid zones
in the equatorial plane versus the distance $d$ from the grid center 
(dots) immediately before the black hole is assumed to form.
The spread of the points at a given radius $d$ reflects the 
deviations from rotational symmetry. For
an axially symmetric configuration all dots at a specific radial 
distance would cluster on top of each other. Also, the motion of the 
gas is clearly sub-Keplerian which indicates the importance of    
pressure support in the object. This can be seen from a comparison
with the bold solid line which represents the local Newtonian Kepler 
velocity,
\begin{equation}
v_{\rm Kepler}^{\rm N}(d) = \sqrt{\frac{GM(d)}{d}} \ ,
\label{eq:vkepN}
\end{equation}
given as a function of the equatorial radius $d$, with $M(d)$ being the 
mass enclosed by the sphere of radius $d$. The situation is different
at the end of the computed torus evolution about 5$\,$ms after the
assumed formation of the black hole (Fig.~\ref{fig:vphiN}
for the Newtonian simulation and Fig.~\ref{fig:vphiP} for the
Paczy\'nski-Wiita case). The spread of the dots has decreased,
indicating that the torus is much more isotropic in $\varphi$ than
the post-merging configuration. For distances $d \la 30\,$km the
Newtonian torus has azimuthal velocities larger than the local Keplerian
value. These allow the gas between 2$\,R_{\rm s}$ and 3$\,R_{\rm s}$
to remain on orbits around the black hole despite of a positive density
and pressure gradient in this region (see Fig.~\ref{fig:phiav}). 
Also for $d > 30\,$km the pressure support is important. This is 
suggested by the significant drop of the azimuthal velocities below
the Keplerian values in Fig.~\ref{fig:vphiN}. In the torus of
Model~${\cal B}$10 the orbital velocities are much closer to 
the Kepler velocity for the Paczy\'nski-Wiita potential. The value of
\begin{equation}
v_{\rm Kepler}^{\rm PW}(d) = \sqrt{\frac{GM(d)d}{(d-R_{\rm s})^2}}\ ,
\label{eq:vkepP}
\end{equation}
is twice as large at $d = 2R_{\rm s}$ than its Newtonian counterpart,
$v_{\rm Kepler}^{\rm PW}(2R_{\rm s}) = 
2v_{\rm Kepler}^{\rm N}(2R_{\rm s}) = c$, and therefore the orbital
velocities within $d\approx 50\,$km are significantly larger in
Model~${\cal B}$10 than in Model~{\bf B}10. 

The specific angular momentum $j(d) = v_{\varphi}(d)d$ 
of the matter in the equatorial planes of both models has typical values
of $(3$--$6)\times 10^{16}\,{\rm cm^2\,s}^{-1}$
(Fig.~\ref{fig:jphi.P}). In the Newtonian torus the lines for $j(d)$ and 
the Keplerian value $j_{\rm Kepler}^{\rm N}(d) = v_{\rm Kepler}^{\rm N}(d)d$
intersect at $d \approx 30\,{\rm km}\approx 3R_{\rm s}$
which is roughly the position of the density and pressure maximum
(Fig.~\ref{fig:phiav}). Inside this radius one has
$j(d) > j_{\rm Kepler}^{\rm N}(d)$. In contrast, for the Paczy\'nski-Wiita 
potential the curves of $j(d)$ and 
$j_{\rm Kepler}^{\rm PW}(d)= v_{\rm Kepler}^{\rm PW}(d)d$
touch at $d \approx 3R_{\rm s}$ where $j_{\rm Kepler}^{\rm PW}(d)$
has a minimum corresponding to the last stable circular orbit.

Figure~\ref{fig:vrad} shows the radial velocities for all grid zones
in the equatorial plane. The Newtonian Model~{\bf B}10 has achieved
a quasi-stationary state with very small radial velocities $v_r$
at the end of the simulation. The average inflow velocity between
25$\,$km and about 130$\,$km is 
$\ave{v_r}\approx -10^8\,{\rm cm\,s}^{-1}$. At smaller distances
from the black hole the gas is rapidly falling in, at distances beyond
130$\,$km the dilute outer parts of the disk are slowly expanding due
to the outward transport of angular momentum in the torus. In contrast,
the Paczy\'nski-Wiita Model~${\cal B}$10 is still evolving and has not
developed stationary conditions. It expands for 
$40\,{\rm km}\la d\la 80\,$km with radial
velocities up to 5\% of the speed of light whereas the gas interior to
$d\approx 3R_{\rm s}\approx 27\,{\rm km}$ is collapsing very rapidly 
into the black hole, and also the dilute gas exterior to 80$\,$km moves
inward with large velocities.

\begin{figure*}
\tabcolsep=2.0mm
 \begin{tabular}{cc}
   \epsfxsize=8.5cm \epsfclipon \epsffile{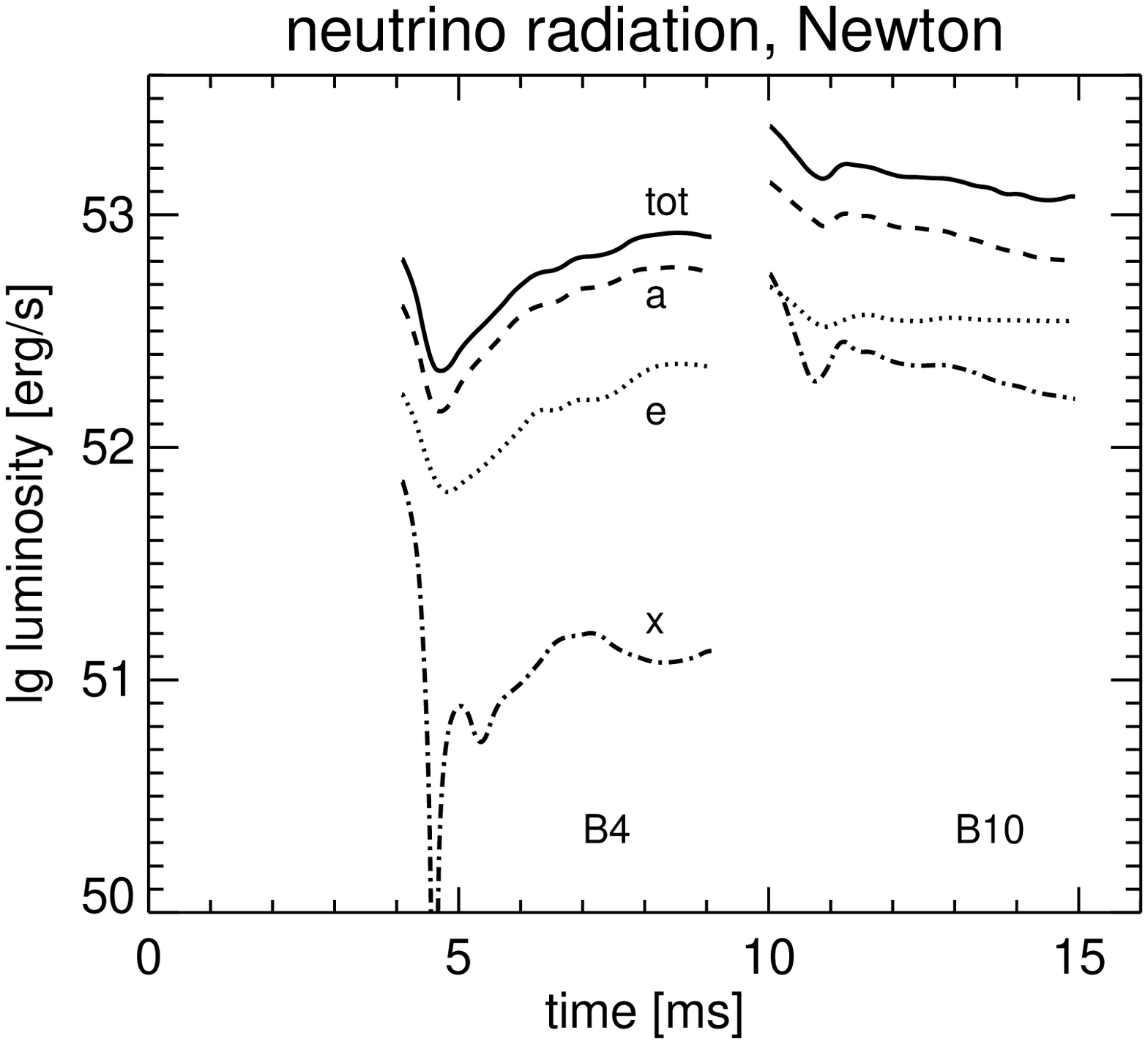} &
   \epsfxsize=8.5cm \epsfclipon \epsffile{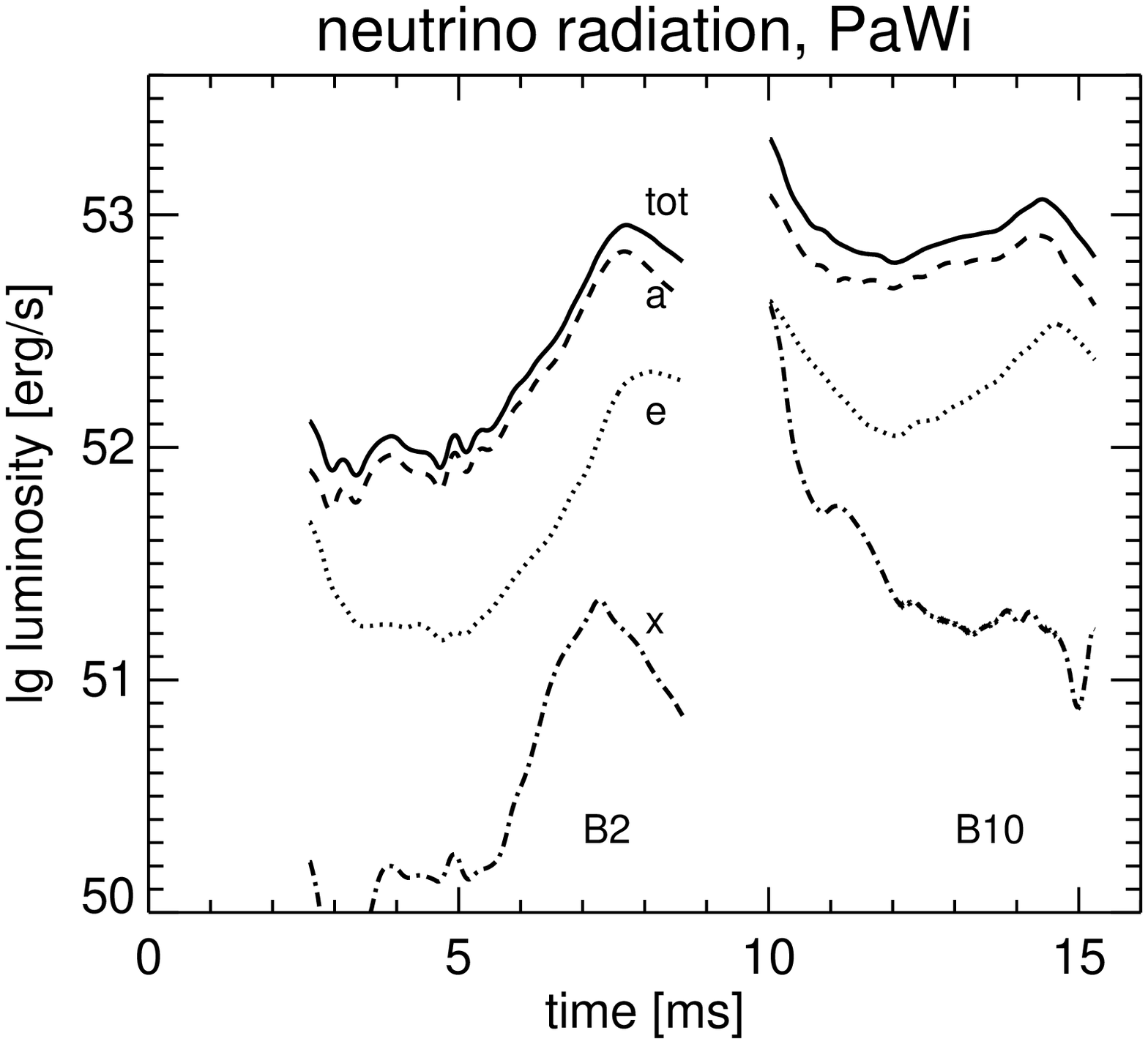} \\
[-3ex]
   \parbox[t]{8.7cm}{\caption[]{Luminosities of electron neutrinos
    (labeled with e), electron antineutrinos (labeled with a), 
    sum of all heavy-lepton neutrinos (labeled with x), and
    total neutrino luminosity (labeled with tot) as functions
    of time for the Newtonian Models~{\bf B}4 and {\bf B}10.}
   \label{fig:neutradN}} &
   \parbox[t]{8.7cm}{\caption[]{Same as Fig.~\ref{fig:neutradN}
    but for the Paczy\'nski-Wiita Models~${\cal B}$2 and ${\cal B}$10.}
   \label{fig:neutradP}} \\
[18ex]
   \epsfxsize=8.5cm \epsfclipon \epsffile{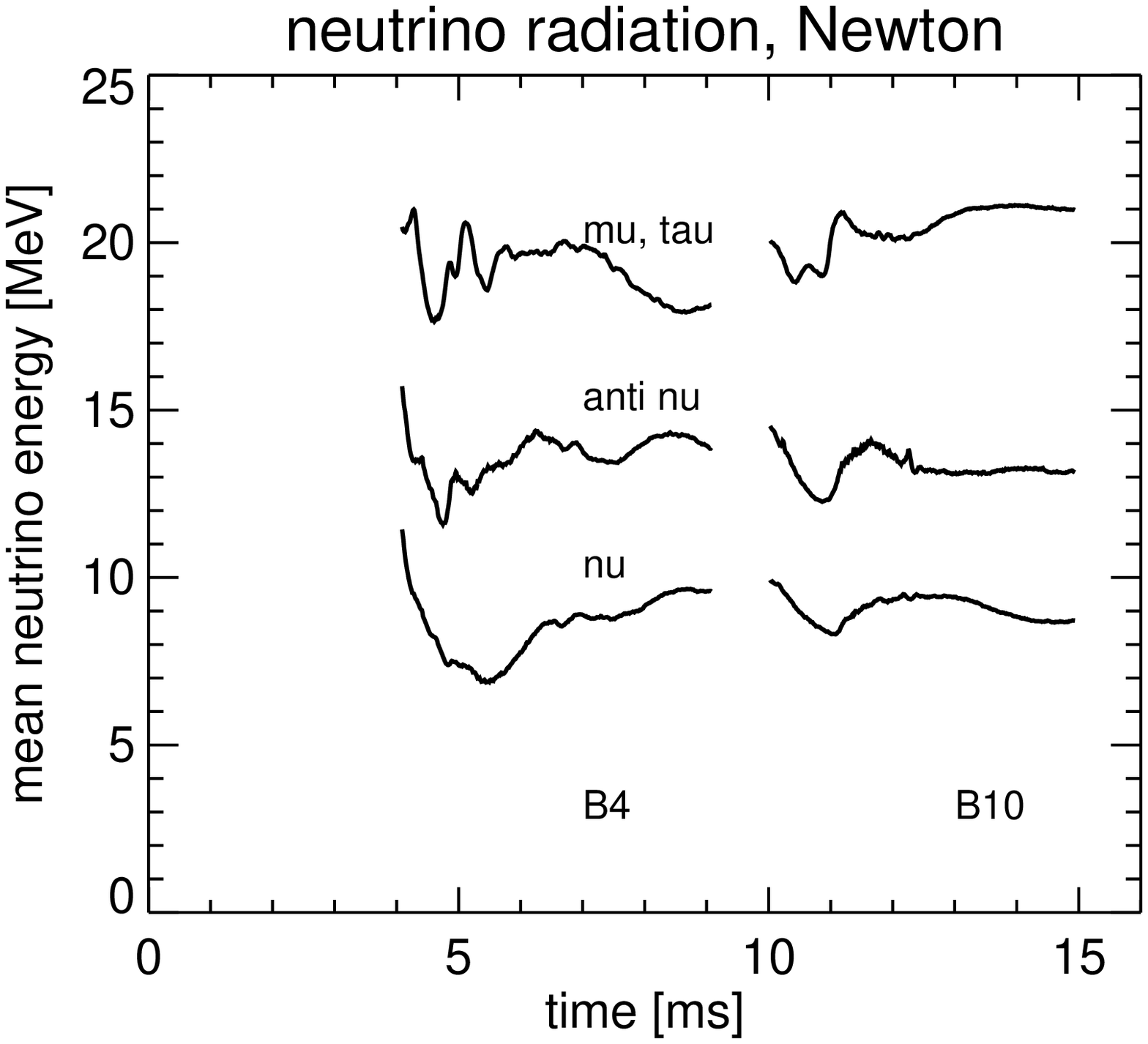} &
   \epsfxsize=8.5cm \epsfclipon \epsffile{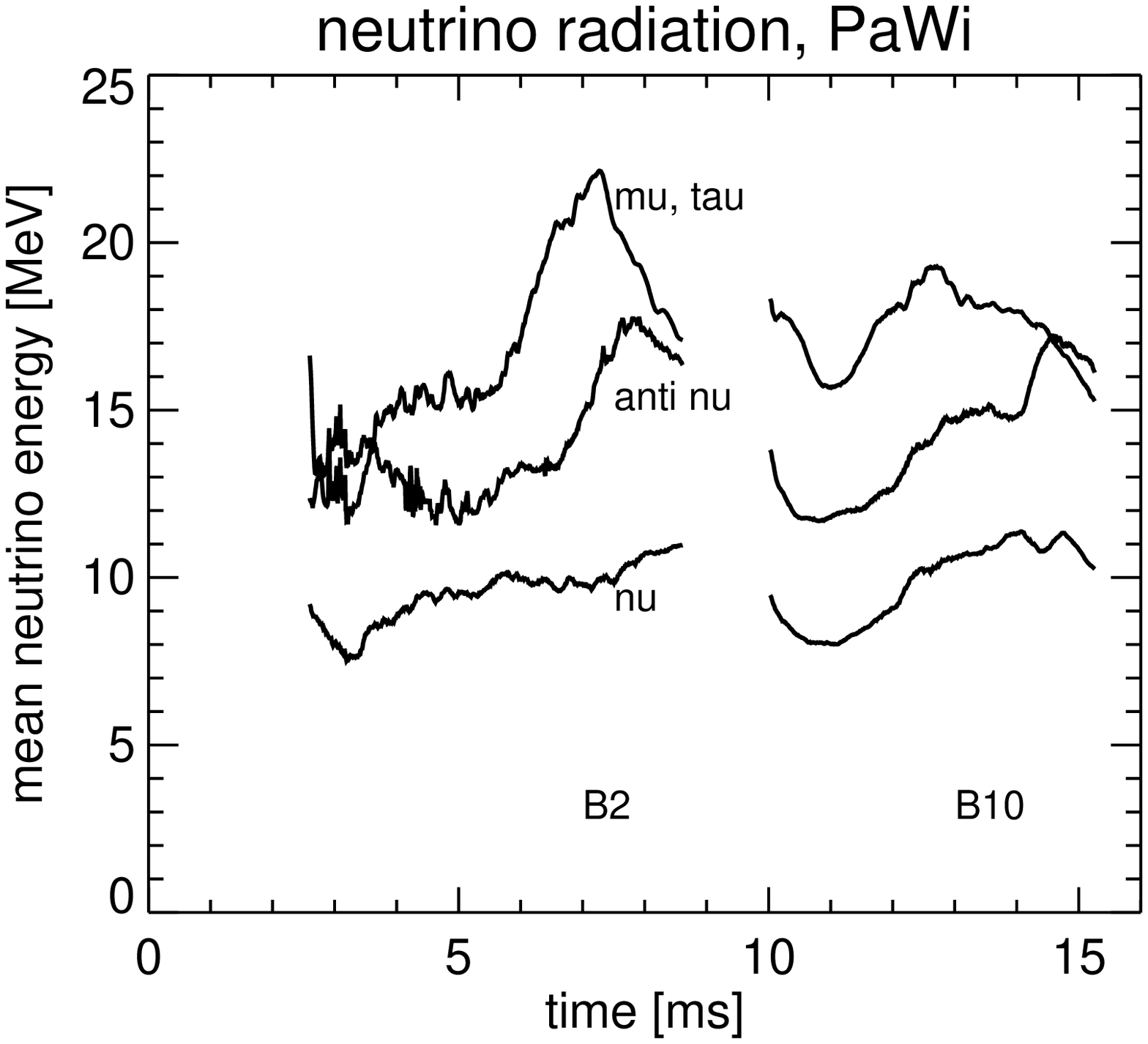} \\
[-3ex]
   \parbox[t]{8.7cm}{\caption[]{Average energies of the emitted
    electron neutrinos (labeled with nu), electron antineutrinos 
    (anti nu), and heavy-lepton neutrinos (mu, tau)
    as functions of time for the Newtonian Models~{\bf B}4
    and {\bf B}10.}
   \label{fig:meneN}} &
   \parbox[t]{8.7cm}{\caption[]{Same as Fig.~\ref{fig:meneN} but 
    for the Paczy\'nski-Wiita Models~${\cal B}$2 and ${\cal B}$10.}
   \label{fig:meneP}}\\
\end{tabular}
\end{figure*}

\begin{figure*}
\begin{tabular}{cc}
 \epsfxsize=8.5cm  \epsfclipon\epsffile{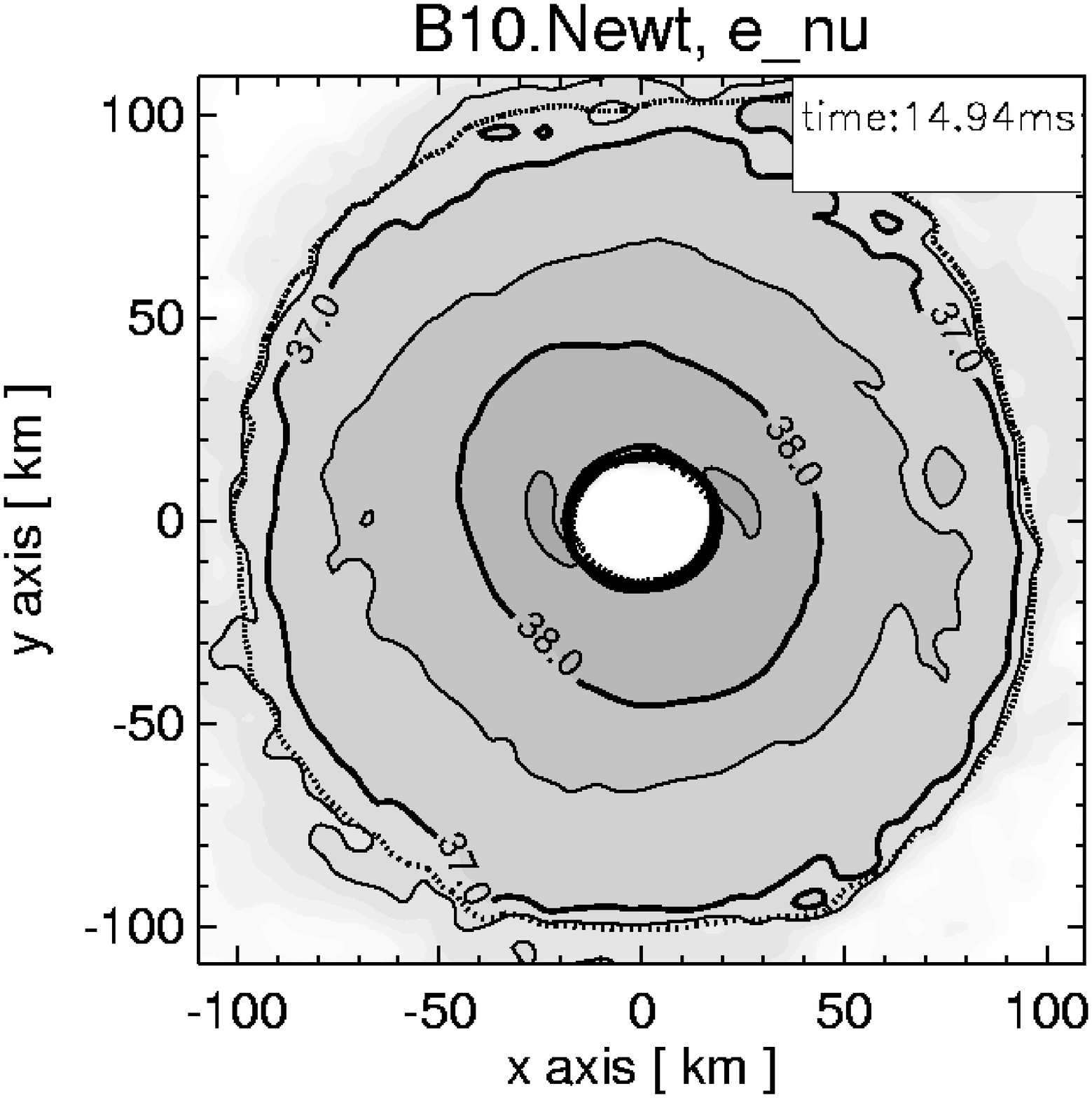} 
                   \put(-0.50,7.12){{\Large \bf \sf a}} &
 \epsfxsize=8.5cm  \epsfclipon\epsffile{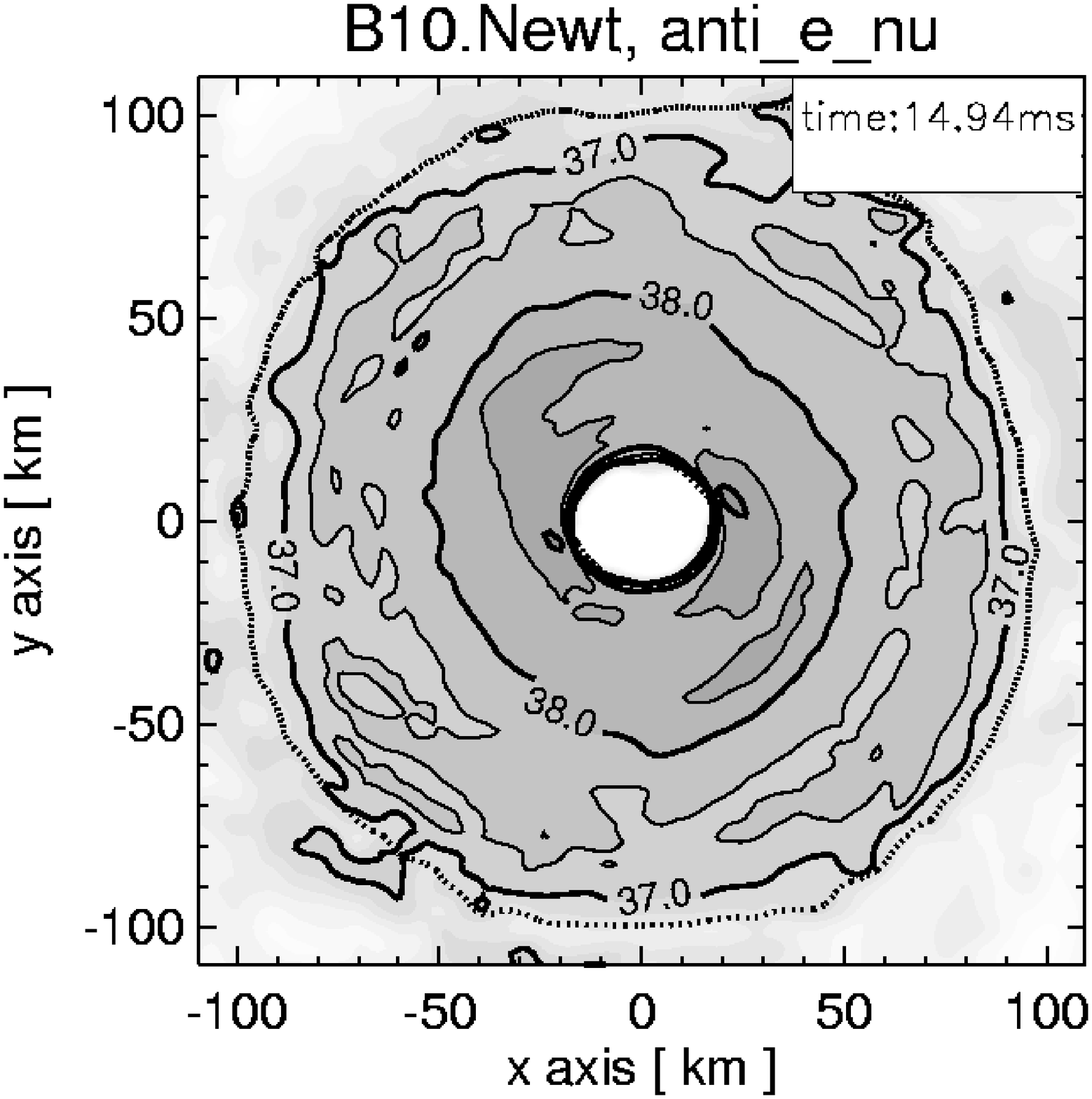}
                   \put(-0.50,7.12){{\Large \bf \sf b}}\\
[2ex]
 \epsfxsize=8.5cm  \epsfclipon\epsffile{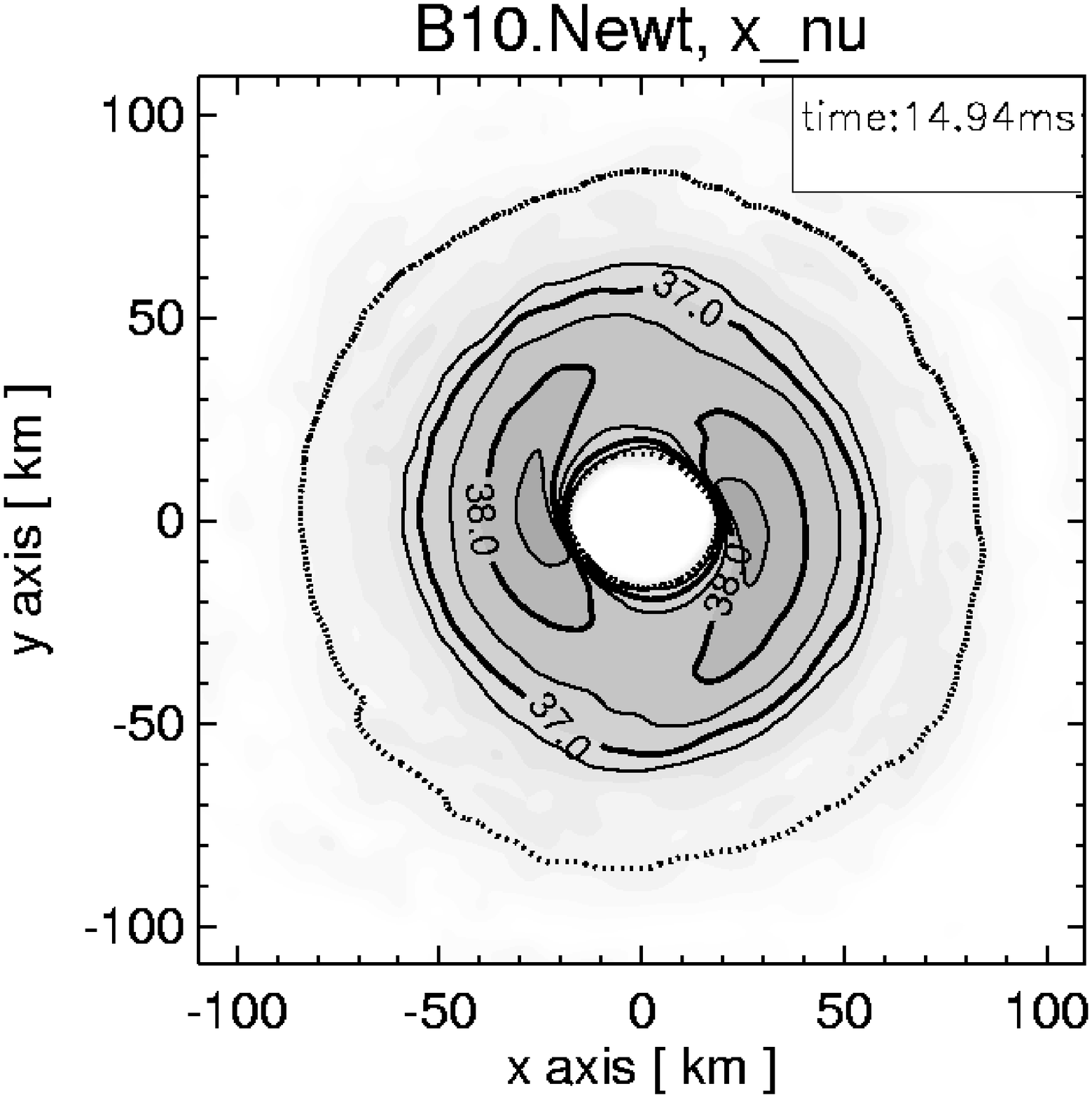} 
                   \put(-0.50,7.12){{\Large \bf \sf c}} &
 \epsfxsize=8.5cm  \epsfclipon\epsffile{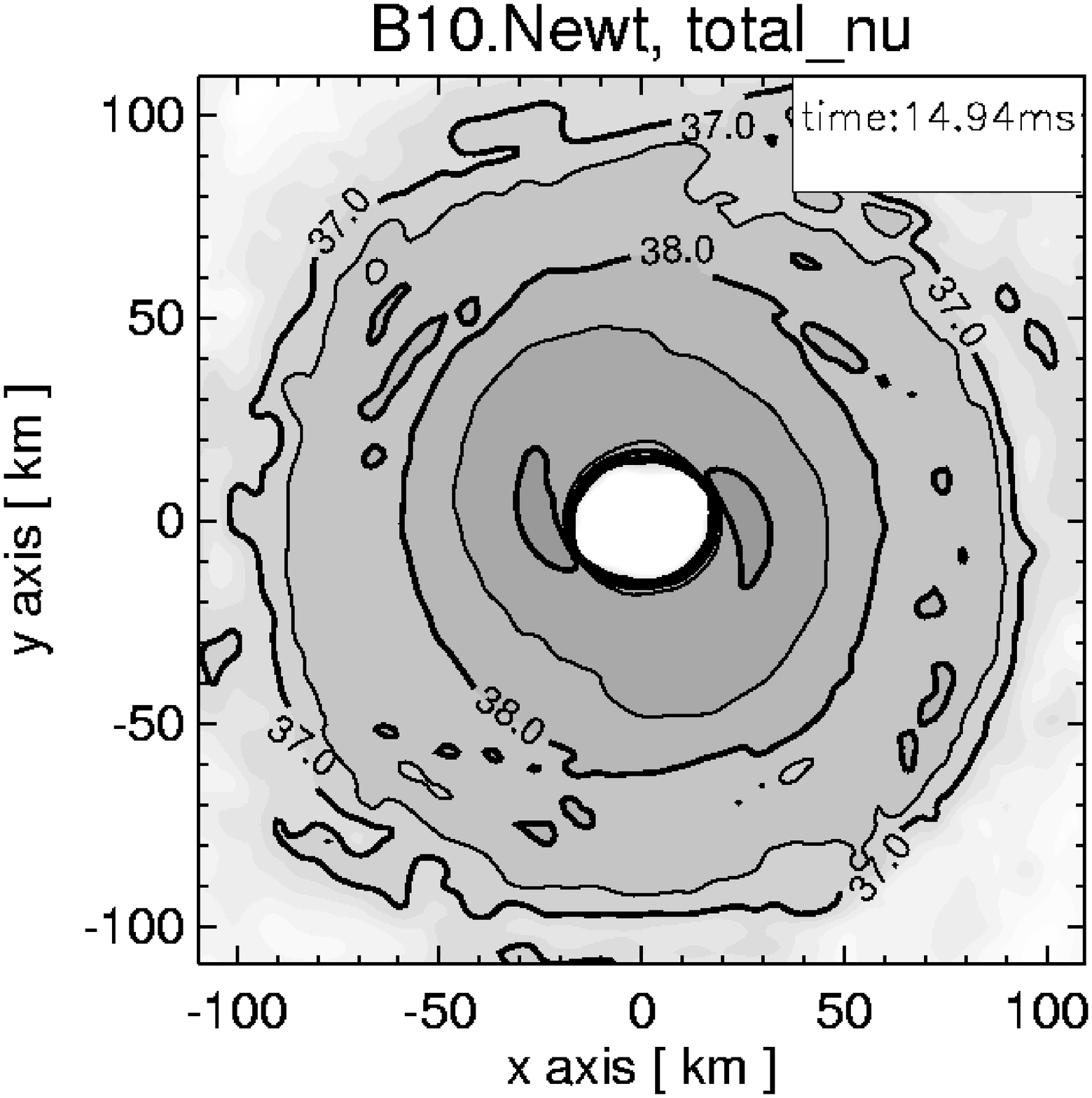}
                   \put(-0.50,7.12){{\Large \bf \sf d}}\\
\end{tabular}
  \caption[]{Surface emissivity of energy in neutrinos 
  (measured in ${\rm erg\,cm}^{-2}{\rm s}^{-1}$) for Model~{\bf B}10 
  at the end of the simulation (time in the top right corner of the panels).
  The plotted contours were obtained by integration of the local energy loss
  rates from $z = 0$ to infinity, in panel~a for electron neutrinos,
  in panel~b for electron antineutrinos, in panel~c for the sum of all
  heavy-lepton neutrinos, and in panel~d for the summed contributions
  from all flavors of neutrinos and antineutrinos. 
  The contours are logarithmically spaced with intervals of
  0.5 dex, bold contours are labeled with their respective values.
  The grey shading emphasizes the emission levels, dark grey
  corresponding to the strongest energy loss by neutrino emission.
  The dotted line marks the intersection of the $z = 0$ plane with
  the surface of optical depth unity where neutrinos start to stream
  off freely.}
\label{fig:neutcontN}
\end{figure*}

\begin{figure*}
\begin{tabular}{cc}
  \epsfxsize=8.5cm  \epsfclipon\epsffile{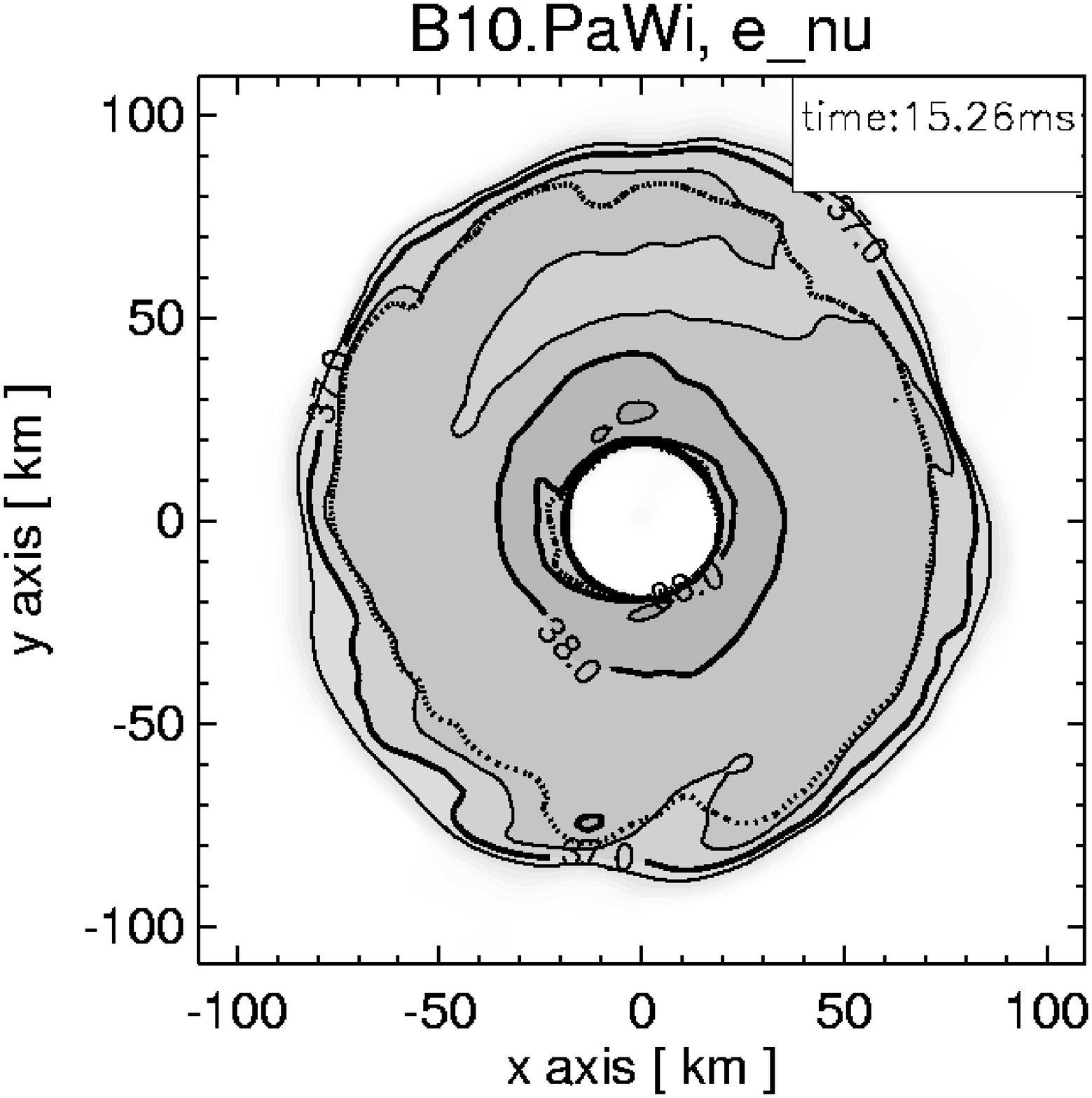} 
                    \put(-0.50,7.12){{\Large \bf \sf a}} &
  \epsfxsize=8.5cm  \epsfclipon\epsffile{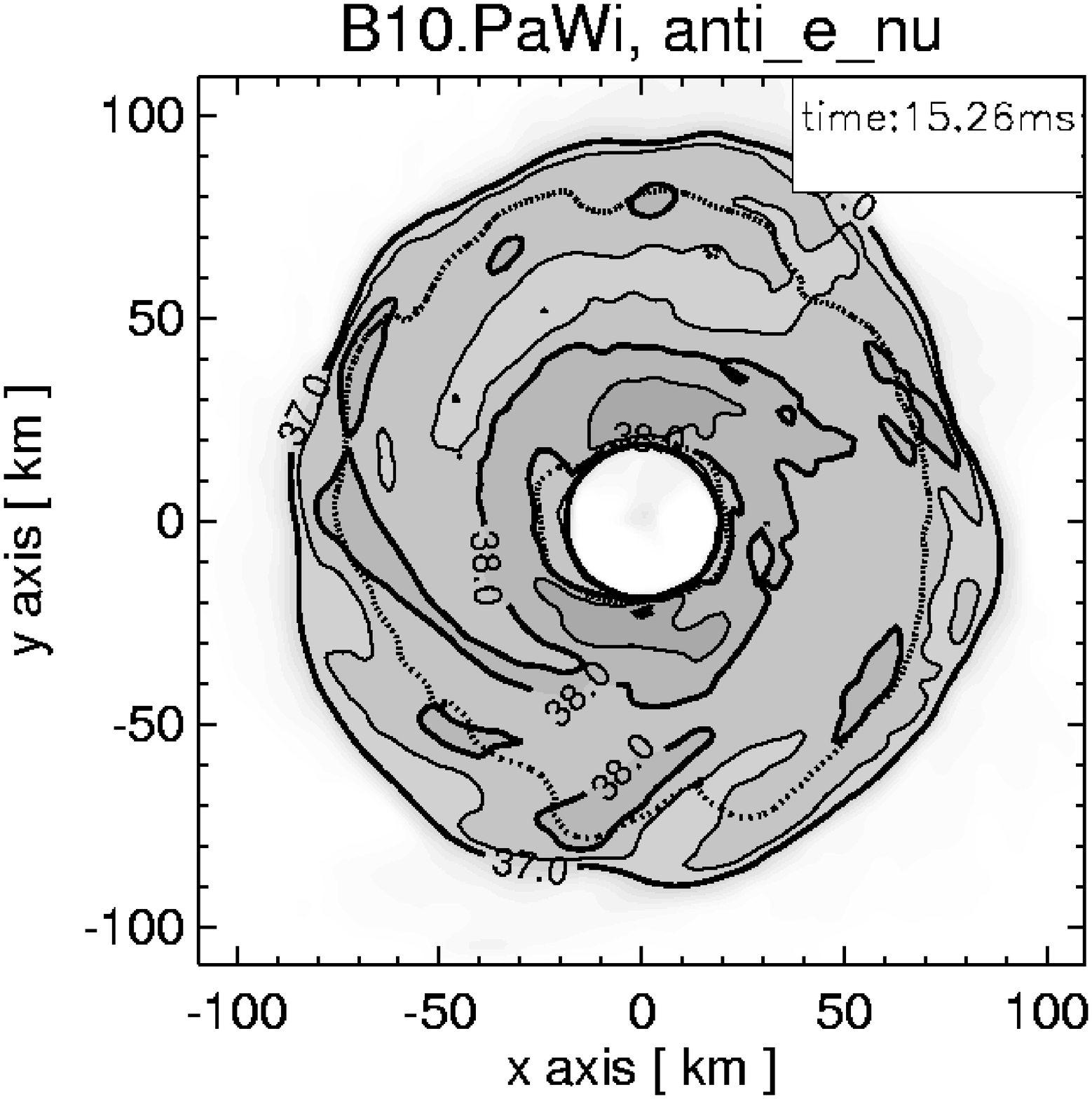}
                    \put(-0.50,7.12){{\Large \bf \sf b}}\\
[2ex]
  \epsfxsize=8.5cm  \epsfclipon\epsffile{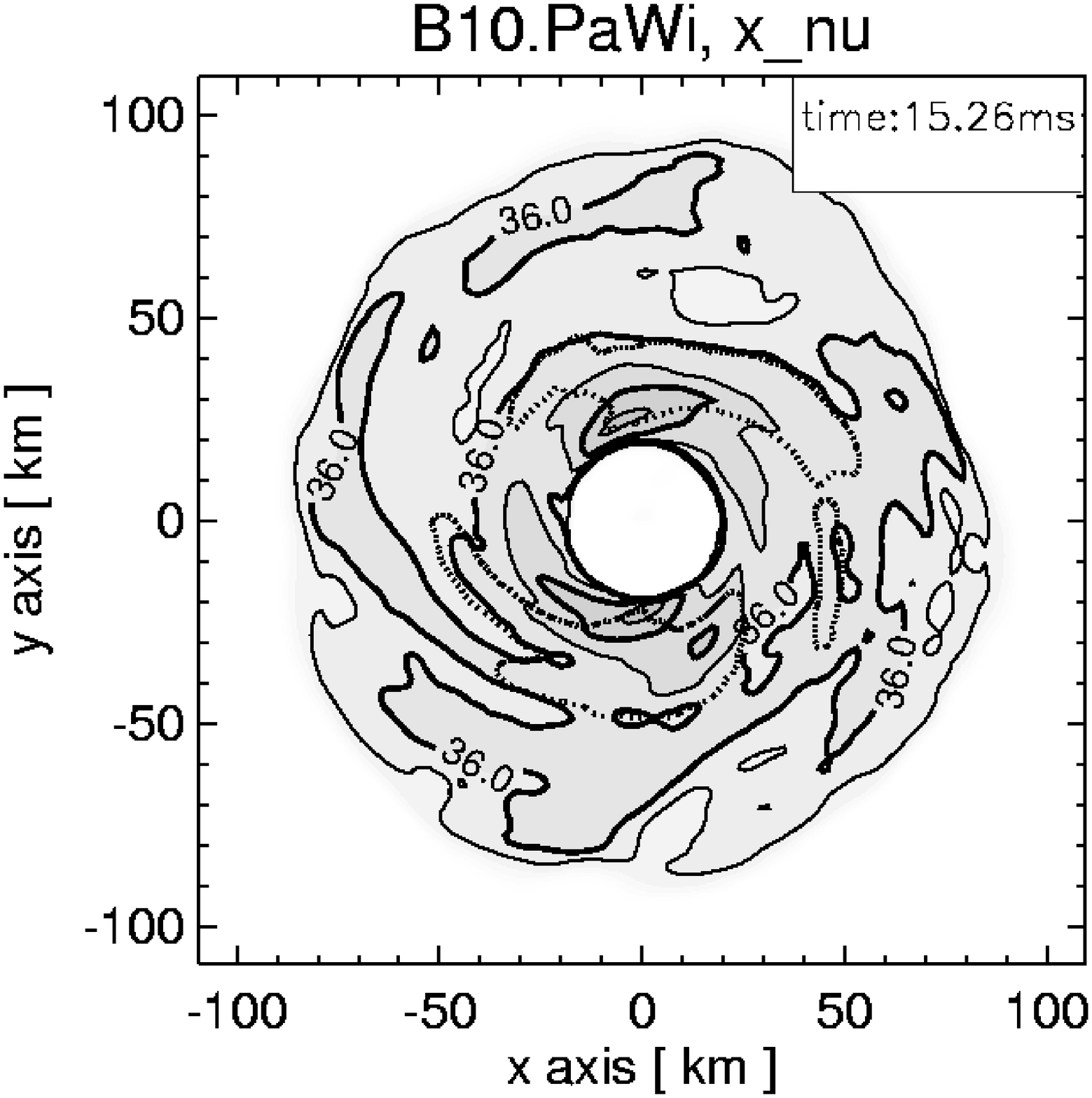} 
                    \put(-0.50,7.12){{\Large \bf \sf c}} &
  \epsfxsize=8.5cm  \epsfclipon\epsffile{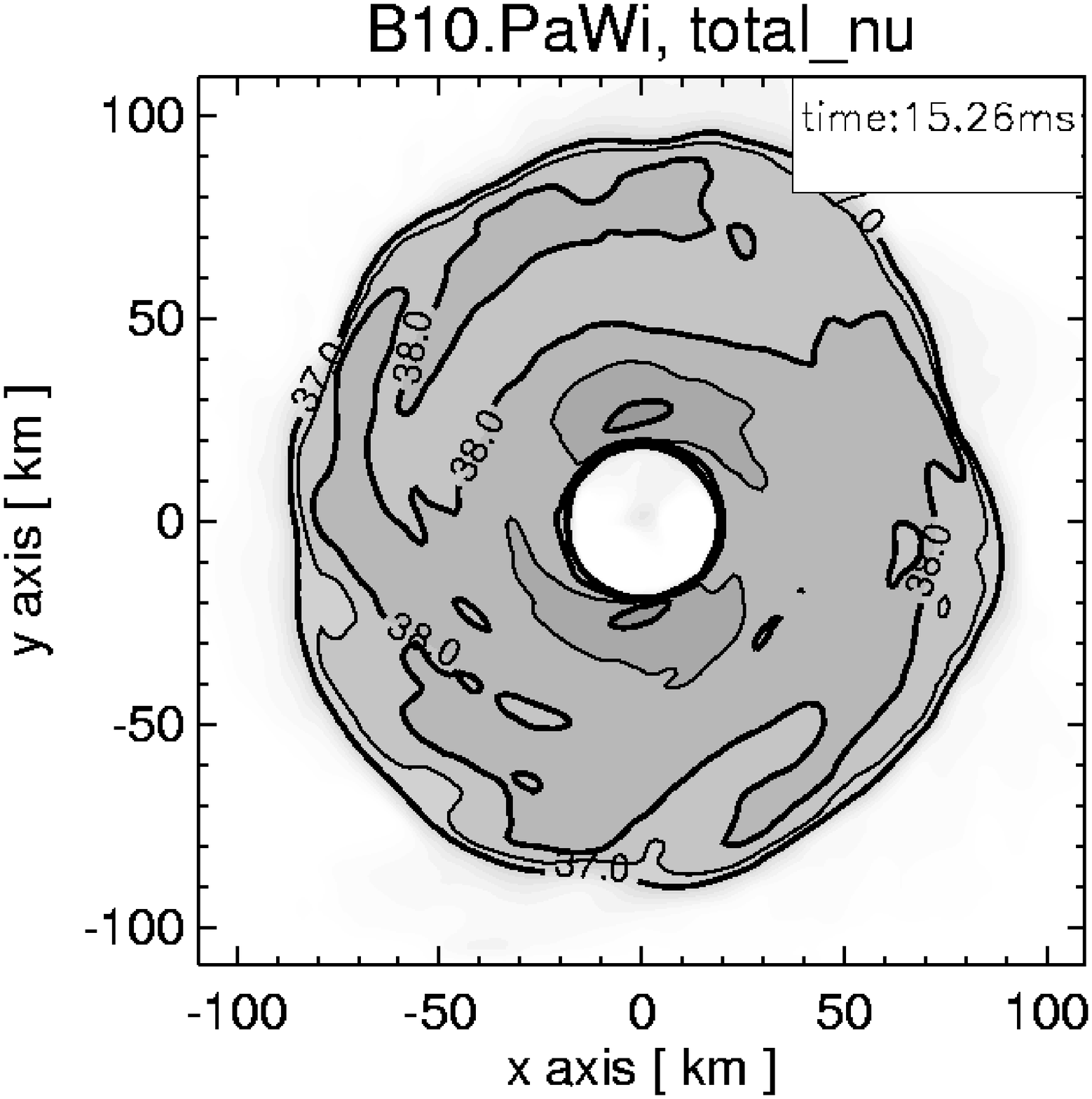}
                    \put(-0.50,7.12){{\Large \bf \sf d}}\\
\end{tabular}
  \caption[]{Same as Fig.~\ref{fig:neutcontN} but for Model~${\cal B}$10
  at time $t = 15.26\,$ms.}
  \label{fig:neutcontP}
\end{figure*}

\begin{figure*}
\begin{tabular}{cc}
   \epsfxsize=8.5cm  \epsfclipon\epsffile{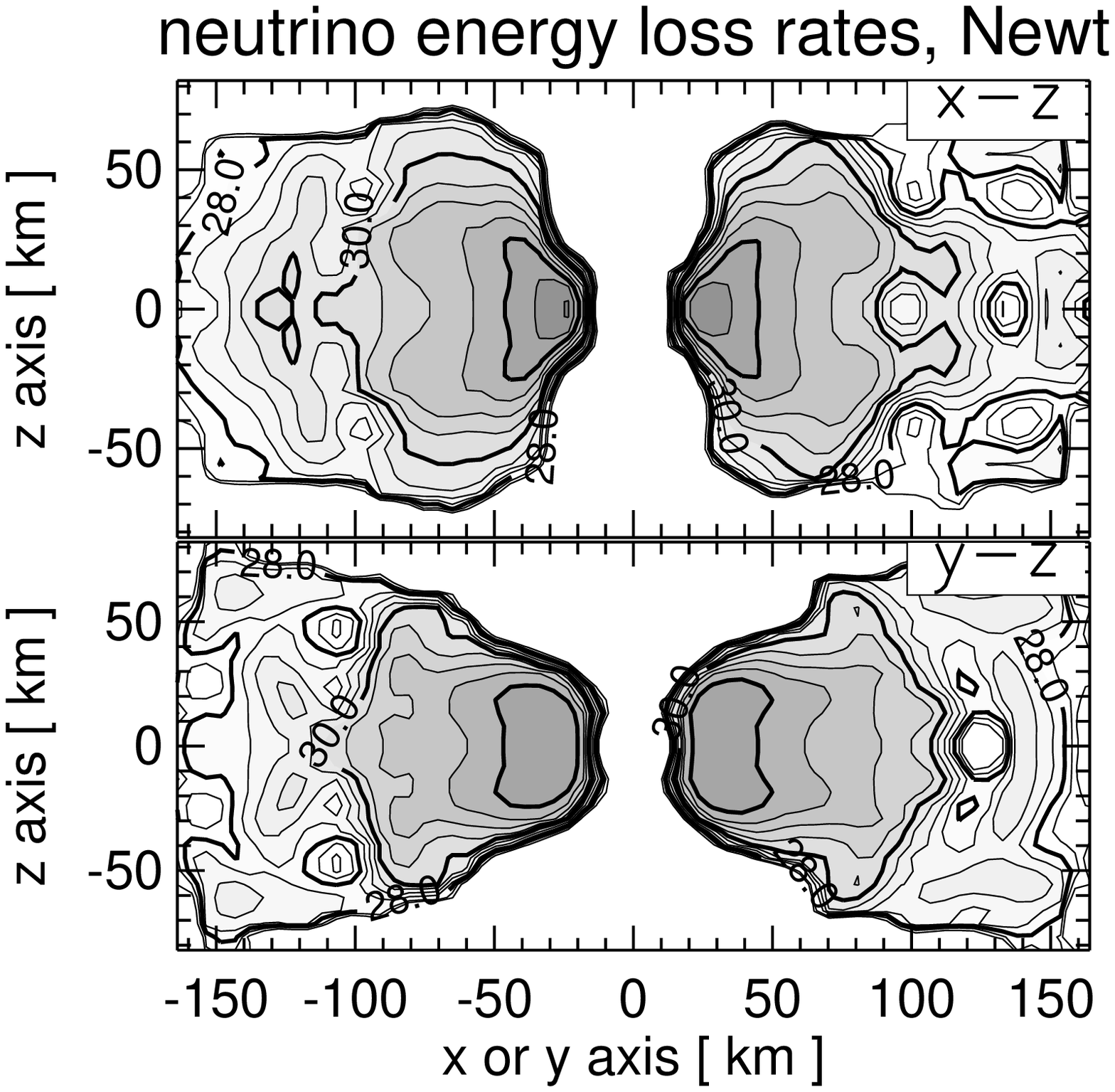} &
   \epsfxsize=8.5cm  \epsfclipon\epsffile{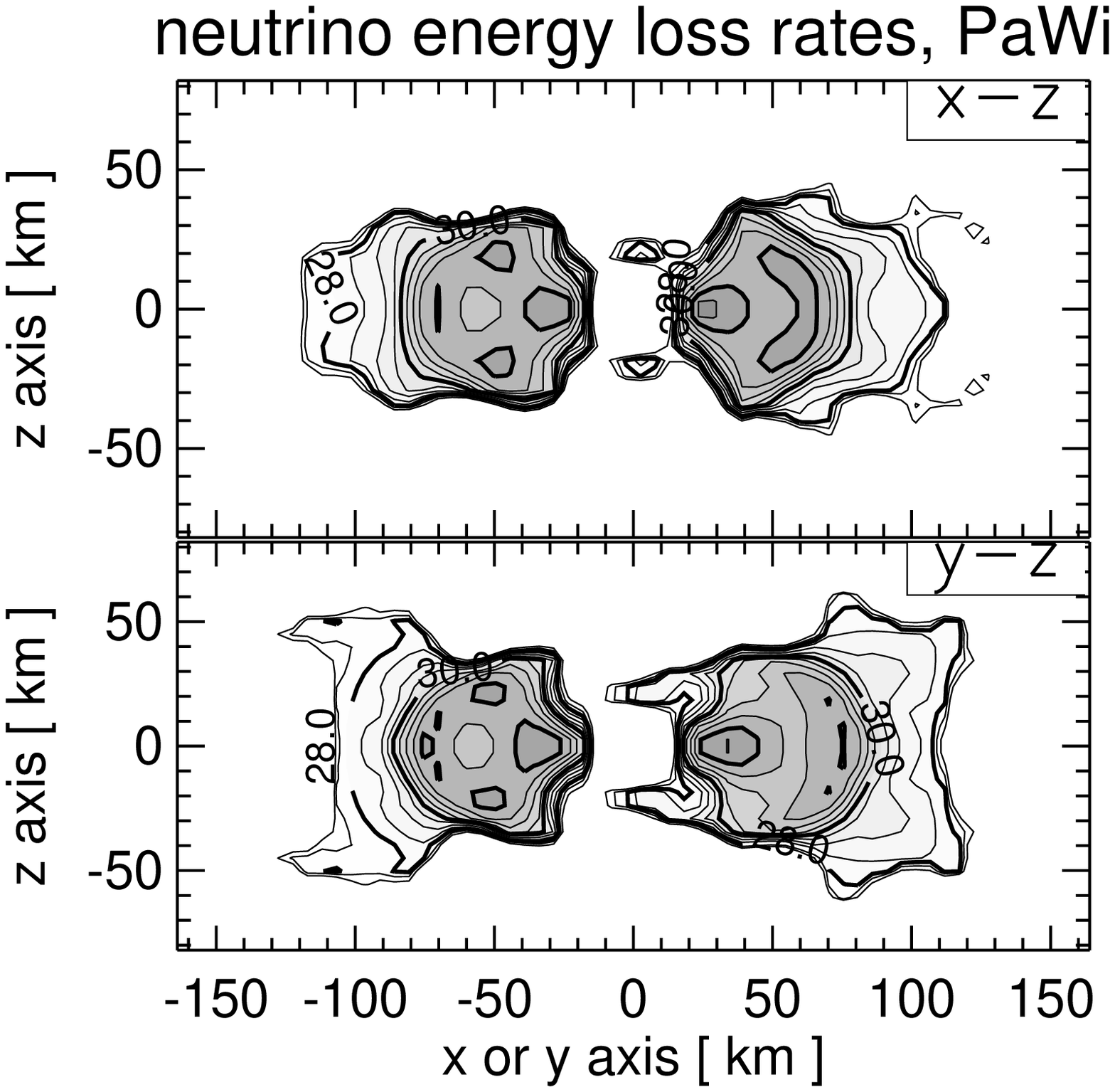}\\
\end{tabular}
  \caption[]{Contour plots of the local energy loss rates (in
  ${\rm erg\,cm}^{-3}{\rm s}^{-1}$ 
  due to the emission of neutrinos and antineutrinos of all
  flavors. The left plot shows cuts in the $x$-$z$- and $y$-$z$-planes
  perpendicular to the equatorial plane of the Newtonian Model~{\bf B}10
  near the end of the simulation, the right plot the corresponding cuts
  for the Paczy\'nski-Wiita Model~${\cal B}$10.
  The contours are spaced logarithmically with intervals of
  0.5 dex, bold contours correspond to the values 28, 30 and 32.
  The grey shading emphasizes the emission levels, dark grey
  corresponding to the strongest energy loss by neutrino emission.}
\label{fig:neutper}
\end{figure*}

\section{Neutrino emission\label{sec:neutrino}}

The neutrino luminosities as functions of time for the Newtonian and
Paczy\'nski-Wiita tori are displayed in Figs.~\ref{fig:neutradN} and
\ref{fig:neutradP}, respectively, and the corresponding mean energies
of electron neutrinos ($\nu_e$), electron antineutrinos ($\bar\nu_e$)
and heavy-lepton neutrinos ($\nu_x\equiv\nu_{\mu},\,\bar\nu_{\mu},\,
\nu_{\tau},\,\bar\nu_{\tau}$) are given in Figs.~\ref{fig:meneN} and
\ref{fig:meneP}. In order to show the sensitivity of the results to the 
moment the black hole is assumed to form, data from two simulations
which were started at different times are plotted in each of the figures:
Models~{\bf B}4 and {\bf B}10 for the Newtonian simulations,
and Models~${\cal B}$2 and ${\cal B}$10 for the Paczy\'nski-Wiita potential.

Despite of its much smaller mass (by a factor 7.5) and lower 
density and temperature (see Fig.~\ref{fig:phiav}), the Paczy\'nski-Wiita 
torus radiates neutrinos with similar luminosities and mean energies
as the Newtonian model. Mainly $\nu_e$ and $\bar\nu_e$ are emitted.
At the end of the computed evolution ($t\approx 15\,{\rm ms}$),
the Newtonian model (Figs.~\ref{fig:neutradN} and \ref{fig:meneN}) has
a total luminosity $L_{\nu} = 12\times 10^{52}\,{\rm erg\,s}^{-1}$,
with contributions of 
$L_{\nu_e} = 3.5\times 10^{52}\,{\rm erg\,s}^{-1}$ from $\nu_e$,
$L_{\bar\nu_e} = 6.5\times 10^{52}\,{\rm erg\,s}^{-1}$ from $\bar\nu_e$, 
and $L_{\nu_x} = 0.4\times 10^{52}\,{\rm erg\,s}^{-1}$ from $\nu_x$
individually. The mean energies of the emitted neutrinos are  
$\ave{\epsilon_{\nu_e}} = 9\,$MeV, $\ave{\epsilon_{\bar\nu_e}} = 13\,$MeV
and $\ave{\epsilon_{\nu_x}} = 21\,$MeV. For the Paczy\'nski-Wiita torus, 
the total luminosity reaches about 50\% of the value obtained in the
Newtonian simulation, $L_{\nu} = 6.7\times 10^{52}\,{\rm erg\,s}^{-1}$,
and the luminosities of $\nu_e$ and $\bar\nu_e$ reach 60--70\% of the
corresponding Newtonian values.
The average energies of $\nu_e$ and $\bar\nu_e$ are
slightly higher (around 10--12$\,$MeV and 15--17$\,$MeV, respectively).
The differences between the simulations are more pronounced in case of 
$\nu_x$. At $t\approx 15\,$ms the $\nu_x$ luminosity is approximately
one order of magnitude smaller for the Paczy\'nski-Wiita model,
and the emitted $\nu_x$ are less energetic with a mean energy of only
16--19$\,$MeV instead of 21$\,$MeV.

These differences result from
the fact that the Paczy\'nski-Wiita torus is essentially transparent
for muon and tau neutrinos whereas a well defined average muon and tau
neutrinosphere (to be more precise: a toriodal neutrinosurface) exists 
in the Newtonian model. This is clearly visible from the dotted lines
in panels~c of Figs.~\ref{fig:neutcontN} and \ref{fig:neutcontP},
which represent the intersections of the equatorial plane at $z = 0$ with
the neutrinosurface. The latter is defined as the two-dimensional hypersurface
where the optical depth of the torus perpendicular to the equatorial plane
is unity, i.e., where
\begin{equation}
\tau_{z,\nu_i}(x,y)\,\equiv\,\int_z^\infty {\rm d}z'\,\kappa_{\nu_i}(x,y,z')
\,=\,1\ ,
\label{eq:tauone}
\end{equation}
with $\kappa_{\nu_i}(x,y,z)$ being the total opacity (defined as the inverse
of the mean free path) for the energy transport
of neutrino $\nu_i$ at a point $(x,y,z)$. In contrast, the very similar 
properties of the $\nu_e$ and $\bar\nu_e$ emission can be understood from
similar thermodynamical conditions (density, temperature, entropy,
see Fig.~\ref{fig:phiav}) at the corresponding neutrinosurfaces. The latter
have an outer radius of about 100$\,$km in case of the Newtonian
torus (panels~a and b in Fig.~\ref{fig:neutcontN}) and of 70--80$\,$km in the
less massive Paczy\'nski-Wiita model (panels~a and b in 
Fig.~\ref{fig:neutcontP}).
The slightly different sizes and thus different areas of the toroidal 
neutrinosurfaces in panels~a and b of Figs.~\ref{fig:neutcontN} and
\ref{fig:neutcontP} account for the moderate differences of the
$\nu_e$ and $\bar\nu_e$ luminosities in both models.

The neutrino emission is primarily determined by the size of the
neutrinosurface and the thermodynamical conditions in the layer where
the average neutrino optical depth is around unity. For this reason the 
torus mass influences the luminosities indirectly through the radius of
the neutrinosurface, until the torus mass and its density 
and temperature become so low that neutrino transparency is reached.
The smaller torus mass and larger relative and absolute accretion rate
in the Paczy\'nski-Wiita simulation
(see Figs.~\ref{fig:massaccBH}, \ref{fig:masstBH} and \ref{fig:diskmass})
lead to a different time evolution of the neutrino luminosities (and mean
neutrino energies) in both models.
Extrapolation of the luminosity decrease towards the end of the simulations
in Figs.~\ref{fig:neutradN} and \ref{fig:neutradP} suggests a longer
decay time scale for the Newtonian model in agreement with the longer
accretion time scale given in Fig.~\ref{fig:masstBH}. 

Comparing the models with early formation of the black hole, 
Model~{\bf B}4 and Model~${\cal B}$2, with those where the black hole
collapse is assumed to happen later, Models~{\bf B}10 and ${\cal B}$10,
shows that the latter have somewhat higher neutrino luminosities.
This is explained by the higher temperatures of the tori of 
Models~{\bf B}10 and ${\cal B}$10, see Fig.~\ref{fig:maxtempT},
and the correspondingly lower densities in the thermally inflated later
states (Fig.~\ref{fig:maxrhoT}). The effect is particularly strong
for $\nu_x$ which are mainly produced by the annihilation of electrons
and positrons into neutrino-antineutrino pairs at the conditions present 
in the tori, because the energy emission rate for this process is 
extremely temperature sensitive and increases proportional to $T^9$. 
Since the accretion rates are time dependent and the temperatures in the
tori show fluctuations, the neutrino luminosities and mean energies are
variable on a time scale of 1--2$\,$ms.

The regions with the strongest neutrino emission are visible as
dark grey shaded areas in Figs.~\ref{fig:neutcontN}, \ref{fig:neutcontP}
and \ref{fig:neutper}. In the former two figures the contours correspond
to levels of constant energy emission rate per unit area as obtained
by integration of the energy loss rate per volume from the equatorial
plane at $z = 0$ to infinity. In Fig.~\ref{fig:neutper} vertical cuts 
through the Newtonian and Paczy\'nski-Wiita accretion tori are displayed 
which show the
total neutrino energy loss rates per unit volume in the $x$-$z$ and
$y$-$z$ planes. The peak values of the emission rates are similar in
both types of models but the main neutrino emitting region is less
extended in case of the Paczy\'nski-Wiita potential. The dotted lines in
Figs.~\ref{fig:neutcontN} and \ref{fig:neutcontP} mark the intersection
of the equatorial plane with the two-dimensional hypersurface where
the transport optical depth for the energy flux is unity 
(Eq.~\ref{eq:tauone}), i.e., where the spectrally averaged mean free 
path of the neutrinos or antineutrinos of a certain flavor is of
the same order as the size of the emitting volume. Outside this 
neutrinosurface the neutrinos stream off essentially freely and interact
by scattering or absorption with the gas particles on average only one 
more time. The muon and tau neutrino opacity is dominated by
neutral-current scatterings off neutrons and protons. Only $\nu_e$ and
$\bar\nu_e$ are also absorbed on neutrons and protons, respectively, via
charged-current inverse beta processes. The neutrinosurface of the 
heavey-lepton neutrinos has a toroidal shape
only in case of the Newtonian calculation (Fig.~\ref{fig:neutcontN}, panel~c)
but splits up into several distinct islands for the less massive and less
dense Paczy\'nski-Wiita torus which is near to neutrino-transparent 
conditions. The highest neutrino
energy loss rates are found in a region between the inner grid boundary
at $2R_{\rm s}\approx 18\,$km and an equatorial radius of approximately 
70$\,$km. In the Newtonian model significant contributions to the neutrino 
luminosity come even from larger distances out to about 100$\,$km where 
the larger volume compensates for the smaller emission rates.
The vertical cuts of 
Fig.~\ref{fig:neutper} confirm an effect which was already visible in
the density plots of Figs.~\ref{fig:vertN} and \ref{fig:vertP}:
In the Paczy\'nski-Wiita model, in contrast to the Newtonian simulation,  
gas flows towards the black hole even from the poles where it shows up 
by its neutrino emission in the right plot of Fig.~\ref{fig:neutper}.

The maximum energy loss rates of neutrinos and antineutrinos of all 
flavors are typically of the order $10^{32}\,{\rm erg}$ ${\rm cm}^{-3}
{\rm s}^{-1}$. In single peaks values of even
$10^{33}\,{\rm erg}\,{\rm cm}^{-3}{\rm s}^{-1}$
can be reached (Fig.~\ref{fig:neutper}). In gas with density
$\rho\approx 10^{11}\,{\rm g\,cm}^{-3}$ (Figs.~\ref{fig:vertN} and 
\ref{fig:vertP}) these rates correspond to a specific energy loss
of 1000$\,{\rm MeV\,s}^{-1}$ per nucleon up to even 
$10^4\,{\rm MeV}$ ${\rm s}^{-1}$ per nucleon just before the gas reaches
the inner grid boundary and disappears in the black hole.
For a total neutrino luminosity of $10^{53}\,{\rm erg\,s}^{-1}$  
from a torus with mass of approximately 0.25$\,M_{\odot}$ in the Newtonian 
simulation, one calculates an average neutrino energy loss rate of 
200 ${\rm MeV\,s}^{-1}$ per nucleon. This means that the binding energy
of a nucleon in the gravitational potential of the 3$\,M_{\odot}$
black hole at the position of the last stable circular orbit at 
$3R_{\rm s}\approx 27\,$km is radiated away in less than a second,
or an energy equivalent of more than 2\% of the nucleon's rest mass 
escapes in neutrinos within only 0.1 seconds. This estimate is 
in agreement with the instantaneous efficiency for the conversion 
of rest-mass energy into neutrino energy,
$q_{\nu}\equiv L_{\nu}/(\dot Mc^2)$, which we calculate at the end of
the simulation to be $q_{\nu}^{\rm N}\approx 1.3$\% for the Newtonian
torus (Table~\ref{tab:efficiencies}). In case of the
Paczy\'nski-Wiita model the corresponding value is about a factor
of 2.5 lower, $q_{\nu}^{\rm PW}\approx 0.5$\%. These numbers are 
significantly smaller than the maximum efficiency of 5.7\% for
relativistic disk accretion onto a Schwarzschild black hole (8.3\% for a
thin Newtonian accretion disk) because the tori are not transparent
for neutrinos. Due to the high densities and temperatures, the 
diffusion time scale for neutrinos becomes longer than the accretion
time scale of the gas into the black hole. Therefore the tori are advection
dominated and cooling does not reach its maximum possible efficiency.

\begin{table*}
\caption[]{
Torus mass $M_{\rm v}$, black hole mass accretion rate $\dot M_{\rm v}$,
typical accretion time scale of the torus, 
$t_{\rm acc}\equiv M_{\rm v}/\dot M_{\rm v}$,
total neutrino luminosity $L_\nu$, efficiency 
$q_{\nu} \equiv L_{\nu}/(\dot Mc^2)$, integral rate of energy
deposition by neutrino-antineutrino annihilation around the accretion
torus, $\dot{E}_{\nu\bar{\nu}}$, 
and efficiency $q_{\nu\bar\nu} \equiv \dot E_{\nu\bar\nu}/L_{\nu}$
for the Newtonian Model~{\bf B}10 and Paczy\'nski-Wiita Model~${\cal B}$10.
All quantities are given at the time $t_{\rm fin}$ when the simulation
was stopped.
$E_{\nu\bar{\nu}}$ gives the estimated total energy deposition by
$\nu\bar\nu$ annihilation in the time interval $t_{\rm fin}+t_{\rm acc}$
which is roughly equal to the duration of neutrino emission.
}
\begin{flushleft}
\tabcolsep=1.5mm
\begin{tabular}{llccccccccc}
\hline\\[-3mm]
model & potential     & $t_{\rm fin}$ & $M_{\rm v}$ & $\dot M_{\rm v}$          
      & $t_{\rm acc}$ & $L_\nu$       & $q_{\nu}$   & $\dot{E}_{\nu\bar{\nu}}$ 
                      & $q_{\nu\bar\nu}$ & $E_{\nu\bar{\nu}}$ \\ 
 & & ms & 
   {\scriptsize$10^{-2}M_\odot$} & {\scriptsize$M_\odot\,{\rm s}^{-1}$} &
   ms &
   {\scriptsize$10^{52}\frac{\rm erg}{\rm s}$} & &
   {\scriptsize$10^{50}\frac{\rm erg}{\rm s}$} & & 
   {\scriptsize$10^{49}{\rm erg}$}
\\[0.3ex] \hline\\[-3mm]
{\bf B}10    & Newt & 14.9 & 26.4 & 5. & 53. & 12. & 0.013 & 4.9 & 0.0041 & 3.3 \\
${\cal B}$10 & PaWi & 15.2 & 3.5  & 7. & 5.  & 6.7 & 0.005 & 3.1 & 0.0046 & 0.6 \\
[0.7ex]
\hline
\end{tabular}
\end{flushleft}
\label{tab:efficiencies}
\end{table*}

\begin{figure*}
\begin{tabular}{cc}
  \epsfxsize=8.5cm \epsfclipon \epsffile{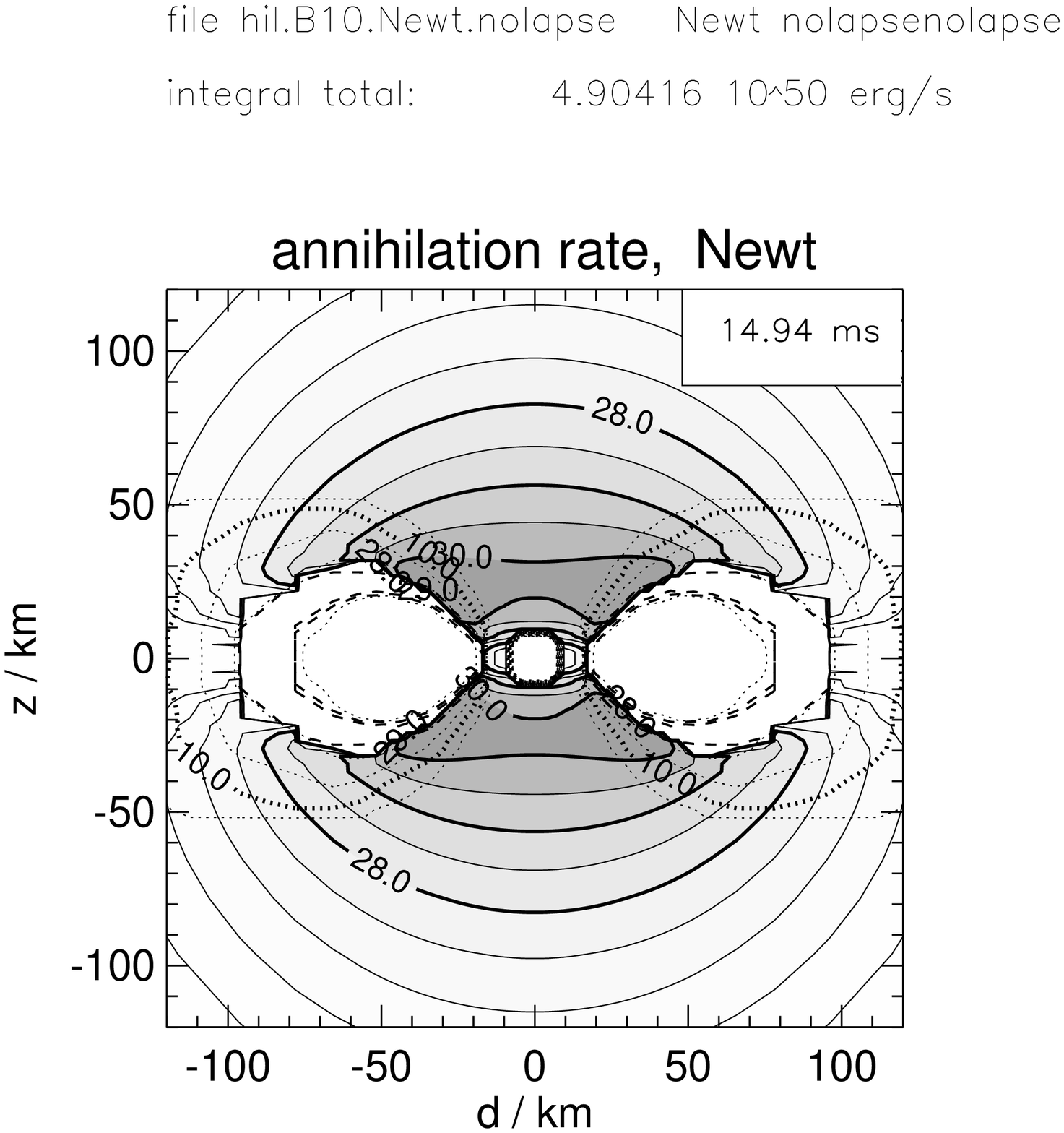} &
  \epsfxsize=8.5cm \epsfclipon \epsffile{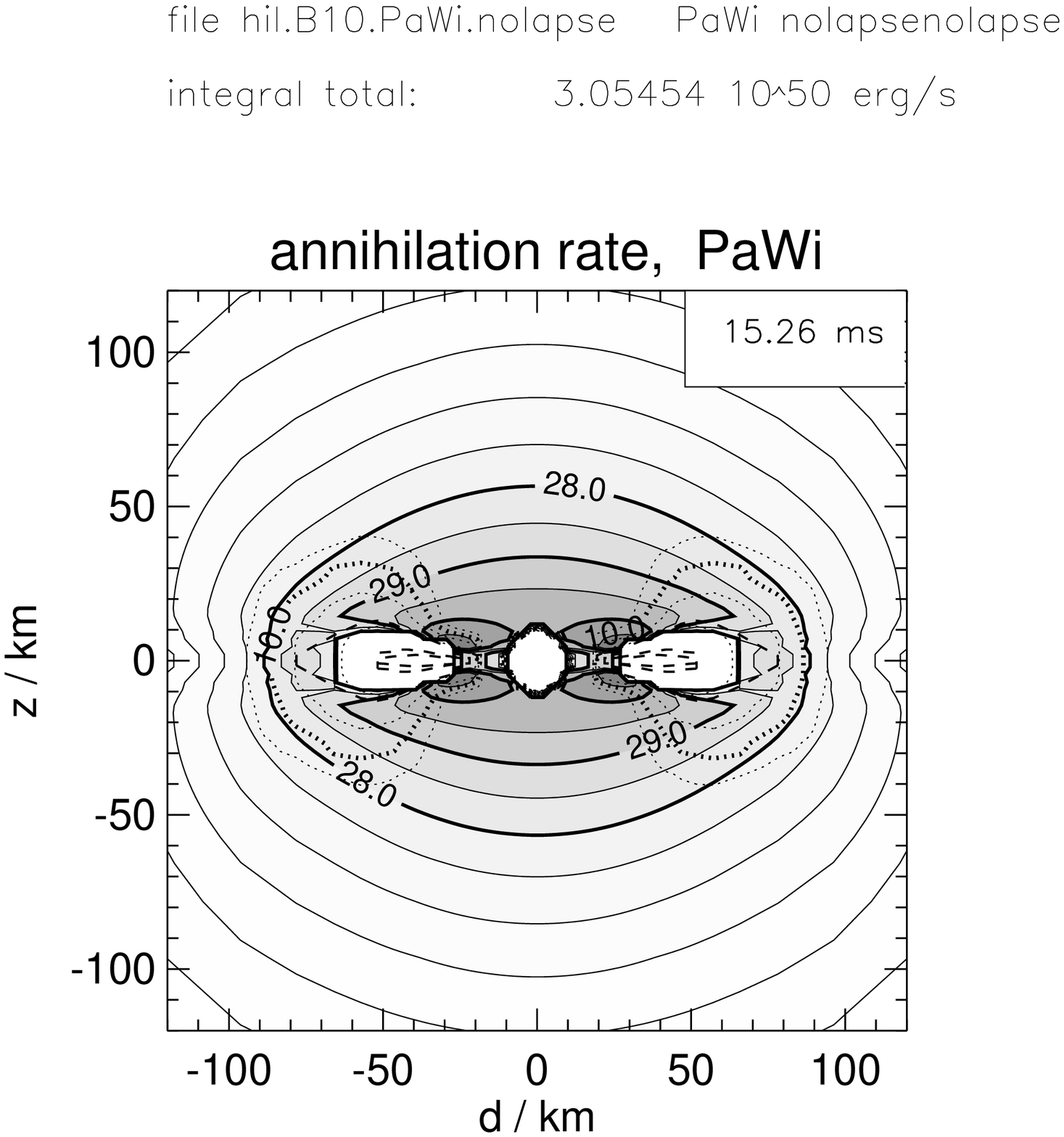} \\
\end{tabular}
  \caption[]{Maps of the local energy deposition rates 
(in~erg$\,$cm$^{-3}\,$s$^{-1}$) by $\nu\bar\nu$ annihilation into
$e^+e^-$ pairs in the surroundings of the accretion torus of
the Newtonian
Model~{\bf B}10 at time $t = 14.94\,$ms (left) and of
the Paczy\'nski-Wiita
Model~${\cal B}$10 at time $t = 15.26\,$ms (right).
The values shown as solid contour lines in a plane perpendicular
to the equatorial plane
were obtained as averages over the azimuthal angle
around the $z$-axis. The abscissa $d$ measures the distance from 
the grid center in the $x$-$y$-plane. The white octagonal area
around the center with a semidiameter of one Schwarzschild radius 
indicates the presence of the central black hole. The contours
are logarithmically spaced in steps of 0.5~dex, the grey shading
emphasizes the levels with dark grey meaning high energy
deposition rate. The energy deposition rate was
evaluated only in that region around the torus where the baryon
mass density is below $10^{11}\,{\rm g\,cm}^{-3}$. The integral
value of the energy deposition rate at the displayed time
is $4.9\times 10^{50}\,{\rm erg\,s}^{-1}$ for Model~{\bf B}10 (left)
and $3.1\times 10^{50}\,{\rm erg\,s}^{-1}$ for Model~${\cal B}$10
(right). The dashed lines mark the (approximate) positions
of the neutrino``spheres'' of $\nu_e$, $\bar\nu_e$, and $\nu_x$
(from outside inward), defined by the requirement that the
optical depths in $z$-direction are $\tau_{z,\nu_i} = 1$
(see also Eq.~(\ref{eq:tauone}) and Figs.~\ref{fig:neutcontN} and 
\ref{fig:neutcontP}). The dotted contour lines
represent levels of constant values of the azimuthally averaged
mass density, also logarithmically spaced with intervals of 0.5~dex,
the bold dotted line corresponding to $\rho=10^{10}\,{\rm g\,cm}^{-3}$.}
\label{fig:annihi}
\end{figure*}

\section{Neutrino-antineutrino annihilation\label{sec:annihilation}}

We evaluate our hydrodynamical models in a post-\-pro\-cess\-ing step for 
the annihilation of neutrinos and antineutrinos into electron-positron
pairs. The corresponding methods were explained in detail in Ruffert
\& Janka (\cite{ruf98a}) and in Ruffert et al.~(\cite{ruf97}).
Neutrinos and antineutrinos emitted from
the hot accretion torus interact with each other in the surroundings
with a finite probability which depends on the number
densities and energies of these neutrinos and on the
angle between the directions of neutrino and antineutrino propagation (see 
Goodman et al. \cite{goo87}, Cooperstein et al. \cite{coo87}, 
Mayle \cite{may90}; also Ruffert et al.~\cite{ruf97}).
Therefore the local energy deposition rate by $\nu\bar\nu$ annihilation
increases proportional to the product of neutrino and antineutrino 
luminosities and the spectrally averaged neutrino energy, times a factor
that accounts for the dependence on the angular distribution of the
neutrinos. The annihilation rate drops rapidly with increasing
distance from the neutrino source because both the neutrino
number densities and the mean angle of neutrino-antineutrino collisions 
decrease. 

\begin{figure}
\begin{tabular}{c}
  \epsfxsize=8.5cm  \epsffile{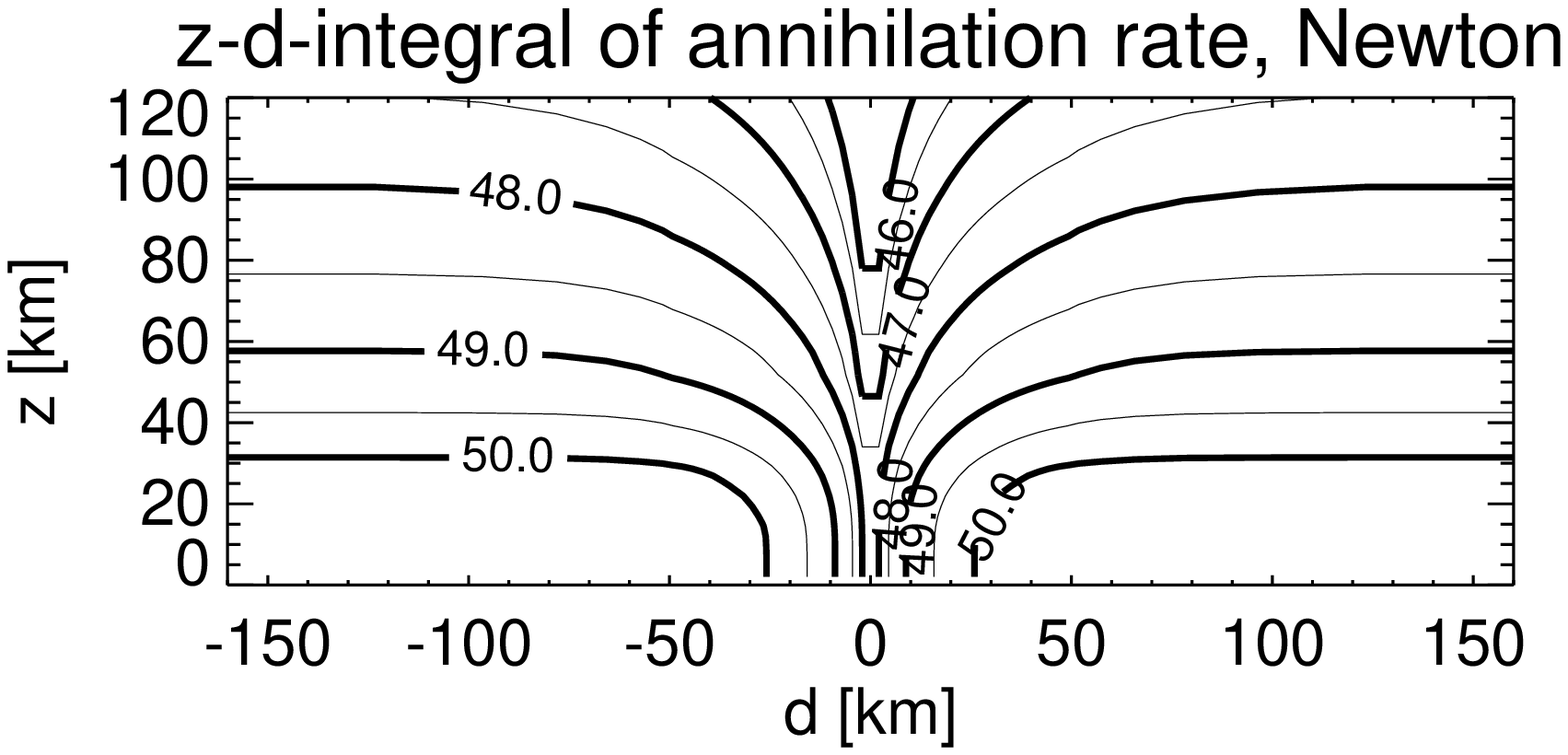} \\
  \epsfxsize=8.5cm  \epsffile{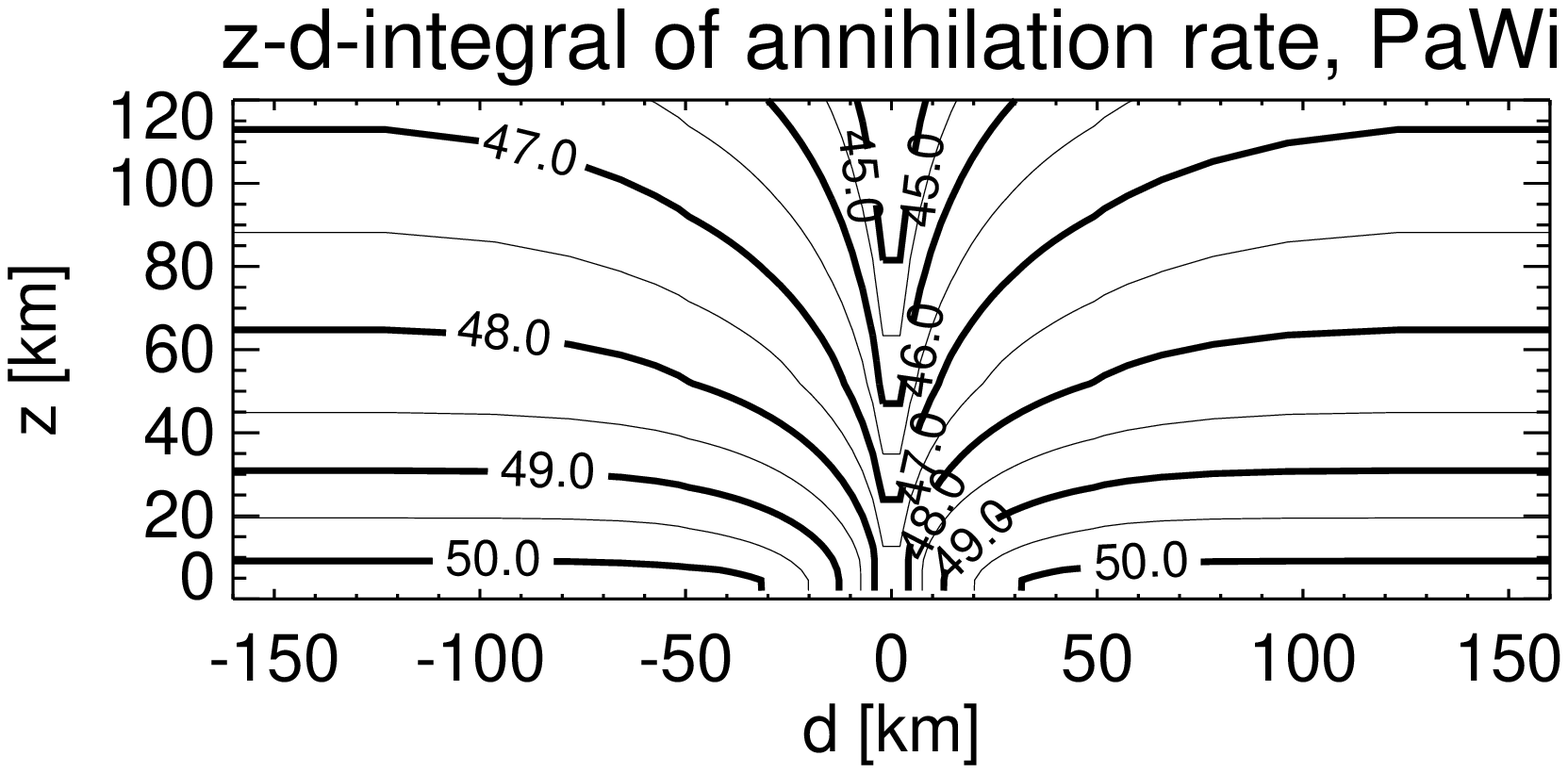} \\
\end{tabular}
  \caption[]{
   Cumulative energy deposition rates by $\nu\bar\nu$ annihilation
   for the Newtonian Model~{\bf B}10 at time $t = 14.9\,$ms (upper
   panel) and for the Paczy\'nski-Wiita Model~${\cal B}$10 at time
   $t = 15.2\,$ms (lower panel). The azimuthally averaged annihilation
   rate per unit volume was integrated along the $z$-direction from 
   a given value of $z$ out to the grid boundary (practically infinity
   because of the rapid decrease of the annihilation rate with distance)
   and in addition from $d = 0$ to the value of $d$ given on the abscissa.
   The contours are spaced logarithmically in steps of 0.5~dex and 
   represent values only for the hemisphere above the equatorial plane
   (measured in erg$\,$s$^{-1}$). They are mirrored along the $z$-axis.
   The two panels should
   be compared with the lower panels in Fig.~\ref{fig:coneNP} and were
   constructed in an analogous way.}
  \label{fig:annconeNP}
\end{figure}

The computational procedure to obtain the energy deposition 
outside the accretion torus can be briefly summarized as follows.
In a first step the hydrodynamical model at a chosen time is mapped from 
the nested grids of the simulation onto an equidistant cartesian grid.
Next, for each type of neutrino or antineutrino the two-dimensional 
surface is determined where the optical depth in $z$-direction is unity
(see Eq.~\ref{eq:tauone}). The neutrino energy loss rates
from the torus volume are projected onto these surfaces and treated as
surface emissivities. In order to compute the energy deposition rate 
per unit volume by $\nu\bar\nu$ annihilation at a point $\vec r$, one has
to sum up the contributions by the neutrino and antineutrino emission
from all parts of the neutrinosurfaces which radiate into the direction
of point $\vec r$. The total energy deposition rate includes contributions
from all three flavors of neutrinos and antineutrinos. Finally, volume 
integrals of the energy deposition rate can be obtained by summation 
over specified regions of the equidistant cartesian grid.
The results of this evaluation
are plotted in Figs.~\ref{fig:annihi} and \ref{fig:annconeNP}
for the final states of our torus simulations. The figures
show azimuthally averaged (around the $z$-axis) quantities
in a plane perpendicular to the equatorial plane.

Figure~\ref{fig:annihi} displays contours of constant 
energy deposition rate per unit volume in those regions 
around the Newtonian (left) and Paczy\'nski-Wiita (right) tori where
the baryonic mass density is less than $10^{11}\,{\rm g\,cm}^{-3}$.
The dotted lines represent density contours, the dashed lines mark
the positions of the average neutrinosurfaces of $\nu_e$, $\bar\nu_e$
and $\nu_x$ (from outside outward). The $\nu_e$ neutrinosurface is very
close to the toroidal surface which corresponds to a density of
$\rho = 10^{11}\,{\rm g\,cm}^{-3}$. The neutrinosurfaces of $\bar\nu_e$
and $\nu_x$ are deeper inside, because $\bar\nu_e$ are absorbed onto
protons which are less abundant in the torus than neutrons, and the
streaming of the $\nu_x$ is inhibited only by neutral-current scatterings
off nucleons. Both the dashed and dotted sets of
contours demonstrate that the Paczy\'nski-Wiita torus is significantly
smaller and nearly transparent for electron antineutrinos and heavy-lepton
neutrinos. 

The maximum energy deposition rates by $\nu\bar\nu$ annihilation 
(solid contours in Figure~\ref{fig:annihi}) exceed
$10^{30}\,{\rm erg\,cm}^{-3}{\rm s}^{-1}$ in the polar regions 
above and below the equatorial plane at heights $|z|$ between 10$\,$km
and 30$\,$km in the Newtonian
model. Such high values occur also in the Paczy\'nski-Wiita simulation,
but only very close to the surface of the torus and to the equatorial 
plane, between the black hole and the last stable circular orbit at
$3R_{\rm s}\approx 27\,$km.
the values at larger distances $|z|$ are typically one order of
magnitude lower than in the Newtonian model at the same $|z|$. 
Within the displayed region, the energy deposition rate along the polar
axis decreases roughly proportional to $z^{-6}$ for $|z|\ga 30\,$km.

The integral rate of heating by $\nu\bar\nu$ annihilation in
the computed volume is $4.9\times 10^{50}\,{\rm erg\,s}^{-1}$ for the
Newtonian model. About 40\% of this energy 
($1.9\times 10^{50}\,{\rm erg\,s}^{-1}$) are deposited in a cylinder
with radius $3R_{\rm s} = 27\,$km around the $z$-axis, 90\%
($4.4\times 10^{50}\,{\rm erg\,s}^{-1}$) in a cone with opening 
half-angle of 30--40 degrees which is approximately bounded by the
isodensity contour for $\rho = 10^{10}\,{\rm g\,cm}^{-3}$ in
Fig.~\ref{fig:annihi}. The corresponding numbers for the 
Paczy\'nski-Wiita torus are $3.1\times 10^{50}\,{\rm erg\,s}^{-1}$ for
the total energy deposition, of which 
$1.3\times 10^{50}\,{\rm erg\,s}^{-1}$ end up in the cylinder along the
axis and $1.9\times 10^{50}\,{\rm erg\,s}^{-1}$ in the cone. 
From these values one computes the total conversion 
efficiencies $q_{\nu\bar\nu}\equiv \dot E_{\nu\bar\nu}/L_{\nu}$
of neutrino energy into energy of the pair-plasma fireball to be
$q_{\nu\bar\nu}^{\rm N} = 4.1\times 10^{-3}$ and 
$q_{\nu\bar\nu}^{\rm PW} = 4.6\times 10^{-3}$ (Table~\ref{tab:efficiencies}).

Figure~\ref{fig:annconeNP} provides information about the spatial
distribution of the cumulative energy deposition rates in both models. 
The plots show $z$-$d$-integrals of the rates per unit volume,
$\dot Q_{\nu\bar\nu}({\bf r})$, which are computed as
\begin{equation}
\dot E_{\nu\bar\nu}(d,z)\,\equiv\,\int_z^\infty{\rm d}\chi\int_0^d
{\rm d}\xi\,\xi\int_0^{2\pi}{\rm d}\varphi\,\dot Q_{\nu\bar\nu}({\bf r})
\ .
\label{eq:zdenergy}
\end{equation}
These plots are thus constructed in analogy to the panels of the 
$z$-$d$ integrals for the mass in Fig.~\ref{fig:coneNP}.
Note that the values correspond to the energy that is deposited only on
one side of the equatorial plane. A large part of the
energy, nearly $10^{50}\,{\rm erg\,s}^{-1}$, 
is dumped within 30~km around the polar axis and at heights larger 
than about $30\,$km above (and below) the equatorial plane in the 
Newtonian model and at heights larger than about $10\,$km in the 
Paczy\'nski-Wiita case. According to Fig.~\ref{fig:coneNP} there are
less than $10^{-4}\,M_{\odot}$ of gas in these regions in both models.
A comparison of the two panels in
Fig.~\ref{fig:annconeNP} reveals that in the Paczy\'nski-Wiita simulation
the energy is more concentrated towards the equatorial plane so that at
heights $|z|\ga 30\,$km the numbers are about one order of magnitude lower
than in the Newtonian model. This difference is explained mainly by the
smaller size of the neutrinospheres in the less massive 
Paczy\'nski-Wiita torus. A minor part of the effect is also due to the lower
neutrino luminosities which reach only little more than half the values of
the Newtonian model.

\section{Analytical estimates\label{sec:analytical}}

\subsection{Viscosity and evolution}

The evolution of the Newtonian torus on longer time scales will be 
governed by the outward-directed viscous transport of angular momentum.
Since the Euler equations which are solved numerically with the PPM
method do not contain viscosity terms, and since this scheme
does not require any artificial viscosity to treat shock waves, 
the most important viscous dissipation in the absence of shocks 
comes from the numerical viscosity of the code. The latter is 
associated with the discretization of the equations and thus depends
on the chosen grid resolution, but is also determined by the input
physics implemented in the hydrodynamics code.
In case of the quasi-stationary Newtonian Model~{\bf B}10 we
shall estimate the size of this numerical viscosity from the torus
properties.

With the average value of the radial inflow velocity, 
$\ave{v_r}\approx -10^8\,{\rm cm\,s}^{-1}$
(see Fig.~\ref{fig:vrad} and Sect.~\ref{sec:torus}), and a mean radius
$\ave{d}$ of about 50$\,$km we estimate an accretion time scale
of $t_{\rm acc} = \ave{d}/\ave{v_r}\approx 50\,$ms in very
good agreement with the independent determination via the mass accretion
rate of the black hole, $t_{\rm acc} = M_{\rm v}/\dot M_{\rm v}\approx 
53\,$ms (see Fig.~\ref{fig:masstBH} and Table~\ref{tab:efficiencies}).
The viscous time scale is given by 
$t_{\rm vis}\approx M_{\rm v}/(6\pi\eta R_{\rm s})$ --- see Eq.~(13) in
Ruffert et al.~(\cite{ruf97}) --- with the dynamic viscosity $\eta \sim \alpha
(\rho H v_{\rm Kepler})$ where $H\sim R_{\rm s}$ is the height of the 
torus, $\rho$ the average density in the torus,
$v_{\rm Kepler} = \Omega_{\rm Kepler}(d)d = \sqrt{GM_{\rm BH}/d}$ 
the Keplerian velocity at a representative radius $d\sim 4R_{\rm s}$,
and $\alpha$ the dimensionless $\alpha$-viscosity parameter. 
Using $\rho = M_{\rm v}/V_{\rm t}$ with the torus volume
$V_{\rm t}\sim 2\pi^2R_{\rm s}^2d$, one finds 
$\alpha\sim \eck{\Omega_{\rm Kepler}(d)t_{\rm vis}}^{-1}$. 
Setting $t_{\rm vis} = t_{\rm acc} = \ave{d}/\ave{v_r}$ we get
a value for the $\alpha$-parameter associated with the numerical viscosity
of $\alpha\sim \ave{v_r}/v_{\rm Kepler}(d)\sim 0.01$. 
It is interesting to note that the torus shapes computed by
Popham \& Gammie (\cite{pop98a}) for such values of the $\alpha$-viscosity
look very similar to the cross sections through the tori of our
simulations plotted in Figs.~\ref{fig:vertN}, \ref{fig:vertP}
and \ref{fig:annihi}.

The numerical viscosity in our simulations therefore corresponds
to a value $\alpha\sim 0.01$, which is a bit higher than the ``optimum'' 
value $\alpha^\ast = \eta^\ast/(\rho H v_{\rm Kepler})$ where the viscous
energy dissipation and the energy emission by neutrinos are balanced,
a requirement which ensures maximum efficiency for the conversion
of rest-mass energy into neutrinos. Ruffert et al.~(\cite{ruf97}) estimated
$\alpha^\ast \sim (1.7\,...\,6.6)\times 10^{-3}
(R_{\rm s}/9\,{\rm km})^{7/2}(M_{\rm v}/0.1\,M_{\odot})^{-3/2}$ 
on grounds of a very simple one-zone model of the torus.
Consistent with the larger value of the viscosity, the mass accretion 
rate of the black hole was found to be 
$\dot M_{\rm v}\approx 5\,M_{\odot}\,{\rm s}^{-1}$ 
in the numerical simulation (Fig.~\ref{fig:massaccBH}
and Table~\ref{tab:efficiencies}), which is somewhat higher --- and the
corresponding lifetime $t_{\rm acc}$ of the torus somewhat shorter ---
than estimated by Ruffert et al.~(\cite{ruf97}) for the optimum value 
$\alpha^\ast$. This implies that the torus does not lose
energy in neutrinos at the maximum theoretical efficiency of about 
8.3\% for disk accretion on a nonrotating black hole in Newtonian gravity.
I.e., the torus in the numerical simulation is advection 
dominated, in agreement with the findings in Sect.~\ref{sec:neutrino}
(see also Table~\ref{tab:efficiencies}).
Because of the high densities and temperatures in the torus,
the matter is not transparent to neutrinos and the neutrino diffusion
time is longer than the accretion time of the gas.
Therefore a sizable fraction of the gravitational binding energy 
that is dissipated into heat is carried into the black hole before
neutrinos are able to transport it away. Although the neutrino luminosities
obtained in the numerical simulations are close to those 
estimated by Ruffert et al.~(\cite{ruf97}), the shorter lifetime of the torus
leads to a smaller time integral of the radiated energy.

Popham et al.~(\cite{pop98b}) have calculated self-consistent stationary models
for neutrino emitting accretion disks of black holes for a wide
range of parameters. With the large accretion rates and high densities,
the tori obtained in our simulations are in an extreme corner of their 
parameter space where the idealized treatment of the neutrino 
cooling by Popham et al.~(\cite{pop98b}), who essentially assumed
neutrino-transparent conditions, reaches its limits of
applicability.

Finally, we point out that neutrino emission is rather inefficient in 
transporting away angular momentum and therefore this is not the 
driving force of the torus evolution in our investigated cases.
Setting the mass accretion rate by the black hole 
in relation to the angular momentum loss rate, 
$\dot J \sim \dot M_{\rm v}\Omega_{\rm Kepler}d^2$,
and requiring $\dot J$ to be equal to the angular momentum carried
off by neutrinos, 
$\dot J = \dot J_{\nu} \sim (L_{\nu}/c^2)\Omega_{\rm Kepler}d^2$,
one finds $\dot M_{\rm v}\sim L_{\nu}/c^2$. For a neutrino luminosity
of $L_{\nu} = 10^{53}\,{\rm erg}\,{\rm s}^{-1}$ this gives a mass
accretion rate of the order of $\dot M_{\rm v}\sim {1\over 20}
M_{\odot}\,{\rm s}^{-1}$ which is two orders of magnitude below our
numerically determined numbers.

\begin{figure*}
\begin{tabular}{cc}
  \epsfxsize=8.5cm  \epsffile{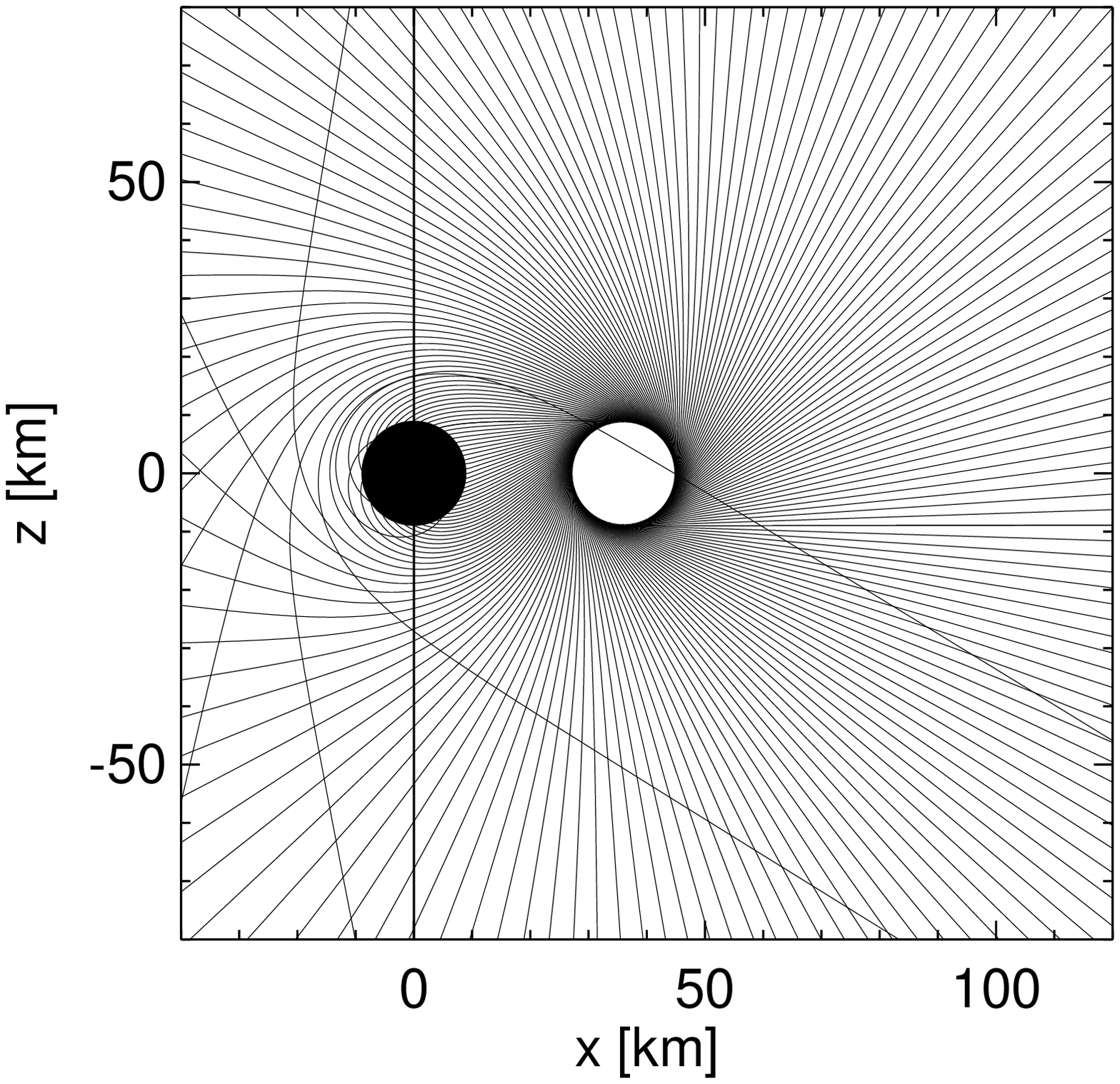} &
  \epsfxsize=8.5cm  \epsffile{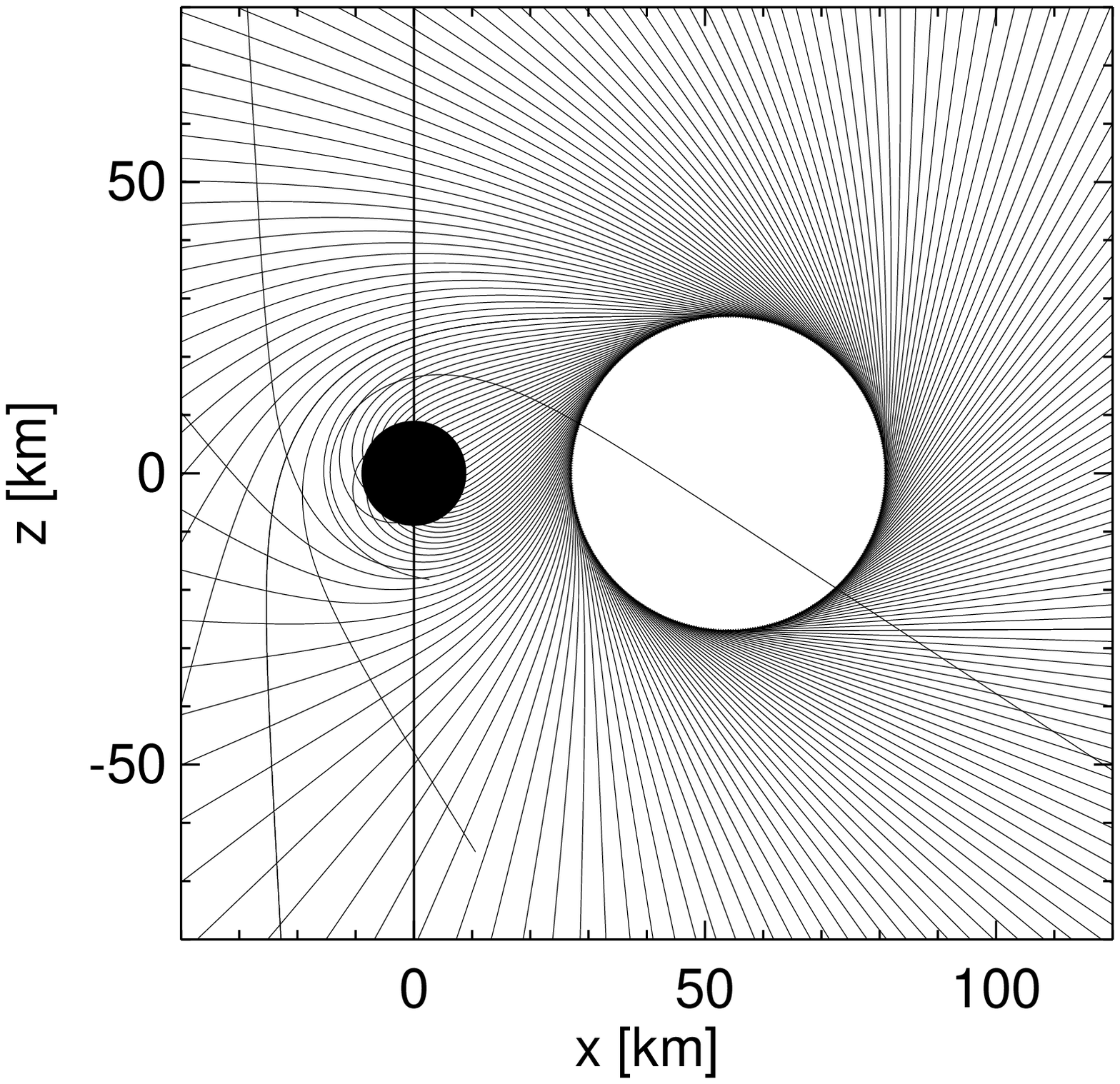} \\
\end{tabular}
  \caption[]{Geodesics of (massless) neutrinos emitted tangentially 
  from a toroidal accretion disk (white circle) with assumed 
  semidiameter $R_{\rm t}$ of one Schwarzschild radius, $R_{\rm s}= 9\,$km,
  (left) and of three Schwarzschild radii (right)
  in the vicinity of a black hole (black circle) with a mass of 
  $3 M_{\odot}$. The torus center is at $x = 3R_{\rm s}+R_{\rm t}$,
  the black circle has a radius of $1R_{\rm s}$.}
  \label{fig:geodesics}
\end{figure*}

\begin{table}
\caption[]{
In the second and third column
ratios of general relativistic (``GR'') to Newtonian (``N'') results 
are given for the integral energy deposited by $\nu\bar\nu$ annihilation for
the different values of $\alpha\equiv R_{\rm t}/R_{\rm s}$ listed in the 
first column. The other columns contain values of the factor
${\cal U}/(\alpha\beta{\cal V})$ in Eq.~(\ref{eq:esctoann}). This factor
decides about the importance of neutrino-electron and 
neutrino-positron scattering
relative to $\nu\bar\nu$ annihilation. It is assumed that the neutrinos
are emitted from a toroidal neutrinosurface with radius $R_{\rm t}$ and 
center at $3R_{\rm s}+R_{\rm t}$. The energy deposition rates by 
annihilation or scattering are integrated either over the volume of a
cylinder with radius $3R_{\rm s}$ around the system axis, or over the
volume of a cone enclosed by the rotated boundary curve
of Eq.~(\ref{eq:balance}), respectively.
}
\begin{flushleft}
\tabcolsep=1.5mm
\begin{tabular}{ccccccc}
\hline\\[-3mm]
                       & $\nu\bar\nu$ & $\nu\bar\nu$ 
                       & $e\nu$ & $e\nu$ & $e\nu$ & $e\nu$ \\
$R_{\rm t}/R_{\rm s}$  &     GR/N     &    GR/N     
                       & N & GR & N & GR                   \\
                       & cylinder & cone
                       & cylinder & cylinder & cone & cone \\[0.3ex]
\hline\\[-3mm]
0.5 & 0.67 & 0.84 & 0.124 & 0.083 & 0.115 & 0.074 \\
1.0 & 0.68 & 0.85 & 0.097 & 0.069 & 0.089 & 0.061 \\
2.0 & 0.72 & 0.89 & 0.064 & 0.049 & 0.057 & 0.043 \\
3.0 & 0.75 & 0.92 & 0.045 & 0.037 & 0.040 & 0.032 \\
4.0 & 0.77 & 0.94 & 0.033 & 0.028 & 0.029 & 0.024 \\[0.7ex]
\hline
\end{tabular}
\end{flushleft}
\label{tab:GRcorrections}
\end{table}

\subsection{General relativistic effects}
\label{sec:GR}

The torus simulations described in this paper were performed with
basically Newtonian physics except that the gravitational
potential was replaced by a Paczy\'nski-Wiita potential in some 
cases. Also the emission of gravitational waves and the corresponding
back-reaction on the hydrodynamic flow were included, which, however,
is not important for the nearly axially symmetric black hole-torus
configurations. 

The influence of general relativistic effects on the neutrino emission
and on the neutrino-antineutrino annihilation are not obvious because 
of several
competing effects which partly cancel each other (see, e.g.,
Shapiro \& Teukolsky \cite{sha83}, Luminet \cite{lum79}, 
Marck \cite{mar96}, Cardall \& Fuller \cite{car97},
Usui et al.~\cite{usu98}). The effects considered here are~: 
\par\noindent
(1) General relativistic blueshifting of those neutrinos which are 
emitted from the accretion torus in the direction of the black hole 
before they annihilate, and redshifting of those neutrinos which
annihilate at distances from the black hole larger than the radius
of their production.
\par\noindent
(2) Gravitational redshifting of the energy as the electron-positron 
pair plasma expands away from the site of its creation by 
$\nu\bar\nu$ annihilation.
\par\noindent
(3) General relativistic ray bending of the neutrino trajectories
which changes the average angles at which neutrinos and antineutrinos
collide and thus also changes the rate of $\nu\bar\nu$ annihilation.
\par\noindent
(4) Gravitational trapping of the energy which is deposited in the close
vicinity of the event horizon by $\nu\bar\nu$ annihilation and 
therefore is not able to escape from the strong gravitational attraction
of the black hole but is ultimately sucked in by the hole.

It is not easy to implement these effects consistently into the 
hydrodynamic simulations and into the post-proc\-essing procedure which
we used to evaluate our models for neutrino-antineutrino 
annihilation. In order to estimate the relative changes due to general
relativity approximately, we therefore consider
a simplified treatment for an idealized model geometry which mimics
sufficiently closely the situation described by the hydrodynamic results.
For this purpose, we assumed that the neutrinos are emitted from a 
neutrinosurface that has the shape of a torus 
centered on the equatorial plane. At least in case of the 
Newtonian accretion disk, which is optically thick for neutrinos,
Fig.~\ref{fig:annihi} shows that this picture is indeed appropriate.
In the following analysis we shall explicitly presume that the 
black hole-torus configuration is axially symmetric. We
neglect effects due to the relativistic Doppler shift of the 
neutrinos emitted from the torus although the gas orbits around the
black hole at a fair fraction of the speed of light.

Integrating the energy deposition rate by $\nu\bar\nu$ annihilation
over a volume $V$ outside of the torus, one can write
\begin{equation}
\dot E_{\nu\bar\nu}\,=\,\int_V{\rm d}^3{\bf r}\,\dot 
Q_{\nu\bar\nu}({\bf r})\,\equiv\,R_{\rm s}^{-1}\cdot
{\cal R}\,{\cal L}\,{\cal V}\,,
\label{eq:dvq}
\end{equation}
where ${\cal R} = 9R_{\rm s}^4R_{\rm t}^{-2}(3R_{\rm s}+R_{\rm t})^{-2}$
is a dimensionless factor that depends on the torus
radius $R_{\rm t}$, the horizon radius of the black hole,
$R_{\rm s}$, and the radius of the torus center, $3R_{\rm s}+R_{\rm t}$.
The factor ${\cal L}$ contains the weak interaction coefficients
$c_{\nu_i\bar\nu_i}$ ($i = e,\mu,\tau$)
and is proportional to the product of the neutrino and antineutrino
luminosities multiplied by the sum of the average neutrino and 
antineutrino energies which determine the $\nu\bar\nu$ annihilation
rate, i.e., ${\cal L}\propto\sum_i c_{\nu_i\bar\nu_i}L_{\nu_i}L_{\bar\nu_i}
(\ave{\epsilon_{\nu_i}}+\ave{\epsilon_{\bar\nu_i}})$.
The term ${\cal V}$ is also a dimensionless factor that results from the
volume integral and carries information about the angular distributions
of the neutrinos and antineutrinos at the positions ${\bf r}$.

Two different axially symmetric volumes are considered. On the
one hand, the integration is performed over a cylinder with radius
$3R_{\rm s}$ around the symmetry axis of the torus-black hole system.
On the other hand, the volume of a cone is considered whose surface is 
described by a rotation of the curve
\begin{equation}
{z\over 3R_{\rm s}}\,=\,\pm\,{x\over 3R_{\rm s}}\,
\sqrt{\rund{{x\over 3R_{\rm s}}}^{\! 2/3}-1}\quad\ {\rm for}\ \,
{x\over 3R_{\rm s}}\,\ge\,1 
\label{eq:balance}
\end{equation}
around the symmetry axis. In all points on this surface the component of
the gravitational force parallel to the equatorial plane is balanced by
the centrifugal force of matter which has angular momentum equal to the 
Keplerian value at radius $3R_{\rm s}$,
$j = j_{\rm Kepler}(3R_{\rm s})= \sqrt{6}GM_{\rm BH}/c$. 
This surface may approximate the boundary of a baryon-free
funnel along the axis. 

If one assumes that the value of the integrand
on the axis at vertical height $z$ is representative of all values in
the plane defined by this $z$, one can make use of the axial symmetry 
which allows one to calculate the $\nu\bar\nu$ annihilation rate 
along the axis particularly easily. The integrations parallel and vertical 
to the system axis can then be separated, and only a one-dimensional
integral for ${\cal V}$ is left for numerical evaluation~:
\begin{equation}
{\cal V}\,=\,\int_{\xi_{\rm min}}^\infty{\rm d}\xi\,W(\xi)\,.
\label{eq:wint}
\end{equation}

First we consider the cases without general relativistic effects.
For the cylindrical volume one sets $\xi \equiv z/R_{\rm s}$ and
starts the integration at $z = R_{\rm s}$, i.e. $\xi_{\rm min} = 1$.
The integrand is found to be
\begin{eqnarray}
W(\xi)&\equiv&W_{\rm cyl}^{\rm N}(\xi)\,=\, \nonumber \\
&=&(\mu_1-\mu_0)^2  
\times\biggl\lbrack{3\over 2}+{1\over 6}\rund{
\mu_1^2+\mu_1\mu_0+\mu_0^2}^2  \nonumber \\
&-&{1\over 2}\rund{\mu_1+\mu_0}^2-{1\over 3}\rund{\mu_1^2
+\mu_1\mu_0+\mu_0^2}\biggr\rbrack
\label{eq:wcylN}
\end{eqnarray}
with
\begin{equation}
\mu_{0,1}\,=\,\mu_{0,1}(\xi)
\,=\,{\xi\sqrt{\beta^2+\xi^2-\alpha^2}\mp \alpha\beta \over 
\beta^2+\xi^2}
\label{eq:mudef}
\end{equation}
where $\alpha \equiv R_{\rm t}/R_{\rm s}$ and 
$\beta\equiv 3+\alpha$. The minus sign
in Eq.~(\ref{eq:mudef}) corresponds to $\mu_0 = \cos\theta_0$, 
the plus sign to $\mu_1 = \cos\theta_1$, if $\theta_0 > \theta_1$ and
$\theta_0$ and $\theta_1$ are the angles between the symmetry axis and
the direction of rays that leave the torus surface tangentially and
go to an observer on the axis. Taking the variable of integration to be
$\xi \equiv x/(3R_{\rm s})$, one can derive in case of the conical volume~:
\begin{eqnarray}
W(\xi)&\equiv&W_{\rm con}^{\rm N}(\xi)\,=\, \nonumber \\
&=&\xi^2\,{4\xi^{2/3}-3\over\sqrt{x^{2/3}-1}}\,
W_{\rm cyl}^{\rm N}(3\xi\sqrt{\xi^{2/3}-1}) \, ,
\label{eq:wconN}
\end{eqnarray}
where the expression in backets is the
argument with which $W_{\rm cyl}^{\rm N}$ of Eq.~(\ref{eq:wcylN})
has to be evaluated. The lower integral boundary is now given
by the value $\xi_{\rm min}$ for which the relation 
$\xi\sqrt{\xi^{2/3}-1} = 1/3$ is fulfilled.

The relativistic effects listed above can be taken into 
account in the following approximate way. General relativistic 
ray bending (point~(3) in the list at the beginning of this section) 
requires to replace the Euclidian values of
$\mu_0$ and $\mu_1$ of Eq.~(\ref{eq:mudef}) by their relativistic
counterparts. These are found by tracing geodesics from an
observer position on the axis to the points where they hit the
torus surface tangentially. Figure~\ref{fig:geodesics} shows two
examples where we calculated the ray trajectories in the gravitational
field of a black hole with approximately $3M_{\odot}$ for a torus 
with radius $R_{\rm t} = 1R_{\rm s}$ and a torus with radius
$R_{\rm t} = 3R_{\rm s}$, respectively. In our analysis we allow
for only one torus image and thus neglect that neutrino trajectories 
may be bent around the black hole to cross the system axis more
than once. Blueshift, redshift and
radiation trapping (points~(1), (2) and (4), respectively) are
accounted for by assuming that they can be split off in a 
separate factor $\ave{S^{\rm GR}C^{\rm GR}}$ in the integrand 
of Eq.~(\ref{eq:wint}). Thus for the cylindrical integration volume 
we take
\begin{equation}
W(\xi)\,\equiv\,W_{\rm cyl}^{\rm GR}(\xi)\,=\,
W_{\rm cyl}^{\rm N}(\xi)\,\ave{S^{\rm GR}\cdot C^{\rm GR}}(\xi)\, ,
\label{eq:wcylGR}
\end{equation}
and for the integration over the cone we use
\begin{eqnarray}
W(\xi)&\equiv&W_{\rm con}^{\rm GR}(\xi)\,=\, \nonumber \\
&=&W_{\rm con}^{\rm N}(\xi)\,
\ave{S^{\rm GR}\cdot C^{\rm GR}}(3\xi\sqrt{\xi^{2/3}-1})\, .
\label{eq:wconGR}
\end{eqnarray}
The factor $\ave{S^{\rm GR}C^{\rm GR}}$ in Eq.~(\ref{eq:wcylGR}) 
depends on $\xi$ and in  Eq.~(\ref{eq:wconGR}) on 
$3\xi\sqrt{\xi^{2/3}-1}$, 
and the brackets $\ave{...}$ indicate that the product 
$S^{\rm GR}C^{\rm GR}$ has been averaged over the plane at constant
$z$. For the latter averaging procedure one has to notice that 
$S^{\rm GR}C^{\rm GR}$, as specified below, has a weakly diverging 
singularity at the event horizon, but its integral has a finite
value and is regular. 

The red- and blueshift factor $S^{\rm GR}$ can be approximated by
\begin{eqnarray}
S^{\rm GR}(r)&=& {\rm e}^{2\phi(r_{\rm o})}\,\rund{{
{\rm e}^{\phi(r_{\rm o})}\over {\rm e}^{\phi(r)}}}^{\! 3}
\nonumber \\
&\approx&
\rund{1-{R_{\rm s}\over 3R_{\rm s}+R_{\rm t}}}^{\! 5/2}
\rund{1-{R_{\rm s}\over r}}^{\! -3/2}
\label{eq:redshift}
\end{eqnarray}
where $r$ denotes the radius where neutrinos and antineutrinos
annihilate, the lapse function is given by 
${\rm e}^{\phi(r)} = \sqrt{1-R_{\rm s}/r}$,
and the second expression is obtained by taking 
$r_{\rm o}\approx 3R_{\rm s}+R_{\rm t}$ as the mean radius
from where neutrinos and antineutrinos originate. The first factor
${\rm e}^{2\phi(r_{\rm o})}$ in Eq.~(\ref{eq:redshift}) 
accounts for the redshift and time dilation from $r_{\rm o}$ to
infinity, and the second factor $({\rm e}^{\phi(r_{\rm o})}/
{\rm e}^{\phi(r)})^3$ for the blueshift or redshift
between $r_{\rm o}$ and $r$. 

Close to the event horizon only photons
moving nearly radially outward can escape from the gravitational
attraction of the black hole whereas photons propagating with large
angles relative to the outward direction are captured by the hole.
We assume that the electron-positron-photon plasma is locally
isotropic and neglect hydrodynamic effects. This allows us to 
estimate the escape probability of energy produced by $\nu\bar\nu$
annihilation at radius $r$ from the fraction of photons on escape
trajectories according to (Shapiro \& Teukolsky \cite{sha83})~:
\begin{equation}
C^{\rm GR}(r)\,\approx\,{1\over 2}\,
\eck{1\pm \sqrt{1-{27\over 4}\rund{{R_{\rm s}\over r}}^{\! 2}
\rund{1-{R_{\rm s}\over r}}}\,\,}\, ,
\label{eq:escape}
\end{equation}
with the minus sign holding for $R_{\rm s}\le r\le {3\over 2}R_{\rm s}$
and the plus sign for $r > {3\over 2}R_{\rm s}$. 
At $r = {3\over 2}R_{\rm s}$ a fraction of 50\% of the pair plasma
is able to escape to infinity. The assumption of local isotropy
probably implies an underestimation of the escape probability because
the non-zero radial momentum of $e^\pm$ pairs produced by $\nu\bar\nu$
annihilation is not taken into account. Ray bending towards the black
hole and positive pressure gradients, on the other hand, act in the 
opposite direction and reduce the expansion of the pair plasma
out of the close vicinity of the event horizon. Therefore we consider
Eq.~(\ref{eq:escape}) as a reasonable zeroth-order estimate.

The expressions of Eqs.~(\ref{eq:redshift}) and 
(\ref{eq:escape}) are now inserted into Eqs.~(\ref{eq:wcylGR}) and
(\ref{eq:wconGR}) to evaluate the integral of Eq.~(\ref{eq:wint}).
Equation~(\ref{eq:dvq}) then yields the energy deposition rate by 
$\nu\bar\nu$ annihilation as measured by an observer at rest at infinity,
when the neutrino luminosities and the mean neutrino energies are taken
as measured at the neutrino source. The annihilation luminosities 
including general
relativistic corrections are reduced relative to the Newtonian results.
The calculated reduction factors are listed
in Table~\ref{tab:GRcorrections} for different values of the parameter
$R_{\rm t}/R_{\rm s}$. Relativistic effects decrease the available
energy in all investigated cases by typically 10--30\%, 
mainly because of redshift to infinity and radiation
capture by the black hole. However, these two effects are nearly 
compensated by the increase of the $\nu\bar\nu$ annihilation probability
due to the facts that the neutrinos emitted towards the black hole are 
blueshifted and the $\nu\bar\nu$ annihilation rate rises with the square
of the neutrino luminosity times the average neutrino energy.
It is interesting to note that general relativistic ray bending
reduces the angle under which the torus surface is seen from a 
position on the system axis near the black hole, whereas the viewing
angle of the torus is increased relative to the Euclidean angle at large
distances. For this reason, ray bending implies a reduction of the
$\nu\bar\nu$ energy deposition rate in the case of the cylindrical 
integration volume, but has the opposite effect for the conical
volume where a larger fraction of the total energy comes from
$\nu\bar\nu$ annihilation far away from the black hole.

\subsection{Additional heating by neutrino-electron scattering}

The pair-plasma cloud which is created around the accretion torus by
$\nu\bar\nu$ annihilation may receive additional energy input by 
neutrino scattering on the abundant electrons and positrons 
(Woosley \cite{woo93a}).
In this subsection we estimate the importance of this
heating process relative to $\nu\bar\nu$ annihilation with and
without relativistic corrections. We adopt again the simplified 
torus-black hole geometry described in Sect.~\ref{sec:GR}.

In the pair-plasma cloud electrons and positrons are present in
equal numbers and thus the electron chemical potential is zero.
For such conditions the summed energy transfer rate by scattering of
neutrinos and antineutrinos of all flavors becomes (Tubbs and 
Schramm \cite{tub75})
\begin{equation}
\dot Q_{e\nu}\,=\,{\sigma_0{\varepsilon_e}\over 6(m_ec^2)^2}\,
{\mu_1-\mu_0\over S_{\rm t}}\,
\sum_{\nu_j} c_{e\nu_j} L_{\nu_j}\rund{\ave{\epsilon_{\nu_j}}-
\ave{\epsilon_e}}
\label{eq:ratescatt}
\end{equation}
with $\nu_j = \nu_e,\,\bar\nu_e,\,\nu_{\mu},\,\bar\nu_{\mu},\,
\nu_{\tau},\,\bar\nu_{\tau}$. Here $S_{\rm t}$ is the surface area 
of the torus, $\sigma_0 = 1.76\times 10^{-44}\,$cm$^2$, and 
$c_{e\nu_j}$ are weak interaction coefficients. In order to estimate
the electron energy density $\varepsilon_e$ and the average electron
energy $\ave{\epsilon_e}\approx 4T$, we assume that the 
$e^\pm \gamma$ plasma stays near the torus for an expansion time
$t_{\rm exp}$. Then the plasma temperature can be estimated by 
equating the energy density of the plasma with the heating rate 
times the expansion time. One gets
\begin{equation}
T\,\sim\,\eck{{15(hc)^3\over 22\pi^5}\,\,
\dot Q_{\nu\bar\nu}t_{\rm exp}}^{1/4}\ \ {\rm and}\ \ 
\varepsilon_e\,\sim\,{7\over 22}\,\dot Q_{\nu\bar\nu}t_{\rm exp}\,.
\label{eq:qtemp}
\end{equation}
Using this in Eq.~(\ref{eq:ratescatt}) and integrating over the 
volume outside of the torus with the same assumptions as made in 
Sect.~\ref{sec:GR}, we find
\begin{equation}
{\dot E_{e\nu}\over \dot E_{\nu\bar\nu}}\,\la\, 
0.075\,{L_{\nu_e,52}\ave{\epsilon_{\nu_e}}_{10}\over R_{\rm s,9}^2}\,
\rund{{t_{\rm exp}\over 1\,{\rm ms}}}\,{{\cal U}\over 
\alpha\beta{\cal V}}\, ,
\label{eq:esctoann}
\end{equation}
where the numerical value was obtained by taking representative
numbers for the neutrino luminosities and mean neutrino energies
from Figs.~\ref{fig:neutradN}--\ref{fig:meneP}. The normalized
quantities are
$L_{\nu_e,52} = L_{\nu_e}/10^{52}\,{\rm erg\,s}^{-1}$,
$\ave{\epsilon_{\nu_e}}_{10} = \ave{\epsilon_{\nu_e}}/10\,{\rm MeV}$
and $R_{\rm s,9} = R_{\rm s}/9\,$km. For the expansion timescale we
assumed 1$\,$ms which is probably an upper limit because the light
crossing time through the main region of neutrino energy deposition 
is only a few $10^{-4}\,$s. The quantity ${\cal V}$ in the denominator
is given by Eq.~(\ref{eq:wint}), and ${\cal U}$ is defined by
\begin{equation}
{\cal U}\,\equiv\,\int_{\xi_{\rm min}}^\infty{\rm d}\xi\,
(\mu_1-\mu_0)W(\xi)\ave{{\widetilde S}^{\rm GR}}(\xi)\,.
\label{eq:udef}
\end{equation}
The factor $\ave{{\widetilde S}^{\rm GR}}$ is a function of $\xi$
and different from unity only when general relativistic corrections
are taken into account. It contains additional terms for the redshift 
between the radius of neutrino emission ($r_{\rm o}$), the radius of
neutrino-electron scattering ($r$), 
and the observer at infinity. The brackets $\ave{...}$ indicate again 
that this factor is averaged on the planes of
constant vertical height $z$. For the function 
${\widetilde S}^{\rm GR}(r)$ we estimate
\begin{eqnarray}
{\widetilde S}^{\rm GR}(r)&=& {\rm e}^{\phi(r_{\rm o})}\,\rund{{
{\rm e}^{\phi(r_{\rm o})}\over {\rm e}^{\phi(r)}}}^{\! 2}
\nonumber \\
&\approx&
\rund{1-{R_{\rm s}\over 3R_{\rm s}+R_{\rm t}}}^{\! 3/2}
\rund{1-{R_{\rm s}\over r}}^{\! -1} \,.
\label{eq:addredshift}
\end{eqnarray}
In the general relativistic case, the time $t_{\rm exp}$ is assumed to
be measured by the observer at rest at infinity. 

In Table~\ref{tab:GRcorrections} numbers for the factor
${\cal U}/(\alpha\beta{\cal V})$ of Eq.~(\ref{eq:esctoann}) 
are listed for both the
cylindrical and conical integration volumes and with and 
without relativistic corrections. Since this factor is always
much less than unity, the contribution to the pair-plasma heating
by neutrino-electron and -positron scattering is unimportant 
relative to $\nu\bar\nu$ annihilation. Certainly it contributes 
not more than about 10\% of the $e^\pm\gamma$ energy for
all parameter combinations suggested by our models. Again,
general relativistic effects lead only to a moderate reduction of 
the results.

\begin{figure*}
  \epsfxsize=18cm \epsffile{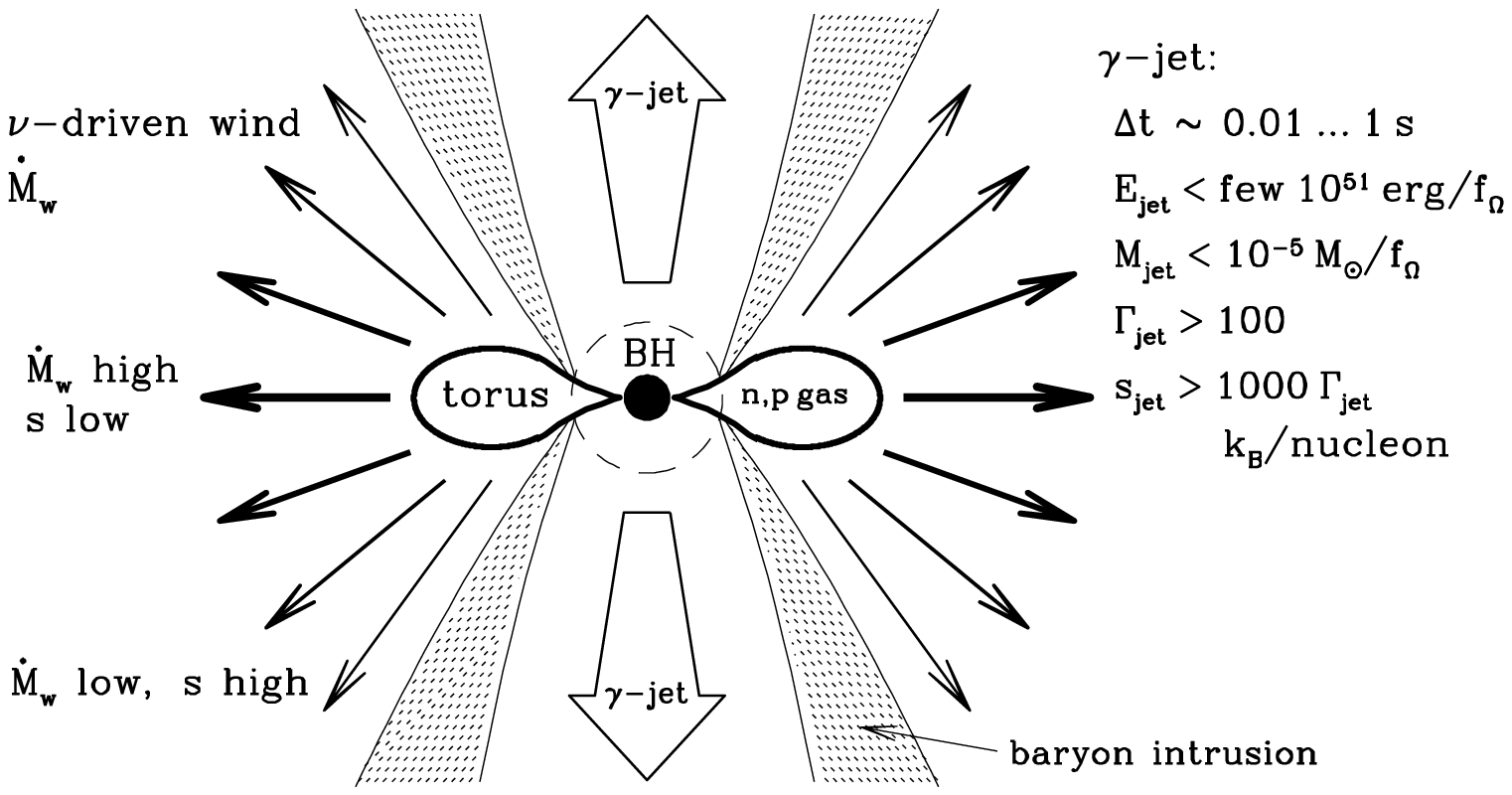} 
  \caption[]{Envisioned gamma-jets and neutrino-driven baryonic wind
  from the massive accretion torus around the BH. The dashed circle
  indicates a region around the black hole with the radius of the 
  last stable circular orbit. The hatched areas
  represent schematically the volume where baryons might be able to 
  penetrate into 
  the jet region; its outer boundary is approximately described by 
  Eq.~(\ref{eq:balance}), its inner boundary may roughly be 
  given by the line where the gravitational force is balanced by
  the component of the centrifugal force in the opposite direction.
  $f_{\Omega}=2\delta\Omega/(4\pi)$ denotes the fraction of the sky 
  into which both jets are beamed.} 
  \label{fig:massloss}
\end{figure*}

\begin{figure*}
  \epsfxsize=18cm \epsffile{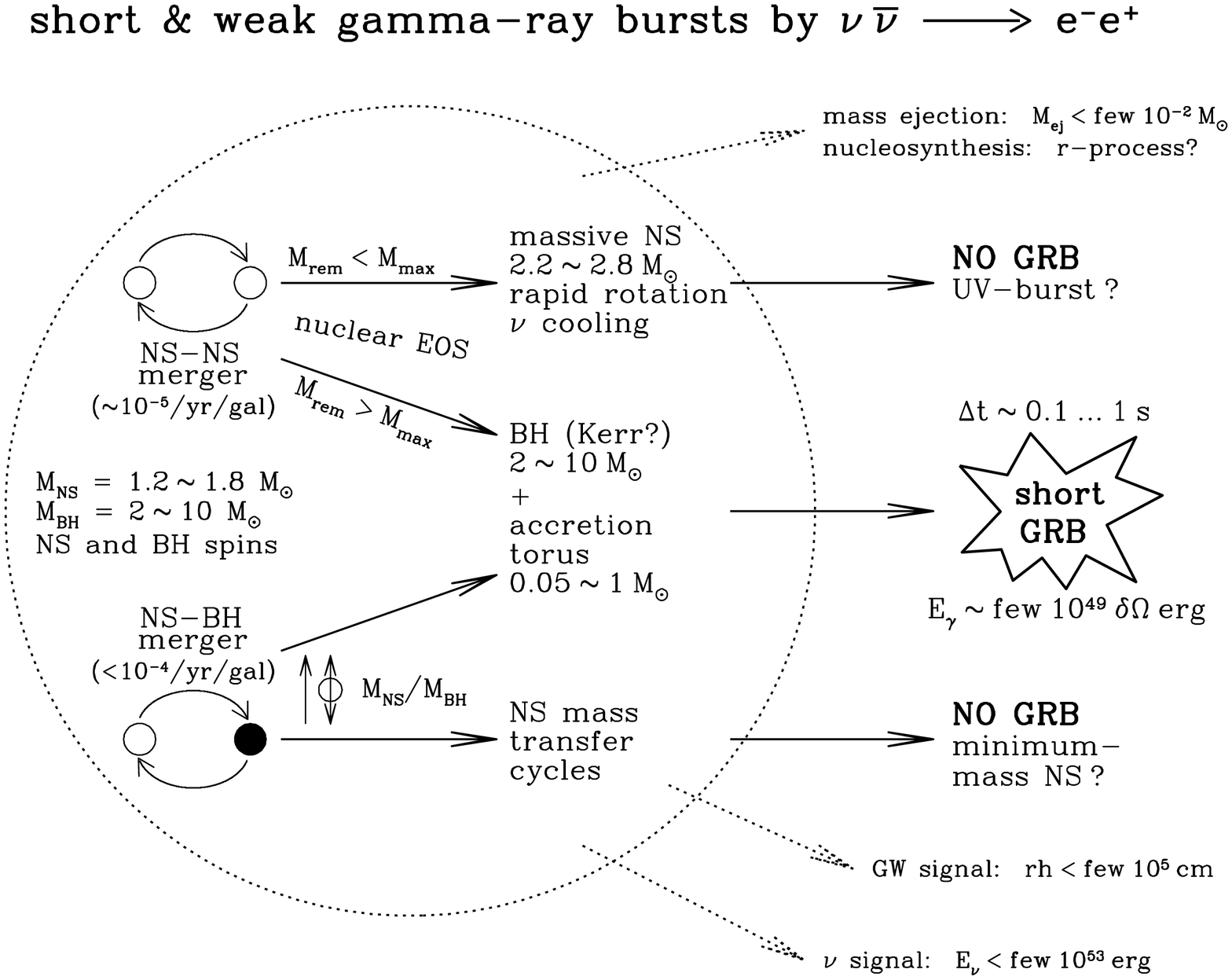} 
  \caption[]{Possible evolution paths from NS-NS and NS-BH mergers to 
  short gamma-ray bursts powered by neutrino-antineutrino annihilation
  into electron-positron pairs. The evolution of the merging binaries 
  depends on the parameters of its stellar components (masses, spins)
  and on the properties of the nuclear equation of state.}
  \label{fig:shortGRB}
\end{figure*}

\section{Summary and discussion\label{sec:summary}}

\subsection{Summary}

With three-dimensional hydrodynamic simulations we continued
our evolution calculations of merging neutron stars
(Ruffert et al.~\cite{ruf96,ruf97}, and in particular
Ruffert and Janka \cite{ruf98b}) 
into the phase when the massive ($\sim 3M_{\odot}$ baryonic mass)
object at the center of the merger remnant has collapsed to a black 
hole, and the surrounding material with high angular momentum is forming
an accretion disk around this black hole. From the available set of merger 
models we chose a case where a massive disk could be expected, and varied
the time after the coalescence when the black hole formation was assumed
to happen. Moreover, we studied the subsequent accretion phase with
two different prescriptions for the gravitational potential of the 
black hole, either of Newtonian or of Paczy\'nski-Wiita type. The latter
potential is deeper and reproduces the existence of the
innermost stable circular orbit of a Schwarzschild black hole at
$3R_{\rm s}$. With these three-dimensional simulations we were not
able to follow the long-time evolution of the accretion disk and the
accretion of its matter until completion, a process which is governed
by the viscous transport of angular momentum. Nevertheless, we could 
study the approach to a nearly quasi-stationary
state where the radial velocities are much smaller than the orbital
velocities. 

We determined the disk properties in the qua\-si-\-sta\-tion\-ary state,
such as its mass, temperature, density, angular momentum,
neutrino emission, and rate of mass loss into the black hole
and thus its lifetime. The numerical results presented in this paper
are in reasonably good agreement with analytical estimates on grounds 
of a very simple torus model by Ruffert et al.~(\cite{ruf97}), and with 
recent calculations of hyper-accreting black holes by 
Popham et al.~(\cite{pop98b}). 

The disk properties and the computed disk evolution do not
depend very much on the exact moment when the black hole formation is
assumed to take place, unless this happens earlier than about 2$\,$ms,
or a very long time after the two neutron stars have merged.
We found, however, that the amount of gas which does not fall into
the black hole immediately on a dynamical timescale, but orbits
around the hole for at least several revolutions, depends strongly
on the employed gravitational potential of the black hole.
Part of this sensitivity is certainly due to the fact that our 
simulations of the neutron star coalescence had been carried out
with Newtonian gravity and therefore the use of the Paczy\'nski-Wiita
potential for the torus models is not fully consistent.

A reliable prediction of the mass which stays in the accretion torus
for several orbital periods is possible by the criterion that the specific
angular momentum of the matter must exceed the Keplerian value at
radius $3R_{\rm s}$. This means that the orbital velocities
in the disk are only slightly sub-Keplerian. 
Pressure effects are not too important to determine the
radial disk structure, but of course are crucial for the vertical 
disk structure. In case of the Paczy\'nski-Wiita potential a gas mass 
of only $3.5\times 10^{-2}\,M_{\odot}$ is able to complete more
than 5 orbits around the black hole. A quasi-stationary state was
not reached within the 5$\,$ms of simulated disk evolution,
and the high accretion rate of $7\,M_{\odot}\,{\rm s}^{-1}$ implies
an estimated accretion timescale of at least another $5\,{\rm ms}$
at the end of the simulation. On the other hand, the final state of
the disk with
Newtonian potential after 5$\,$ms shows very small radial
velocities. The accretion rate is $5\,M_{\odot}\,{\rm s}^{-1}$ and
the accretion of the disk
mass of $2.6\times 10^{-1}\,M_{\odot}$ can be estimated to continue
for more than 50$\,$ms (Table~\ref{tab:efficiencies}). 

The quasi-stationary evolution
of the torus is determined by numerical viscosity. For the chosen grid
resolution and the input physics implemented in our hydrodynamics code
we estimate a viscosity parameter of $\alpha\sim 0.01$.
Maximum temperatures
in the tori are around 10$\,$MeV, and maximum densities are between
several $10^{11}$ ${\rm g\,cm}^{-3}$ and
$10^{12}\,{\rm g\,cm}^{-3}$ with somewhat higher values in
the Newtonian case. Due to the larger mass, the 
Newtonian torus is optically thick to neutrinos whereas in the 
Paczy\'nski-Wiita calculation the disk is nearly transparent for
$\bar\nu_e$ and heavy-lepton neutrinos.
For this reason the total neutrino luminosities in both cases
are similar, around $10^{53}\,{\rm erg\,s}^{-1}$. The
$\bar\nu_e$ luminosity is about twice as high as the $\nu_e$
luminosity, and heavy-lepton neutrinos contribute only a minor
fraction to the total energy loss. The 
mean energies of the emitted neutrinos are around 10$\,$MeV for
electron neutrinos, 13--16$\,$MeV for electron antineutrinos
and 15--20$\,$MeV for the heavy-lepton neutrinos
(Table~\ref{tab:models}).

Because of the rapid radial infall of the torus gas in
the Paczy\'nski-Wiita simulation on the one hand, and the rather
large neutrino opacity of the Newtonian torus on the other, 
the efficiency $q_{\nu}$ for transforming rest-mass energy into 
neutrino emission is below the maximum theoretical limit of 8.3\%
for the radiation efficiency of a Newtonian accretion disk 
(5.7\% for relativistic disk accretion onto a nonrotating black 
hole). We found values for $q_{\nu}$ of 0.5\% for the 
Paczy\'nski-Wiita model and of 1.3\% for the Newtonian computation
(Table~\ref{tab:efficiencies}). Neutrino-antineutrino annihilation 
in the surroundings of the black hole deposits energy at a rate 
of (3--$5)\times 10^{50}\,{\rm erg\,s}^{-1}$ in the region where
the baryonic mass density is below $10^{11}\,{\rm g\,cm}^{-3}$.
This corresponds to an efficiency of 0.4--0.5\% for the
conversion of neutrino energy into $e^+e^-$ pairs. A fraction
of 10--30\% of this $\nu\bar\nu$ annihilation energy or an 
estimated time integral during the accretion phase of
$E_{\nu\bar\nu}\approx 10^{48}$--$10^{49}\,{\rm erg}$ are
dumped in a cone around the system axis
where the total baryon mass is below $10^{-5}\,M_{\odot}$.
A comparison of Figs.~\ref{fig:coneNP} and \ref{fig:annconeNP}
shows that the radius of this low-density cone at its
base is 15--20$\,$km.

General relativistic effects like gravitational
redshift and blueshift, radiation capture and ray bending reduce 
the observable energy in the expanding pair-plasma only
insignificantly, mainly because the different effects act in 
opposite directions and thus partly cancel each other. However,
because the black hole has an initial relativistic rotation 
parameter $a = Jc/(GM_{\rm BH}^2)\approx 0.4$ and is spun up to
$a\approx 0.5$ by the accretion of the torus material, 
effects associated with the Kerr character of the black hole
may lead to an increase of the lifetime of the torus and may
raise the energy dumped in $e^+e^-$ pairs above the values
obtained in this work. This was discussed by Jaroszy\'nski (\cite{jar96})
and Popham et al.~(\cite{pop98b}), although the results in these papers
cannot be directly applied to the very high mass accretion rates
of our models where the tori start to become opaque against neutrinos.

The high baryon density and integral baryon mass outside of an 
axis-near cone suggest that relativistic expansion of an $e^\pm\gamma$
plasma can only develop in jets along the axis. From the shape of the
low-density funnel above the poles of the black hole (approximately
bounded by the contour of density $10^9\,{\rm g\,cm}^{-3}$ in 
Figs.~\ref{fig:vertN} and \ref{fig:vertP} or by the contour 
corresponding to $10^{-5}\,M_{\odot}$ in the plots of the 
$z$-$d$-integrals of Fig.~\ref{fig:coneNP}) we expect that
the relativistic pair plasma will break out with rather wide opening 
half-angle $\theta$ between about 10 degrees and several 10 degrees,
which implies a moderate beaming factor for the two jets of
$f_{\Omega}=2\delta\Omega/(4\pi)=1-\cos\theta$ between 1/10 and 1/100.
This means that the determined $\nu\bar\nu$ annihilation energies may
be sufficient to explain gamma-ray bursts with observed energies of
$E_{\gamma}\approx 4\pi E_{\nu\bar\nu}/
(2\delta\Omega)\la 10^{50}$--$10^{51}\,{\rm erg}$, if the observer
assumes isotropy of the source.
For short $\gamma$ bursts with typical durations $t_{\gamma}$
between 0.1$\,$s and 1$\,$s, luminosities 
$L_{\gamma}\approx E_{\gamma}/t_{\gamma}$ between
$10^{50}\,{\rm erg\,s}^{-1}$ and $10^{52}\,{\rm erg\,s}^{-1}$
may be within the reach of our models. If Kerr effects play a role
even larger energies and luminosities may be possible.

\subsection{Discussion of implications and outlook}

Figure~\ref{fig:massloss} sketches the envisioned geometry of the
formation of the pair-plasma jets and their properties as suggested
by our models. Energy deposition by electron
neutrino and antineutrino absorptions in the outer layers of the
accretion disk is likely to drive a baryonic mass flow off the
disk surface similar to the neutrino-driven winds of newly-formed
hot neutron stars (Woosley \& Baron \cite{woo92}, 
Qian \& Woosley \cite{qia96}).
This wind expands with subrelativistic velocities and has much
lower entropies than the $\gamma$-jets in which the entropy is
extremely high: $s_{\rm jet}\approx (\varepsilon+P)/(Tn_b)\sim
{4\over 3}m_{\rm u}\dot E_{\nu\bar\nu}/(T\dot M_{\rm jet})
\approx {4\over 3}(m_{\rm u}c^2/T)\Gamma \sim 1000\,\Gamma\,
k_{\rm B}$ per nucleon ($\varepsilon$, $P$, $n_b$, $T$ are energy 
density, pressure, baryon density and temperature, respectively, at the
base of the jet, $m_{\rm u}$ is the atomic mass unit, and $\Gamma$ the
Lorentz factor of the jet). In and near the boundary layers 
between the wind region and the jets (hatched in
Fig.~\ref{fig:massloss}) a significant amount of dilute gas
($\sim 10^{-3}\,M_{\odot}$) may be present with such high temperatures
that the entropies could be much larger than in the neutrino-driven
winds of nascent neutron stars. These ejecta from 
hot accretion tori around stellar mass black holes might therefore be
an ideal site for high-entropy r-processing. Future simulations will have
to determine the mass of this material and its contribution to the 
galactic nucleosynthesis of heavy elements.

Despite of the promising results for energy and luminosity of the gamma
jets, our models still suffer from an unpleasant problem. The maximum 
relativistic Lorentz factors
$\Gamma-1 = E_{\nu\bar\nu}/(Mc^2)$, which can be estimated from the
integral energy deposition rates given in Fig.~\ref{fig:annconeNP}
(multiplied by the torus lifetime) and the integral masses of
Fig.~\ref{fig:coneNP}, are only around 0.1--0.5. Even if the
total energy deposited by $\nu\bar\nu$ annihilation 
gets somehow focussed into the jets, this estimate for $\Gamma-1$
increases only to values between 1 and 5. This is roughly two
orders of magnitude below the desired Lorentz factors
of at least 100 which would require a baryon loading of at most
$10^{-7}\,M_{\odot}$ within the low-density cone around the system axis.
The mass of roughly $10^{-5}\,M_{\odot}$ that fills the region
near the axis in our models, however, is only one order of magnitude
above the numerical lower limit of the mass resolution of our 
simulations. The mass resolution is limited because we set a minimum
baryonic mass density of $\rho_{\rm min} = 5\times 10^7\,{\rm g\,cm}^{-3}$
in the cells of our grid. Therefore we
cannot exclude that the large mass of dilute gas
in the surroundings of our tori has purely numerical reasons.
As a consequence, the polar regions of the black hole might not 
clear up in our models as fast as they actually would if the baryons 
were allowed to deplete unlimited because they are sucked in rapidly 
along the axis by the black hole.

There is another, even more important effect which can help opening up
a clean funnel for relativistic plasma jets. In a short but extremely 
luminous outburst of neutrinos right after the neutron star merging,
peak rates of the energy deposition by $\nu\bar\nu$ annihilation
(and additional $\nu_e$
and $\bar\nu_e$ absorption) of more than $10^{52}\,{\rm erg\,s}^{-1}$
are reached. This produces very high energy
densities above the polar caps of the compact massive object at the 
center of the merger remnant (for details, see Ruffert
\& Janka \cite{ruf98b}). The maximum energy deposition
rates are several $10^{32}\,{\rm erg\,cm}^{-3}{\rm s}^{-1}$ at 
heights $|z|\approx 30\,{\rm km}$ in gas with densities between
$10^9\,{\rm g\,cm}^{-3}$ and $10^{10}\,{\rm g\,cm}^{-3}$. This means that
the equivalent of the gravitational binding energy is transferred to
the matter within only one millisecond. The heated plasma must expand 
extremely rapidly, because it experiences only little resistance from
the gas farther out due to 
the very steep density gradient and the correspondingly small density
scale height above the poles of the forming black hole.
The expanding hot gas will push overlying baryons away and
the baryon densities must drop quickly. Recently MacFadyen
\& Woosley (\cite{mac98}) have performed simulations of collapsing,
rapidly rotating and massive stellar cores (``collapsars'').
Although the situation there is less favorable because of the 
huge mass that surrounds the black hole and the accretion torus,
their simulations
demonstrate impressingly the depletion of the axis-near region
due to the outward expansion of the gas in response to
powerful energy input. This creates a much cleaner 
funnel for the pair-plasma jet which is driven by the annihilation
of neutrinos and antineutrinos emitted from the disk around the 
black hole during the subsequent, longer period of accretion.
For all these reasons we do not consider the current problem of baryon 
pollution in the jets as seriously worrying as yet. This problem is 
most likely surmountable by simulations which achieve a better
mass resolution and take into account the feedback effects of the 
$\nu\bar\nu$ annihilation around the torus self-consistently.

Figure~\ref{fig:shortGRB} provides an overview over 
different possible evolution paths of merging binary neutron stars 
and neutron star black hole binaries. Whether these systems lead
to short gamma-ray bursts by $\nu\bar\nu$ annihilation or not
depends on the parameters of the components (masses and spins) and
on the properties of the nuclear equation of state. All events are
accompanied by the emission of gravitational waves (see, e.g., 
Shibata et al.~\cite{shi92,shi93}; Rasio \& Shapiro \cite{ras94};
Zhuge et al. \cite{zhu95}; Ruffert et al.~\cite{ruf96}) and neutrinos
(Ruffert et al.~\cite{ruf97}), and by the ejection of mass with possible 
implications for heavy-element nucleosynthesis (Rosswog et al.~\cite{ros98}
and references therein). Although Kerr
effects associated with the rotation of the black hole can increase
the energies relative to our results, $\nu\bar\nu$ annihilation 
is most likely not able to yield enough energy for long complex
gamma-ray bursts. Magnetic fields and magnetohydrodynamic conversion 
of gravitational energy into radiation may be required to explain
the long and complex gamma bursts by NS-NS and NS-BH mergers 
(Thompson \cite{tho94}, M\'esz\'aros \& Rees \cite{mes97},
M\'esz\'aros et al.~\cite{mes98}), unless this subclass of bursts
is connected with events described by the
``failed supernova'' or ``collapsar'' models (Woosley \cite{woo93a},
Popham et al.~\cite{pop98b}, Paczy\'nski \cite{pac98}, MacFadyen
\& Woosley \cite{mac98}).

An option also not included in the scheme
of Fig.~\ref{fig:shortGRB} is the one that the neutron stars 
collapse to black holes prior to their merging.
This scenario was suggested by Mathews \& Wilson (\cite{mat98a}) and
Mathews et al.~(\cite{mat98b})
on grounds of results of general relativistic simulations which
yield a compression of the stars as they spiral in towards each 
other, instead of tidal stretching as suggested by Newtonian and
post-Newtonian simulations. The possibility of a collapse depends
on the question how close the neutron star
is below the maximum stable mass for the considered
nuclear equation of state. It is currently a matter of vivid discussions
whether the numerical results by 
Mathews \& Wilson (\cite{mat98a}) and Mathews et al.~(\cite{mat98b})
are correct and in agreement with analytical understanding of the effects of
general relativity on the pre-merging evolution 
(see Thorne \cite{tho97} and references therein).

Similarly extreme phases of mass accretion by a black hole
as studied in this paper occur during the 
merging of neutron star black hole binaries. Three-dimensional
hydrodynamic simulations of the dynamical interaction of these binary
stars and of the post-merging evolution were performed by Eberl (\cite{ebe98a})
with the same input physics as described in this paper and in
Ruffert et al.~(\cite{ruf96},\cite{ruf97}), i.e., gravitational-wave and neutrino
emission were taken into account. The results will be published separately
(Eberl et al.~\cite{ebe98b}). They reveal that the accretion disks can be more 
massive ($\la 0.5\,M_{\odot}$) than those obtained in neutron
star neutron star mergers. Since the average accretion rates are
significantly higher, the neutrino luminosities are up to 10 times 
larger, but the accretion phase is shorter, so that effectively
the gamma jets receive about 10 times more energy. The
neutrino luminosities show considerable variation because of
large temporal fluctuations of the accretion rate.

Klu\'zniak \& Lee (\cite{klu98}) and Lee \& Klu\'zniak (\cite{lee98})
have discovered the possibility
of perhaps long periods of episodic mass transfer from the orbiting 
neutron star to the black hole. These are connected with
cycles of orbital decay due to grav\-ita\-tion\-al-\-wave emission and
subsequent widening of the orbit after mass has been donated
to the black hole. The existence of these cycles
might be sensitive to the properties of the nuclear equation
of state (assumed to be a polytropic law in the simulations by
Klu\'zniak \& Lee \cite{klu98} and Lee \& Klu\'zniak \cite{lee98})
and on the binary parameters like the neutron star spin and the 
mass ratio of the neutron star to the black hole.

If neutron stars near the minimum stable mass were
formed after dozens or hundreds of these cycles 
(Klu\'zniak \& Lee \cite{klu98}) and then explode 
(Colpi \& Shapiro \cite{cop89,cop91,cop93};
Sumiyoshi et al.~\cite{sum98}), it is very unlikely that gamma-ray bursts
are produced because large amounts of baryonic 
matter will be ejected into the surrounding space. Even if the mass
donating neutron star were heated by viscous dissipation due to the 
action of tidal forces during the cyclic
phases of decreasing and increasing orbital distance, there is 
hardly a chance to make gamma-ray bursts by the associated 
neutrino emission. 
The heating of the neutron star surface layers by neutrino absorption
would drive a slow baryonic wind rather than create a
relativistically expanding pair-plasma with a low enough baryon
loading (Paczy\'nski~\cite{pac90} and, in particular, Woosley \cite{woo93b}). 

The same argument also applies if the stiffness of the 
supranuclear equation of state prevented the rapidly rotating, 
hot remnant of the binary neutron star merger from collapsing to a
black hole on the dynamical timescale. In that case the remnant would
cool by emitting its huge gravitational binding energy in neutrinos,
which are able to escape from the dense object only on the 
diffusion timescale of several seconds. Again, the surroundings
of the merger would be polluted with the baryons which are driven off
the surface of the remnant in a continuous flow. 

In contrast, the black hole-accretion disk systems provide ideal
conditions for efficient $\nu\bar\nu$ annihilation and the creation
of relativistic outflow (Woosley \cite{woo93a}, 
Moch\-ko\-vitch et al.~\cite{moc93,moc95}).
On the one hand, the neutrino source is 
very compact which ensures large neutrino number densities. On the
other hand, the axis-near region above the poles of the black hole
is an environment with naturally low baryon loading where neutrinos 
and antineutrinos can collide at large angles and annihilate with high
efficiency. Our simulations have confirmed that accreting black holes from
coalescing binary neutron stars can indeed power short ($\sim 0.1\,$s)
gamma-ray bursts with luminosities up to $10^{52}\,{\rm erg\,s}^{-1}$,
provided the jets are beamed into 1/100--1/10 of the sky.
Our models need to be improved and extended with respect to general
relativistic effects, in particular associated with the possible Kerr 
character of the black hole. In addition, it would be interesting to
study cases with different disk viscosities and to include the 
effects of $\nu\bar\nu$ energy deposition self-consistently in the
hydrodynamic calculations.

\bigskip\bigskip

\begin{acknowledgements}
We would like to thank S.~Woosley, C. Fryer and A.~MacFadyen for
inspiring conversations and their motivating interest in our work.
Discussions with M.~Camenzind, W.~Hillebrandt, F.~Meyer, M.~Rees
and R.~Wijers are acknowledged. We apologize to all whose important
work on gamma-ray bursts we have not adequately appreciated by 
references. 
MR is grateful for support by a PPARC
Advanced Fellowship, HTJ acknowledges support by the
``Sonderforschungsbereich 375-95 f\"ur Astro-Teilchenphysik''
der Deut\-schen For\-schungs\-ge\-mein\-schaft.
The calculations were performed at the Rechenzentrum Garching on a
Cray-Jedi(J90) and an IBM-SP2.
\end{acknowledgements}

\end{document}